\documentclass[twocolumn]{article}

\usepackage{amsmath,amssymb,amsfonts}
\usepackage{bm}
\usepackage{algorithmic}
\usepackage{graphicx}
\usepackage{dblfloatfix} 
\usepackage{adjustbox}
\usepackage{multirow}
\usepackage{tabularx}
\usepackage{makecell}
\usepackage{hyperref}
\usepackage{subcaption}
\usepackage{comment}
\usepackage[utf8]{inputenc}
\usepackage[left=2cm, right=2cm, top=3cm, bottom=3cm]{geometry}
\usepackage{authblk} 
\usepackage[margin=10pt,font=small,labelfont=bf,labelsep=endash]{caption}

\hypersetup{
    colorlinks=true,
    linkcolor=blue,
    filecolor=magenta,      
    urlcolor=cyan,
    pdfpagemode=FullScreen,
    }

\usepackage{lastpage}
\usepackage{fancyhdr}
\DeclareCaptionLabelFormat{subfigure}{#1~\thefigure#2}
\title{Thermal investigation of buffer and transition layers in AlGaN HEMT stacks}

\begin{document}

\author{%
\begin{minipage}[t]{0.45\textwidth}
\centering
\textbf{Simon Fernbach\thanks{Corresponding author: simon.fernbach@live.at}}\\
\textit{Microelectronics}\\
\textit{Materials Center Leoben}\\
Leoben, Austria\\
ORCID: 0009-0008-2169-3869
\end{minipage}
\hfill
\begin{minipage}[t]{0.45\textwidth}
\centering
\textbf{Lisa Mitterhuber}\\
\textit{Microelectronics}\\
\textit{Materials Center Leoben}\\
Leoben, Austria\\
ORCID: 0000-0001-7734-5603
\end{minipage}

\par\vspace{1.5em}

\begin{minipage}[t]{0.45\textwidth}
\centering
\textbf{Egbert Zojer}\\
\textit{Institute of Solid State Physics}\\
\textit{Technische Universität Graz}\\
Graz, Austria\\
ORCID: 0000-0002-6502-1721
\end{minipage}
\hfill
\begin{minipage}[t]{0.45\textwidth}
\centering
\textbf{Natalia Bedoya-Martínez}\\
\textit{Simulation}\\
\textit{Materials Center Leoben}\\
Leoben, Austria\\
ORCID: 0000-0001-9824-6129\vspace{20pt}
\end{minipage}
}

\date{\today\vspace{20pt}}

\maketitle

\vspace{60pt}


\begin{abstract}

In this work, we investigate thermal transport in a set of thin-film layers used in modern HEMT designs. Our theoretical analysis employs three phonon-transport models of increasing sophistication, ranging from a relatively simple Debye-approximation model to a first-principles framework that incorporates Density Functional Theory (DFT)–derived harmonic and anharmonic phonon properties. To describe ternary semiconductors, alloy scattering is also considered, and to account for the properties of non-ideal, polycrystalline thin films, scattering at grain and film boundaries is taken into account. For modeling the specific samples at hand, the latter are fitted to thermal conductivity data measured with Time-Domain Thermoreflectance (TDTR). Overall, we find that the first-principles model performs best and produces the most physically consistent results. Special emphasis is placed on establishing a reliable procedure for determining intrinsic transport properties dominated by Umklapp scattering. Building on these, it is then possible to systematically add additional scattering terms to the models and assess how model-inherent factors, such as the treatment of scattering rates or the description of the phonon dispersion, affect model accuracy and overall predictive capabilities.

\end{abstract}


\section{Introduction}

GaN-based high-electron-mobility transistors (HEMTs) are among the most intensively researched technologies in microelectronics \cite{AlGaN_HEMT_review} due to their promising properties for power and optoelectronics applications, such as high breakdown field strength, high operating frequency, large band gap, good thermal conductivity, and small thermal expansion coefficient. Moreover, the high mobility of electrons in the two-dimensional electron gas (2DEG) at the AlGaN/GaN junction enables very high switching frequencies, making these transistors well suited for high-frequency amplifiers \cite{AlGaN_HEMT_review}. However, the small dimensions of modern devices and the use of layered materials introduce additional thermal resistances arising from geometric confinement of phonon states (phonon confinement), reduced phonon mean free paths due to boundary scattering (finite-size effects), and thermal boundary resistance at material interfaces. These effects collectively reduce the effective thermal conductivity. Therefore, a detailed understanding of the thermal properties of the involved materials is essential. The present study combines experiments and simulations to investigate phonon thermal transport in three samples containing thin-film layers commonly used in AlGaN high-electron-mobility transistors \cite{AlGaN_HEMT_review}\cite{AlGaN_buffer_layer_growth}. The thermal conductivity of the samples was measured at different temperatures using time-domain thermoreflectance (TDTR) and then described using three different phonon transport models of increasing complexity.  These include (1) the relatively simple Debye-approximation model \cite{Callaway1959}, which assumes a Debye-like linear dispersion; (2) a model that uses the ab initio calculated dispersion for the three acoustic phonon branches branches, but still ignores optical phonons; and (3) a fully ab initio-based model that incorporates both the full phonon dispersion and phonon scattering rates derived from first principles. We compare the three models to identify their shortcomings, rank their performance, and contrast our findings with experimental data from the literature. To that aim, additional scattering processes like alloy scattering and boundary scattering at grain boundaries and thin-film surfaces are also considered. The measurements and modelling show that thermal conductivities of the investigated AlGaN HEMT layers are strongly affected by their microstructure, alloy composition, and thickness. The low conductivity of the AlN transition layer is attributed primarily to its highly polycrystalline structure, whereas the AlGaN layers exhibit pronounced size effects arising from the combined influence of alloy and boundary scattering. Comparison of different transport models further demonstrates that simplified descriptions of scattering can, to some extent, compensate for their inherent limitations through fitted parameters, whereas a fully ab initio treatment, supplemented by appropriate sample-specific corrections, provides a more consistent and robust description across the investigated layers.

\begin{figure*}[htp!]
\begin{subfigure}{0.329\textwidth}
\centering
\includegraphics[width=\textwidth,trim=0 0 380 120,clip]{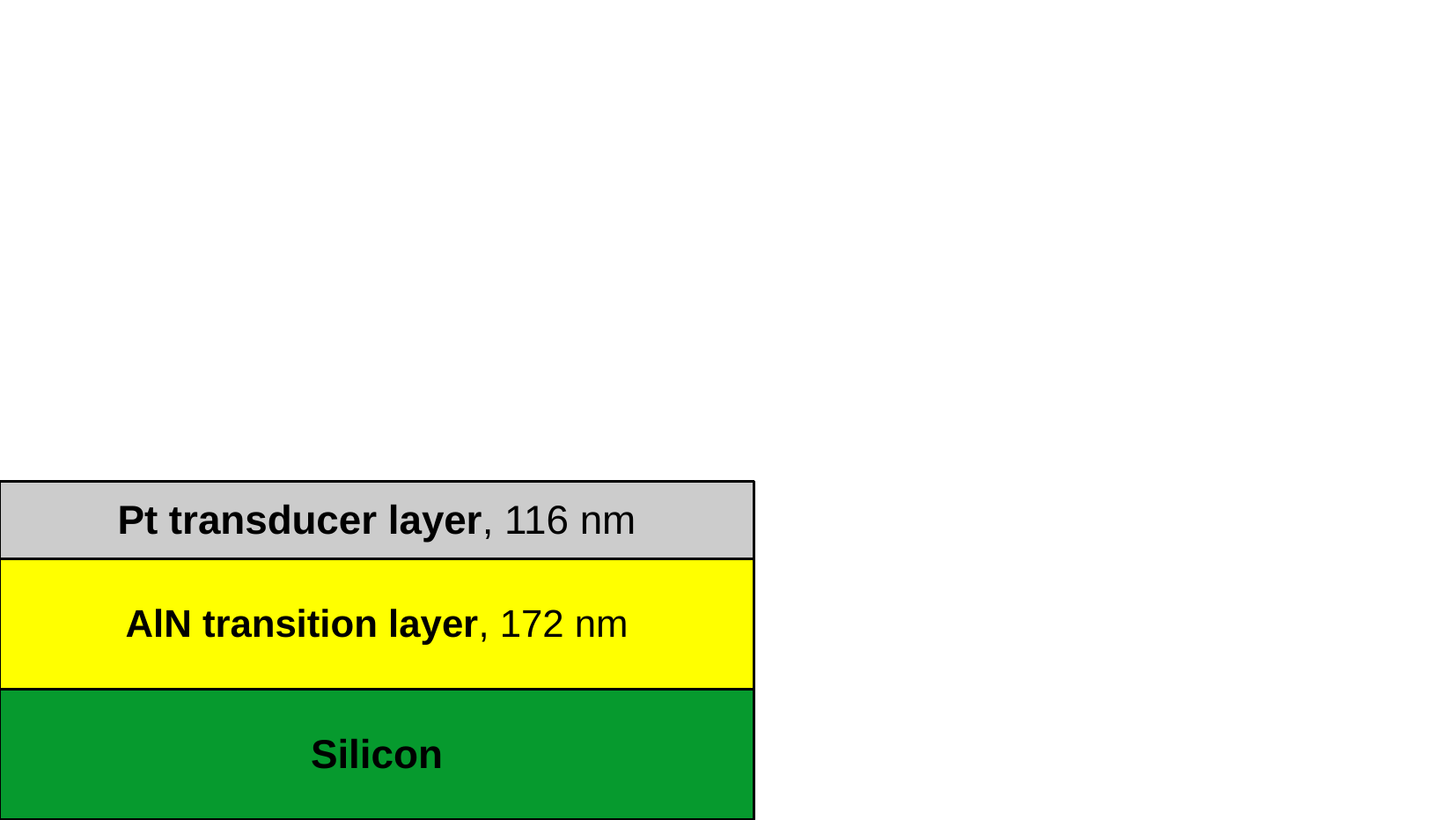}
\caption*{\Large\textbf{Transition layer}}
\end{subfigure}
\begin{subfigure}{0.329\textwidth}
\centering
\includegraphics[width=\textwidth,trim=0 0 380 120,clip]{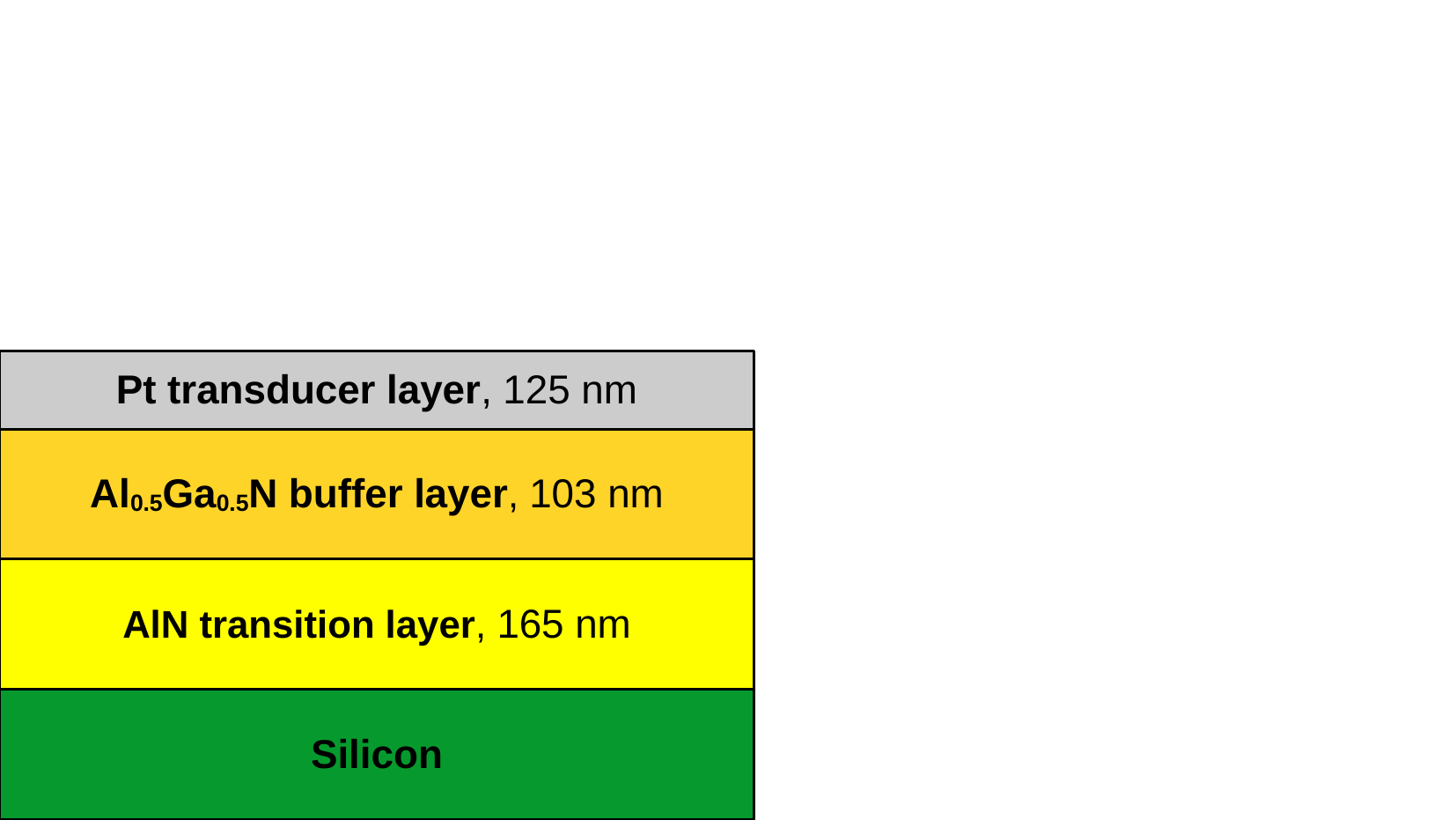}
\caption*{\Large\textbf{buffer layer}}
\end{subfigure}
\begin{subfigure}{0.329\textwidth}
\centering
\includegraphics[width=\textwidth,trim=0 0 380 120,clip]{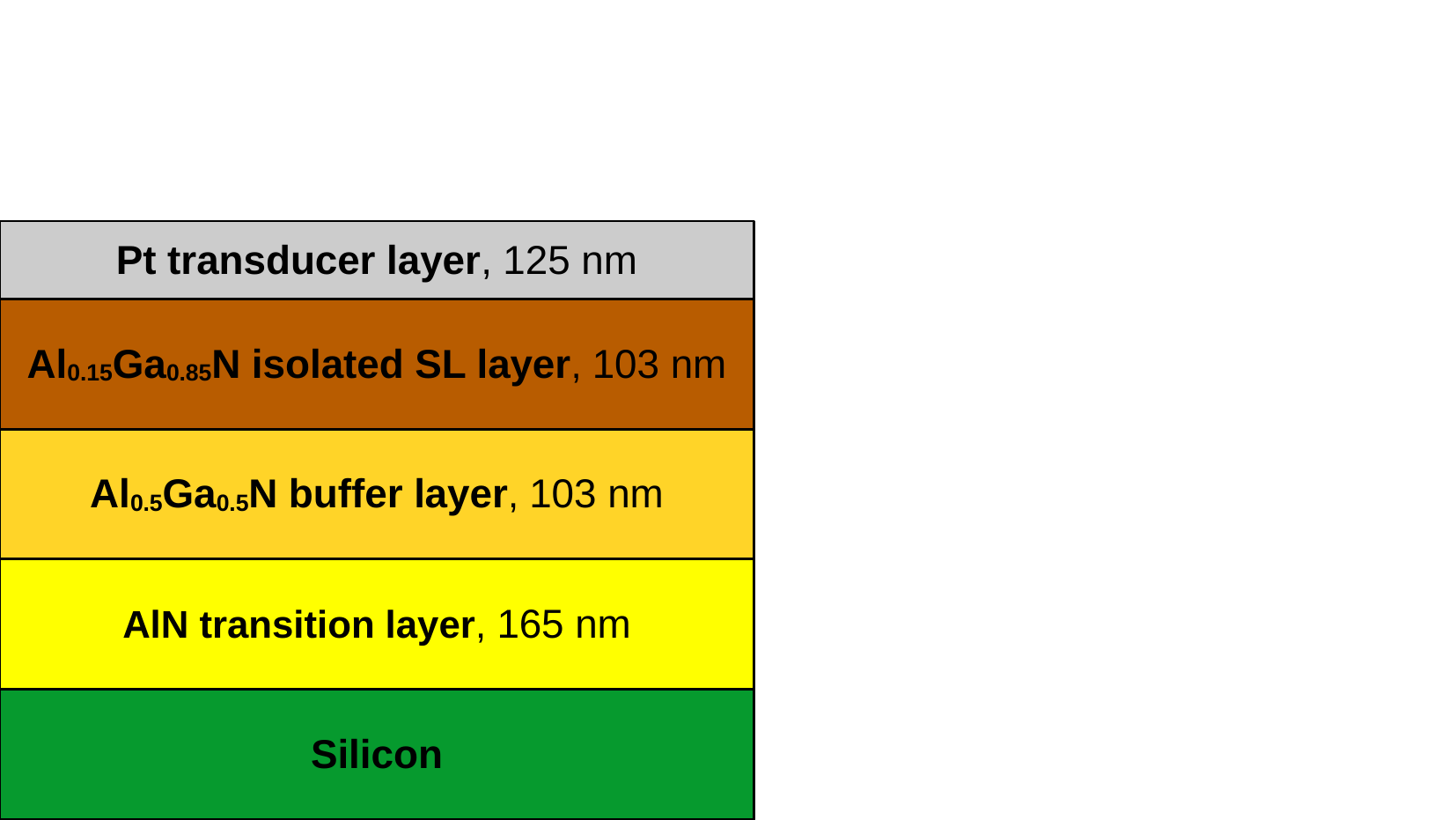}
\caption*{\Large\textbf{isolated SL layer}}
\end{subfigure}
\caption{Schematic illustration of the samples investigated in this work. From left to right, each sample stack adds one additional layer on top of the previous configuration. The left panel shows the AlN transition layer deposited on a Si substrate. The middle panel includes the Al$_{0.5}$Ga$_{0.5}$N buffer layer, and the right panel contains and Al$_{0.15}$Ga$_{0.85}$N layer that is a part of an AlN/AlGaN superlattice. The platinum transducer layer added on top, is required for TDTR measurements. Layer thicknesses were measured using scanning electron microscopy (SEM).}
\label{fig:samples_schematics}
\end{figure*}

\section{Methodology}
\subsection{Samples}\label{sec:samples}
\autoref{fig:samples_schematics} shows the samples analyzed in this study. By design, each new sample adds one additional layer on top of the previous sample. This design allows us to determine the thermal properties of each layer individually instead of fitting the parameters of multiple layers simultaneously, making the evaluation more reliable. All samples are deposited on a Si substrate and are covered with a platinum layer required for TDTR measurements. The reported layer thicknesses were determined by SEM, as described in Sec. S1. The first sample consists of a 172 nm AlN transition layer on top of the Si substrate. The second sample adds a 103 nm thick Al$_{0.5}$Ga$_{0.5}$N buffer layer, and the third sample adds a 407 nm thick Al$_{0.15}$Ga$_{0.85}$N layer. These layers are commonly used in AlGaN/GaN HEMTs. The AlN transition layer helps alleviate the lattice-mismatch-induced stress between the Si substrate and the GaN channel layer. The Al$_{0.5}$Ga$_{0.5}$N buffer layer further reduces residual stress, improves the crystalline quality of the channel layer, traps defects, and increases the breakdown voltage \cite{AlN_transition_layer}\cite{AlGaN_buffer_layer_growth}\cite{AlGaN_HEMT_review}. The Al$_{0.15}$Ga$_{0.85}$N layer is part of a superlattice proposed as an alternative to the conventional AlGaN buffer layer in HEMT designs and will be referred to as the isolated SL layer throughout this work.


\subsection{Time-Domain Thermoreflectance}\label{sec:TDTR}

The thermal conductivity of the thin films was measured using Time-Domain Thermoreflectance (TDTR) with the commercially available PicoTR system by Netzsch \cite{PicoTR}. The setup employs a pump-probe configuration, in which the pump beam heats the sample surface, and a delayed probe pulse measures the resulting temperature response. This measurement is performed by detecting changes in surface reflectance as a function of temperature. Because the signal level is often comparable to the noise floor, lock-in detection is used. For this purpose, the pump beam is modulated at a frequency of 200 kHz. Thermal properties are extracted by fitting a heat conduction model \cite{Cahill2004} to the signal recorded by the lock-in detector. This analysis was performed with a software tool developed at Materials Center Leoben (MCL) documented in \cite{Krainz2021}. The setup controls the delay between the pump and probe beams using an electrical delay stage. The pump and probe beams operate at wavelengths of 1550 nm and 775 nm, respectively, with spot sizes of 45 $\mu m$ and 25 $\mu m$. The measurement chamber contains an integrated oven for temperature-dependent experiments. During these expriments, the chamber is flushed with argon to prevent oxidation. Measurements were performed at temperatures of 298, 373, 474, 573, and 673 K. The AlN transition layer was additionally measured at 773 K.

\subsection{Lattice thermal conductivity}\label{sec:lattice_thermal_conductivity}

Several methods are available for calculating the thermal conductivity of crystalline materials \cite{Ziman2001}. A common strategy, that is often employed for modelling thermal conductivity measurements of crystalline systems that contain various defects, starts from a closed-form analytical expression for the lattice thermal conductivity tensor $\kappa_{\alpha\beta}$, \autoref{eq:lattice_thermal_conductivity} \cite{Turney2009}\cite{Togo2015}. It can be derived as a steady-state solution of the linearized Peierls-Boltzmann equation under the relaxation time approximation (RTA) and reads \cite{Karamitaheri2013}
\begin{equation}\label{eq:lattice_thermal_conductivity}
\resizebox{0.9\columnwidth}{!}{$
\kappa_{\alpha\beta}=\frac{1}{V}\sum\limits_{\vec{q}}\sum\limits_{\nu=1}^{3N_{unit}} C_\text{ph}(\vec{q},\nu,T)\,{v}_\text{g}^{\alpha}(\vec{q},\nu)\,{v}_\text{g}^{\beta}(\vec{q},\nu)\,\tau_{\text{tot}}(\vec{q},\nu),$}
\end{equation}
where $\text{v}^{\alpha}_{\text{g}}(\vec{q},\nu)$ are the group velocities, $\alpha$ and $\beta$ are Cartesian indices, $V$ is the volume of the crystal, $\vec{q}$ and $\nu$ label the phonon wave vectors and the individual phonon bands, $N_{\text{unit}}$ is the number of atoms in the unit cell, and the sums run over all wave vectors in the Brillouin zone and over all phonon branches. The phonon heat capacity $C_{\text{ph}}(\vec{q},\nu)$ in \autoref{eq:lattice_thermal_conductivity} is given by

\begin{equation}\label{eq:phonon_heat_capacity}
\resizebox{0.9\columnwidth}{!}{$
C_{\text{ph}}\left(\vec{q},\nu\right) = k_b^{-1}\left(\frac{\hbar\,\omega\left(\vec{q},\nu\right)}{k_BT}\right)^2\frac{\exp\left(\frac{\hbar\,\omega\left(\vec{q},\nu\right)}{k_BT}\right)}{\left(\exp\left(\frac{\hbar\,
\omega(\vec{q},\nu)}{k_BT}\right)-1\right)^2}$}
\end{equation}

and the total phonon relaxation times $\tau_{\text{tot}}(\vec{q},\nu)$ in \autoref{eq:lattice_thermal_conductivity} are determined from contributions of different scattering sources using Matthiessen's rule \cite{Ziman2001}:

\begin{equation}\label{eq:matthiessens_rule}
\text{w}_{\text{tot}}(\vec{q},\nu) =\sum_{s} \text{w}_{\text{s}}(\vec{q},\nu) = \sum_{\text{s}}\tau^{-1}_{\text{s}}(\vec{q},\nu) = \tau^{-1}_{\text{tot}}(\vec{q},\nu),
\end{equation}

This empirical rule assumes independent scattering sources. It states that the total scattering rates $\text{w}_{\text{tot}}\left(\vec{q},\nu\right)$, and consequently, the total inverse scattering times $\tau^{-1}_{\text{tot}}\left(\vec{q},\nu\right) = \text{w}_{\text{tot}}\left(\vec{q},\nu\right)$, can be calculated as the sum of the individual scattering rates $\text{w}_{\text{s}}\left(\vec{q},\nu\right)$ from different sources, where $s$ labels each source. Equation \autoref{eq:lattice_thermal_conductivity} serves as the starting point for the models considered in this work, which provide a straightforward and computationally efficient description of phonon-mediated thermal transport in semiconductors when combined with suitable descriptions of the phonon band structure $\omega(\vec{q},\nu)$ and scattering rates $\text{w}_(\vec{q},\nu)$ (or phonon lifetimes). This approach has been successfully applied to model thermal conductivity measurements in a wide variety of systems, including single-crystalline bulk materials \cite{Callaway1959}\cite{Tiwari1971}, thin films \cite{Park2014}, alloys \cite{Abeles1963}, and samples containing various types of defects \cite{Williams1987}\cite{Srivastava2009}. \\

Compared with the full solution of the Peierls-Boltzmann equation, \autoref{eq:lattice_thermal_conductivity} omits the collective redistribution of phonon momentum and thus incorrectly treats Normal (N) scattering processes. This is, however, no serious issue when Umklapp, impurity and, other scattering types dominate, as is the case here (see Sec. S2 in the SI).\\

In the following sub-sections, we summarize the models used in this work and their assumptions regarding the phonon band structure and scattering rates.


\subsection{Debye-approximation model (Deb)}\label{sec:Deb}

The simplest and most widely used approach is the Callaway model \cite{Callaway1959}. In its original formulation, it assumes a linear, Debye-like dispersion for the acoustic phonon branches and expresses the thermal conductivity as the sum of two terms: the first accounts for resistive scattering processes, while the second corrects the effects of incorrectly treating Normal scattering processes as resistive in the relaxation-time approximation. In the literature, however, the term Callaway model is also commonly used for the corresponding formulation without the Normal-process correction. Here, as discussed above, the impact of Normal scattering is negligible compared with the other scattering mechanisms considered. We therefore use the single-term form of the Callaway model, making the results directly comparable to those obtained from the lattice thermal conductivity expression given in \autoref{eq:lattice_thermal_conductivity} and the models derived from it. To avoid confusion, we will refer to this model as the Debye-approximation model (Deb).\\

Within this model, the phonon band structure is approximated by a single linear, Debye-like, three-fold degenerate dispersion for the three acoustic branches, while the optical branches are neglected. The phonon dispersion is expressed as $\omega(\vec{q},\nu) = \text{v}_{\text{s}}\vert\vec{q}\vert$, where v$_{\text{s}}$ is the average sound velocity of the acoustic branches. The isotropic nature of this dispersion allows the lattice thermal conductivity to be reformulated in terms of the density of states (DOS), $D(\omega)$, as an integral over phonon frequencies.
\begin{equation}\label{eq:LTC_callaway}
\kappa_{\text{Deb}}(T) = \frac{1}{3}\int_{\omega=0}^{\omega_D} D\left(\omega\right)C_{\text{ph}}\left(\omega,T\right)v_s^2\tau_{\text{tot}}\left(\omega,T\right)d\omega,
\end{equation}
here, $D(\omega)=\frac{3\omega^2}{2\pi^2v_s^3}$ is the phonon DOS in the Debye approximation, $C_{\text{ph}}(\omega,T)$ is the phonon heat capacity (\autoref{eq:phonon_heat_capacity}), and $\omega_D = v_s(6\pi^2 n)^{1/3}$ is the Debye frequency. A detailed derivation of \autoref{eq:klemens_model} is provided in Sec. S3. The key quantity characterizing the phonon band structure in this model is the sound velocity, which corresponds to the slope of the isotropic linear dispersion. This parameter is commonly taken from the literature or treated as an additional fit parameter. The latter approach is undesirable because it adds model flexibility without a clear physical basis, making it harder to assess the model's predictive capability. For this reason, and to facilitate comparison with the first-principles-informed models, we determine the sound velocity from the DFT-calculated phonon dispersion. For each acoustic branch, the magnitude of the phonon group velocity is averaged over the approximately linear region near the Brillouin-zone center. Specifically, only grid points within a distance of 10\% of the magnitude of the largest reciprocal lattice vector from the zone center are included in the average. Scattering rates are calculated using the analytical expressions described in \autoref{sec:scattering_rates} and combined according to Matthiessen's rule \autoref{eq:matthiessens_rule}.


\begin{table*}[htbp]
\centering
\caption{Summary of the models utilized in this study. The table summarizes the assumed phonon dispersion, including the functional form and number and type of independent branches considered (acc. = acoustic, opt. = optical; in the Deb model, all three acoustic branches are assumed to be degenerate), the approaches used to describe intrinsic scattering processes such as Umklapp and alloy (point-defect) scattering, the corresponding fit parameters, and the expression used to calculate the thermal conductivity.}
\resizebox{\textwidth}{!}{
\begin{tabular}{|c|c|c|c|c|c|c|c|}
\hline
\multirow{2}{*}{\textbf{Model}} & \multicolumn{2}{c|}{\textbf{Phonon dispersion}} & \multicolumn{2}{c|}{\textbf{Scattering rates}} & \multicolumn{2}{c|}{\textbf{fit param.}} & \textbf{Thermal} \\
\cline{2-7}
& \textbf{description} & \textbf{branches} & \textbf{Umklapp} & \textbf{Alloy} & \textbf{Umklapp} & \textbf{alloy} & \textbf{conductivity} \\
\hline
Deb (\autoref{sec:Deb}) & linear & 1 $\times$ acc. & \autoref{eq:umklapp_scattering} & \autoref{eq:klemens_model} & $\gamma$, $b$ & - & \autoref{eq:LTC_callaway}\\
AcB (\autoref{sec:AcB}) & DFT & 3 $\times$ acc. & \autoref{eq:umklapp_scattering} & \autoref{eq:klemens_model} & $\gamma$, $b$ & - & \autoref{eq:lattice_thermal_conductivity}\\
AbI (\autoref{sec:AbI}) & DFT & 3 $\times$
acc. + opt. & DFT (3-ph) & DFT (\autoref{eq:tamura_model}) & - & - & \autoref{eq:lattice_thermal_conductivity}\\
\hline
\end{tabular}}
\label{tab:model_overview}
\end{table*}


\subsection{Acoustic band model (AcB)}\label{sec:AcB}

This model employs first-principles-calculated data (see \autoref{sec:first_principles_calculations}) for the acoustic phonon branches, while retaining the analytical expressions for the scattering rates described in \autoref{sec:scattering_rates}. The lattice thermal conductivity is then calculated according to \autoref{eq:lattice_thermal_conductivity}. The key differences of this approach compared to a fully first-principles-based evaluation (see below) are that optical phonons are neglected (for reasons explained at the end of the paragraph) and that scattering rates do not need to be derived from first-principles calculations. This can be an advantage, as scattering rates (as intrinsically anharmonic quantities) are much more costly to compute with DFT than the phonon dispersions. Moreover, a direct comparison between this model and the fully first-principles-based model provides insight into the relative importance of phonon dispersion and explicitly calculated scattering rates in determining the accuracy of the predicted thermal conductivity. Since the actual phonon dispersions in the studied materials are anisotropic, the thermal conductivity tensor $\kappa_{\alpha\beta}$ does not reduce to a scalar quantity $\kappa$, like in the Debye-approximation model. Instead, the average thermal conductivity of the isotropic crystal is calculated as the trace of the thermal conductivity tensor $\kappa = \frac{1}{3}\sum_{\alpha}\kappa_{\alpha\alpha}$. Both Umklapp and alloy scattering are described using the expressions proposed by Klemens, given in \autoref{eq:umklapp_scattering} and \autoref{eq:klemens_model}, respectively. These expressions have been  derived under the assumptions of low frequencies and linear dispersion, which makes them unsuitable for the optical branches \cite{Klemens1955}\cite{Callaway1959}. Therefore, optical phonon branches still cannot be included in the description.


\subsection{Ab-initio-derived model (AbI)}\label{sec:AbI}

Finally, the ab initio-derived model (AbI) employs the full first-principles phonon dispersion, accounting for both acoustic and optical branches, along with Umklapp and alloy scattering rates calculated from first principles using the ShengBTE code \cite{ShengBTE}. Consequently, when Eq. 1 is evaluated using only the scattering mechanisms also accounted for in ShengBTE, the model reproduces the thermal conductivity calculated directly by ShengBTE within the relaxation time approximation (see Fig. S5 in the SI). \\

An overview of the models used in this work and their underlying assumptions is provided in \autoref{tab:model_overview}.


\subsection{First-principles calculations} \label{sec:first_principles_calculations}

The lattice thermal conductivity and harmonic and anharmonic phonon properties are calculated using the ShengBTE software package \cite{ShengBTE} and DFT-derived second-order (FC2) and third-order (FC3) force constants. The calculated phonon properties and scattering rates directly inform the ab-initio-derived model described in the preceding section. The acoustic phonon bands are used in the AcB model, while their low-wavevector slopes provide the sound velocities required by the Debye-approximation model. All ShengBTE calculations are performed within the relaxation-time approximation (RTA), using a \(30\times30\times30\) phonon wavevector grid. The convergence of the grid was verified (see Fig. S4 in the SI).\\

Second-order force constants are obtained within density functional perturbation theory using the open-source density functional theory (DFT) package Quantum Espresso (QE) \cite{QE1}\cite{QE2}. They are calculated on a 5x5x5 q-point grid, which corresponds to a 5x5x5 real-space supercell after a Fourier transform. Finally, the force constants are converted to a format compatible with Phonopy \cite{Phonopy} for use in ShengBTE. The third-order force constants are calculated with the finite displacement method for a $2\times2\times2$ supercell, with anharmonic interactions included up to the third-nearest-neighbor shell. The choice of supercell size was verified by comparing the resulting $k_{zz}$ with that obtained using a $2\times2\times3$ supercell, for which a negligible difference was observed (see Fig. S3 in the SI). This is consistent with previous first-principles calculations by Lindsay et al. \cite{Lindsay2013}, who observed a weak dependence of the thermal conductivity of Si on supercell size. Furthermore, their calculations for GaAs, another III–V semiconductor like AlN and GaN, showed that the thermal conductivity is relatively insensitive to extending the anharmonic interaction cutoff beyond the third-nearest-neighbor shell, despite the polar nature of these materials. Forces in the displaced supercells are obtained with QE. For the QE calculations, we used the PBESol exchange-correlation functional \cite{PBESol} together with pseudopotentials from the Standard Solid-State Pseudopotentials library (\href{https://www.materialscloud.org/discover/sssp/table/precision}{SSSP}, precision version). Electronic-structure calculations were performed using a \(10\times10\times10\) grid for Brillouin-zone sampling, with the electronic self-consistency converged to an energy threshold of \(10^{-12}\) Ry. The wavefunction cutoff was set to 90 Ry (1225 eV), while the charge-density cutoff was set to four times the wavefunction cutoff for aluminium and nitrogen and to eight times the wavefunction cutoff for gallium. These cutoff values are recommended by the authors of the pseudopotential library based on convergence tests reported on their webpage. Additional convergence tests for the wavefunction cutoff and electronic reciprocal-space sampling grid are given in Sec. S4 of the SI. They demonstrate that the chosen values yield well-converged results. 

\subsection{Scattering rates}\label{sec:scattering_rates}

Following the description of the different levels of approximation employed to calculate the thermal conductivity, an overview of the scattering mechanisms relevant to the systems studied in this work and their technical implementation is provided. The scattering processes comprise anharmonic phonon scattering (3-phonon Umklapp and Normal scattering), which is always present,  boundary scattering from both thin films and grain boundaries, and point-defect scattering in alloys (Al$_{x}$Ga$_{1\!-\!x}$N).\\

\subsubsection{Normal scattering}\label{sec:normal_scattering}

Normal scattering refers to anharmonic phonon-phonon interactions that conserve the total phonon quasi-momentum. Thus, it does not directly affect the thermal conductivity through the scattering rates, but influences it indirectly by redistributing phonon momenta and energies to states that are more or less likely to scatter via Umklapp processes \cite{Callaway1959}. At the relatively high temperatures considered in this work ($\geq 300,$K), normal processes are assumed to be negligible. Our ab initio calculations, indeed, indicate that their contribution is small, resulting in a relative change in thermal conductivity between the RTA-based solution and a full iterative solution of 7\% for AlN and (due to the significant alloy scattering) less than 0.1\% for AlGaN. This is insofar relevant, as solving the Peierls-Boltzmann equation in the RTA (as done throughout the rest of this manuscript) does not correctly account for N-processes (see above).\\ 
 
\subsubsection{Umklapp scattering}\label{sec:umklapp_scattering}
Thus, Umklapp scattering is the dominant phonon–phonon scattering mechanism in the materials considered in this work and plays a key role in determining their thermal conductivity. Here, we consider two approaches to describe Umklapp scattering. The first involves explicitly calculating the scattering rates for all phonon modes using second- and third-order force constants obtained from DFT, as detailed in \autoref{sec:first_principles_calculations}. The second approach employs an analytical expression originally proposed by Klemens \cite{Klemens1955}\cite{Klemens1956}. This expression is based on the RTA and assumes a linear phonon dispersion. Consequently, it applies only to acoustic phonon branches and, strictly speaking, to the low-wavevector regime where the linear dispersion approximation holds. The corresponding expression is given by
\begin{equation}\label{eq:umklapp_scattering}
\resizebox{0.9\columnwidth}{!}{$
\text{w}^{\text{\tiny Klemens}}_{\text{\tiny Umklapp}}(\vec{q},\nu) =\frac{\hbar\gamma^2}{Mv_{g}(\vec{q},\nu)^2T_\text{D}}\omega^2(\vec{q},\nu)\,T\,\exp\left(-\frac{T_\text{D}}{bT}\right),$}
\end{equation}
with $\gamma$ denoting the macroscopic (bulk averaged) Grüneisen parameter, $M$ the average atomic mass, $v_g(\vec{q},\nu)$ the group velocity, $T_\text{D}$ the Debye temperature and $b$ serves as a fit parameter. \\

The group velocity and Debye temperature are determined from the first-principles phonon dispersion. Following \cite{Slack2006}, the Debye temperature is defined in terms of the maximum vibrational frequency of the respective phonon branch. The specific functional form of \autoref{eq:umklapp_scattering} can be understood from the underlying Debye-type approximation: the ($\omega^2$) dependence reflects the frequency dependence of the three-phonon scattering phase space and interaction strength, resulting in an increasing scattering rate with phonon frequency. The exponential factor describes the thermal activation of Umklapp processes, which are suppressed at low temperatures due to the limited population of phonons with sufficiently large wave vectors.\\


\subsubsection{Alloy and point-defect scattering}\label{sec:alloy}

While the modelling of phonon transport in periodic crystalline materials such as AlN is comparatively straightforward, the situation is considerably more complex for the disordered alloys like Al$_{0.5}$Ga$_{0.5}$N and Al$_{0.15}$Ga$_{0.85}$N, as they naturally lack the long-range periodicity required to properly define phonons in the first place. For these systems, we adopted the following strategy: first, the disordered systems are replaced by an effective periodic material, a procedure known as the virtual crystal approximation (VCA) \cite{Abeles1963}. The effective material is described by linearly interpolating the lattice constants, atomic masses, and force constants of the constituent compounds according to the alloy fraction. This strategy should not be confused with the procedure in which the pseudopotentials of the alloying elements are interpolated, which is also commonly referred to as the virtual crystal approximation \cite{Vanderbilt2000}. \\

However, the approach described above neglects disorder-induced fluctuations in the lattice constants, masses, and force constants present in the actual material. Since these fluctuations can be a significant source of phonon scattering, they must be incorporated as a perturbation on top of the VCA description. Two commonly used approaches are the Klemens model \cite{Klemens1955}, which was originally developed for use with a Debye-like description of the phonon spectrum, and the Tamura model \cite{Tamura1983}, which can be applied to realistic phonon dispersions and, thus, can also be applied when considering the optical branches. Although neither approach is fully rigorous, both   have been successfully applied to describe thermal transport in a variety of alloys. The Klemens model is commonly used in conjunction with a Debye-type description of thermal transport to interpret thermal conductivity measurements \cite{Abeles1963,Balandin2005,Gurunathan2020_2,DQTran2020,Filtaova2020}, while the Tamura model is regularly employed in first-principles calculations of thermal transport in alloys \cite{Garg2011,Lindsay2012_1,Lindsay2012_2,Tian2012}. Because of their perturbative nature, these expressions are valid only within certain limits. They tend to overestimate scattering rates at high frequencies, while providing fairly accurate predictions in the low-frequency regime \cite{Malakkal2024,Thebaud2022}. This likely explains their success in describing alloys, where most heat is carried by low-frequency phonons.\\

Another factor worth discussing is which aspects of the disorder are captured by these models. Introducing chemical disorder into a crystal leads to two kinds of perturbations: a mass perturbation and a perturbation of the force constants, which stems from changes in bond lengths and bond strengths. While mass perturbations are captured by both models and the Klemens model can, in principle, account for force-constant perturbations, the Tamura model cannot \cite{Protik2024}. However, rigorous non-perturbative investigations have shown that scattering contributions from force-constant perturbations are largely frequency independent \cite{Malakkal2024}. Therefore, they cannot be adequately represented by the strongly frequency-dependent scattering expression used in the Klemens model. As a consequence, the impact of force-constant fluctuations is entirely neglected in the present work.\\

The decision to neglect force-constant disorder is further supported by an analysis by Arrigoni et al. \cite{Arrigoni2018}, who investigated under which conditions a first-principles description of thermal transport in alloys based on the VCA and the Tamura model is adequate and when it breaks down. Not surprisingly, they found that the approach performs well in predominantly mass-disordered systems, while it fails in materials with significant bond-length and bond-strength disorder. This is in particular the case for alloys with polar bonds, where long-range Coulomb interactions combined with large random bond-length fluctuations can lead to substantial force-constant perturbations. In fact, a material for which an approach combining the VCA with the Tamura model is known to fail is InGaAs \cite{Arrigoni2018}. There, the large difference in bond length between Ga--As and In--As ($\sim 7.2\%$) leads to strong scattering contributions from force-constant disorder. In contrast, for Al$_{x}$Ga$_{1\!-\!x}$N (relevant to this study), the difference in bond length between Al--N and Ga--N is much smaller ($\sim 2.4\%$), suggesting that force-constant perturbations are comparatively weak. At the same time, the mass contrast between Al and Ga is exceptionally large, implying that mass-disorder scattering should dominate alloy scattering. Consequently, a VCA-based description that neglects force-constant disorder is expected to remain applicable in the alloys studied here. \\

The Klemens expression for the defect scattering rate is given by

\begin{equation}
\begin{aligned}\label{eq:klemens_model}
\text{w}^{\text{\tiny Klemens}}_{\text{\tiny defect}}(\vec{q},\nu) &= \frac{\text{V}_{\text{atom}}\Gamma}{4\pi v_p(\vec{q},\nu)v_{g}(\vec{q},\nu)^3}\omega^4(\vec{q},\nu) \\
\Gamma &= \sum_{i} f_i\left(\frac{M_i-\overline{M}}{\overline{M}}\right)^2 
\end{aligned}
\end{equation}
Here, $V_{\text{atom}}$ is the atomic volume, $i$ labels the atomic species in the material, $f_i$ denotes their concentrations, and $M_i$ their masses. $\overline{M}$ is the average atomic mass in the primitive cell, and $\Gamma$ determines the strength of scattering. As seen from the expression for $\Gamma$, the Klemens model treats the entire unit cell as an effective mass, so deviations are defined relative to the average atomic mass of the unit cell. We use the Klemens model for the Debye-approximation model \autoref{sec:Deb} and the Acoustic band model \autoref{sec:AcB}. \\

The scattering rates in the Tamura model are given by
\begin{flalign}\label{eq:tamura_model}
\text{w}^{\text{Tamura}}_{\text{\tiny defect}}(\vec{q},\nu) &= \frac{\pi\omega^2(\vec{q},\nu)}{2N}\sum_{n}\sum_{(\vec{q}^{\,\prime},\nu^{\prime})}g_{n}\vert\vec{e}_n(\vec{q},\nu) &&\notag\\
&\boldsymbol{\cdot}\vec{e}_n(\vec{q}^{\,\prime},\nu^{\,\prime})\vert^2\delta\left(\omega(\vec{q},\nu)-\omega(\vec{q}^{\,\prime},\nu^{\prime})\right) &&\\
g_n &= \sum_i f_{n,i}\left(\frac{M_{n,i}-\overline{M}_{n}}{\overline{M}_{n}}\right)^2 &&\notag
\end{flalign}

where $n$ labels the lattice sites in the primitive cell, $i$ the atomic species, and $f_{n,i}$ and $M_{n,i}$ are their concentrations and masses. $\overline{M}_n = \sum_i f_{n,i}M_{n,i}$ is the average mass at lattice site $n$, $N$ is the number of unit cells in the crystal, $\vec{e}_n(\vec{q},\nu)$ are the phonon eigenvectors, and $\delta\left(\omega(\vec{q},\nu)-\omega(\vec{q}^{,\prime},\nu^{\prime})\right)$ is a Dirac delta function that ensures energy conservation. In contrast to the Klemens model, the parameter $g_n$, which determines the strength of the scattering, is lattice site resolved. The Tamura model can account for the full phonon band structure of a material, as reflected by the phonon eigenvectors appearing explicitly in the above expression. It is therefore combined with the ab initio model to describe thermal transport of alloys. In practice, the scattering rates corresponding to the Tamura model are computed using ShengBTE \cite{ShengBTE}, by providing the scattering strength parameters $g_n$ for the alloy under investigation.\\


\subsubsection{Boundary scattering}\label{sec:boundary_scattering}

To account for phonon scattering at thin-film boundaries, we use a modified Casimir expression \cite{Casimir1938}. In the Casimir limit, the scattering rates are determined by the average distance between scattering events, described by a characteristic length. The Casimir limit assumes perfectly diffuse scattering at all boundaries, which may not accurately reflect the actual situation. To account for the possibility that some phonons are not scattered resistively, either because they do not interact with the boundary or because they are scattered specularly, we introduce an average specularity parameter $\text{p}$. The scattering rates due to thin-film boundaries are then given by
\begin{equation}\label{eq:boundary_scatteringA}
\text{w}_{\text{\tiny FB}}(\vec{q},\nu) = (1-\mathrm{p}_{\text{\tiny FB}})\frac{\text{v}_{g}(\vec{q},\nu)}{\text{L}_{\text{\tiny FB}}},
\end{equation}
where L$_{\text{\tiny FB}}$ is the characteristic length, set to the film thickness for heat transport parallel to the film boundaries. The subscript refers to film boundary scattering (in contrast to grain boundary; see below). Setting p$_{\text{\tiny FB}}=0$ recovers the Casimir limit, while $\text{p}_{\text{\tiny FB}}=1$ corresponds to a completely non-resistive situation, such that boundary scattering does not contribute to the thermal resistance. The parameter $\text{p}_{\text{\tiny FB}}$ is treated as a fit parameter.

Finally, grain boundaries are assumed to scatter phonons diffusely, such that the phonon direction is randomized upon interaction with a boundary. Since the grains are assumed to consist of the same material and to be in thermal equilibrium, the Diffuse Mismatch Model \cite{Pohl1989} predicts a transmission probability of 50\%, with the remaining 50\% of phonons being reflected. Thus a value of p$_{\text{\tiny GB}}=0.5$ is used for the specularity of grain boundaries. The characteristic distance between successive grain-boundary scattering events is approximated by the average grain size. A geometry-dependent factor relating the grain size to the mean scattering length is neglected due to the lack of detailed experimental information on the samples. Under these conditions, the grain-boundary scattering rates can be expressed as
\begin{equation}\label{eq:boundary_scatterinB}
\text{w}_{\text{\tiny GB}}(\vec{q},\nu) = 0.5\frac{\text{v}_{g}(\vec{q},\nu)}{\text{L}_{\text{\tiny GB}}},
\end{equation}
In this equation, L$_{\text{\tiny GB}}$ represents the average grain size and is treated as a fit parameter in the present context.


\begin{figure*}[ht]
\centering
\begin{subfigure}{0.32\textwidth}
\caption*{\large\textbf{AlN}}
\includegraphics[width=\textwidth,trim=0 0 0 0,clip]{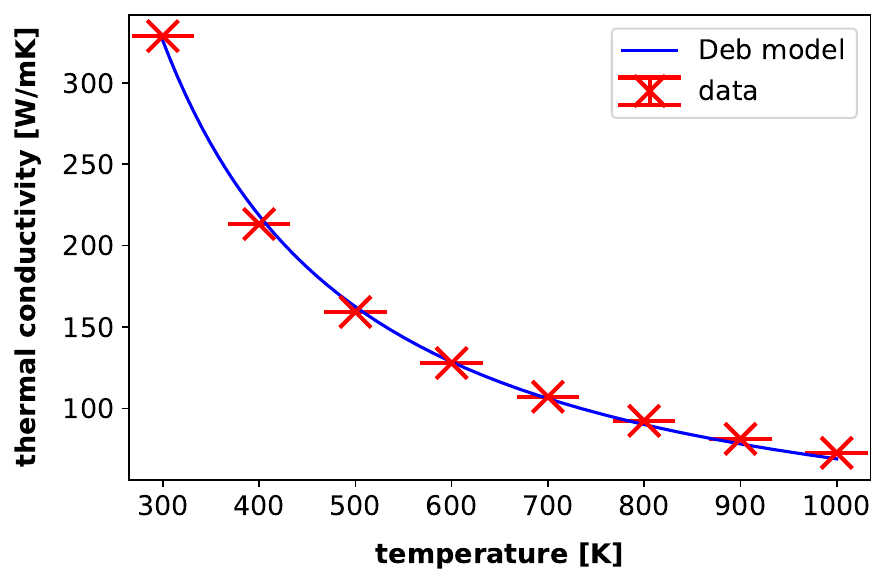}
\includegraphics[width=\textwidth,trim=0 0 0 0,clip]{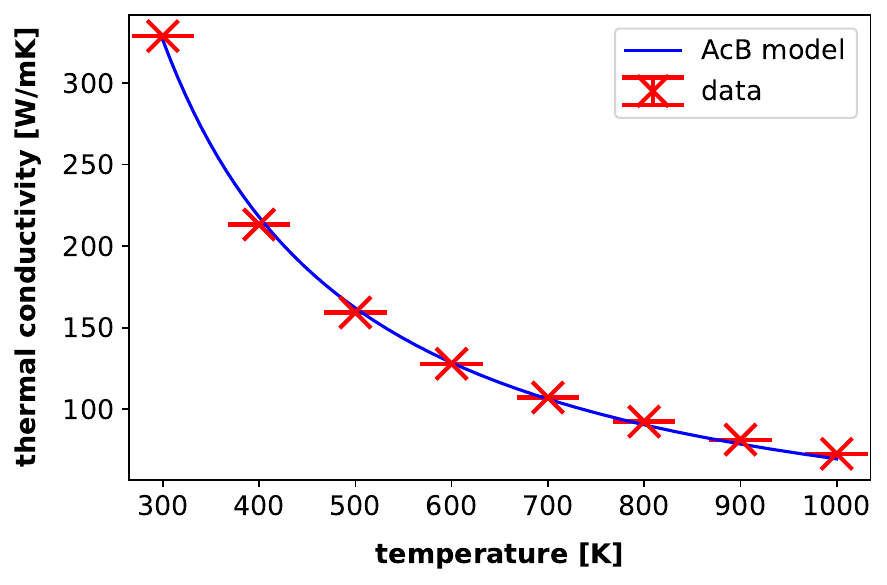}
\end{subfigure}
\hfill
\begin{subfigure}{0.32\textwidth}
\caption*{\large\textbf{Al$_{0.5}$Ga$_{0.5}$N}}
\includegraphics[width=\textwidth,trim=0 0 0 0,clip]{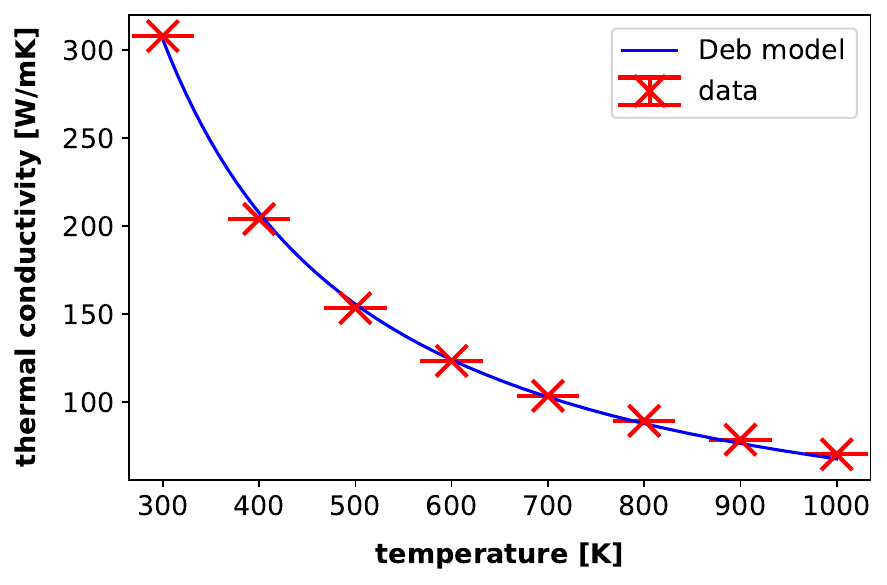}
\includegraphics[width=\textwidth,trim=0 0 0 0,clip]{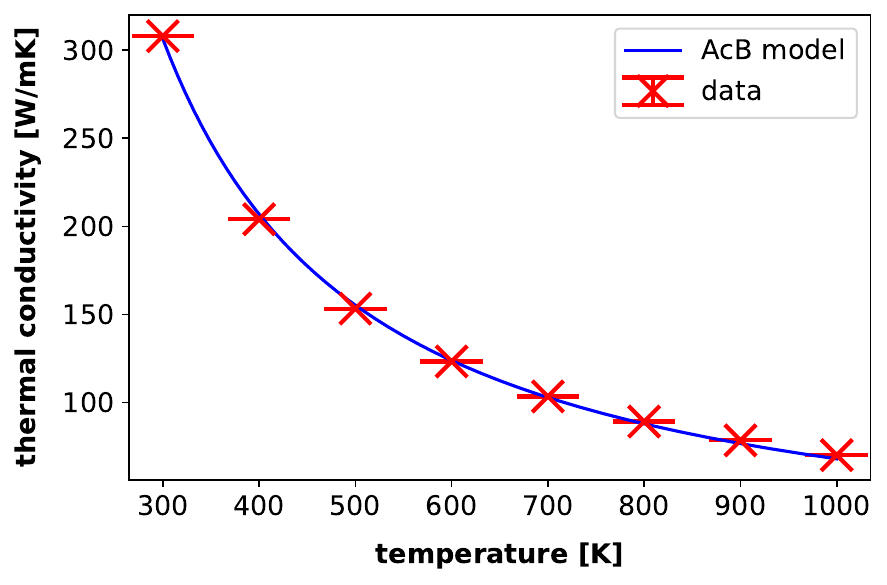}
\end{subfigure}
\hfill
\begin{subfigure}{0.32\textwidth}
\caption*{\large\textbf{Al$_{0.15}$Ga$_{0.85}$N}}
\includegraphics[width=\textwidth,trim=0 0 0 0,clip]{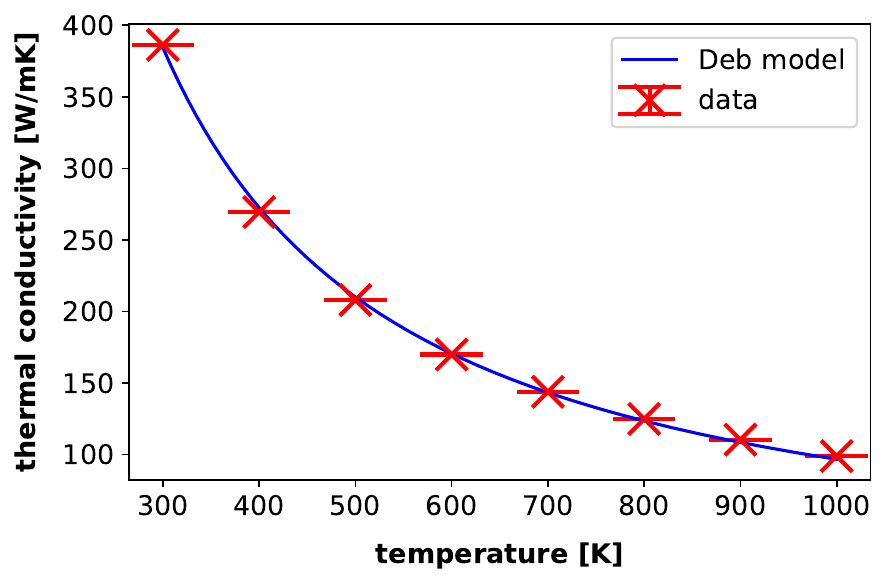}
\includegraphics[width=\textwidth,trim=0 0 0 0,clip]{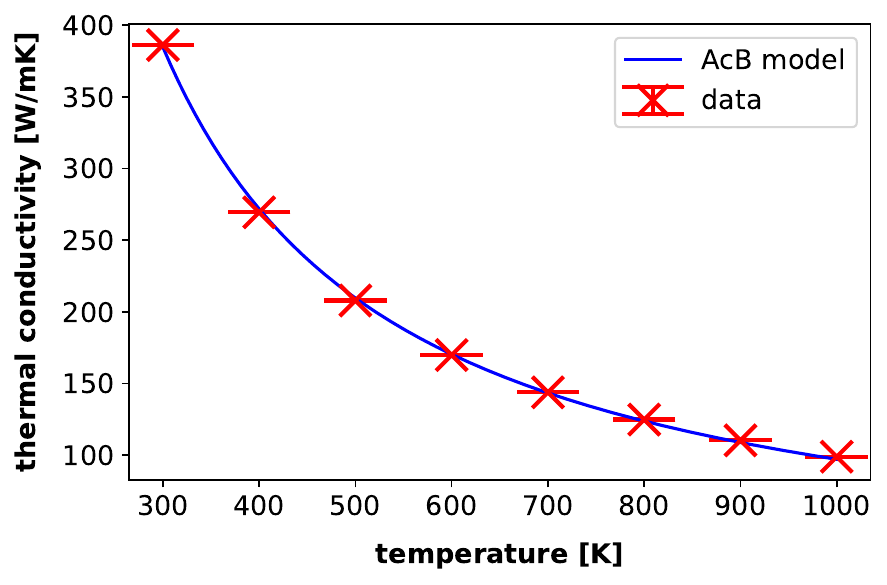}
\end{subfigure}
\caption{Comparison of the temperature-dependent thermal conductivity predicted by our models (blue lines) with first-principles reference data (red markers) for ideal bulk single-crystalline materials, calculated using ShengBTE. The left, middle, and right columns show AlN, Al$_{0.5}$Ga$_{0.5}$N, and Al$_{0.15}$Ga$_{0.85}$N, respectively. The top row depicts the results for the Debye-approximation model \autoref{sec:Deb} and the bottom row those for the acoustic band model \autoref{sec:AcB}. In the case of AlGaN, alloy scattering is neglected, thus these results describe a ficticous, idealized, and effective material. The root-mean-square errors of the respective fits are reported in \autoref{tab:results_single_crystal}.}
\label{fig:res_single_crystal}
\end{figure*}



\subsection{Fitting procedure}\label{sec:fitting_methodology}

The parameters of the models described above are fitted using the Nelder–Mead optimization algorithm \cite{Nelder1965}, by minimizing the root-mean-square error (RMSE) between the model predictions and the target temperature-dependent thermal conductivity data discussed in the following. The corresponding cost function is defined as

\begin{equation}\label{eq:cost_function}
\mathrm{S}_{\text{RMSE}} = \sqrt{\frac{1}{\text{N}}\sum_{i}^{\text{N}} \biggl(\kappa_{\text{model}}\left(\text{T}_{i}\right)-\kappa_{\text{exp}}\left(\text{T}_i\right)\biggr)^2}
\end{equation}
The sum runs over the fitted data points and $\kappa_{\text{exp}}$ and $\kappa_{\text{model}}$ are the experimental and model-predicted thermal conductivities. The resulting parameter values for the different models and investigated thin films are presented in \autoref{sec:results}. All parameters were constrained to physically meaningful ranges. Since none of the fitted values approached the imposed bounds, these constraints did not affect the fitting results.


\section{Results}\label{sec:results}


\subsection{Bulk-single Crystals}\label{sec:results_single_crystal}


Before analyzing the actual samples, we first assess whether our models can reproduce the thermal conductivities of the corresponding single-crystalline bulk materials. For the Debye-approximation and acoustic band models, which describe Umklapp scattering using \autoref{eq:umklapp_scattering}, this requires determining the macroscopic Grüneisen parameter $\gamma$ and the dimensionless parameter $b$. Since these parameters describe phonon–phonon scattering, which is dominated by Umklapp processes, they should not be determined by fitting experimental temperature-dependent thermal conductivity data. Such measurements also reflect the contributions of other scattering mechanisms, particularly alloy scattering in disordered materials, as well as isotope and defect scattering and, in thin films, boundary scattering. Instead, we fitted the parameters for the Deb and AcB models to first-principles reference data obtained using ShengBTE. The fitted parameters are listed in \autoref{tab:results_single_crystal}. The resulting fits reproduce the first-principles results exceptionally well, see \autoref{fig:res_single_crystal}, suggesting that \autoref{eq:umklapp_scattering} provides an accurate description of Umklapp scattering in the materials considered here. For completeness, we note that our implementation of the ab-initio-derived model exactly reproduces the ShengBTE results (see Fig. S5 in the SI). \\


\begin{table}[htbp]
\caption{Shows the fitted Umklapp scattering related parameters $\gamma$ (Grüneisen parameter), and b. These values are obtained by fitting the thermal conductivities predicted by the Debye-approximation and acoustic band models to the first-principles results calculated using ShengBTE for ideal bulk single crystals (represented within the virtual crystal approximation for the alloy compositions). As discussed in the main text, only Umklapp scattering is considered in all models, i.e., alloy scattering is neglected for Al$_{x}$Ga$_{1\!-\!x}$N. RMSE refers to the root-mean-square error between model- and first-principles-calculated thermal conductivities and is given in [W/mK].}
\centering
\resizebox{\columnwidth}{!}{
\begin{tabular}{|c|c|c|c|c|c|c|}
\cline{2-7}
\multicolumn{1}{c}{} & \multicolumn{3}{|c|}{Deb (\autoref{sec:Deb})} & \multicolumn{3}{c|}{AcB (\autoref{sec:AcB})}\\
\cline{2-7}
\multicolumn{1}{c|}{}  & $\gamma$ & b & RMSE & $\gamma$ & b & RMSE \\
\hline
AlN & 1.03 & 3.34 & 9.49 & 0.82 & 3.36 & 7.00 \\
\hline
Al$_{0.5}$Ga$_{0.5}$N & 0.91 & 3.22 & 4.43 &  0.65 & 2.92 & 3.10 \\
\hline
Al$_{0.15}$Ga$_{0.85}$N & 0.69 & 4.63 & 3.57 & 0.47 & 4.25 & 2.23 \\
\hline
\end{tabular}}
\label{tab:results_single_crystal}
\end{table}



\begin{figure*}[ht!]
\begin{subfigure}{0.329\textwidth}
\centering
\textbf{AlN}
\includegraphics[width=\columnwidth,trim=0 0 0 0,clip]{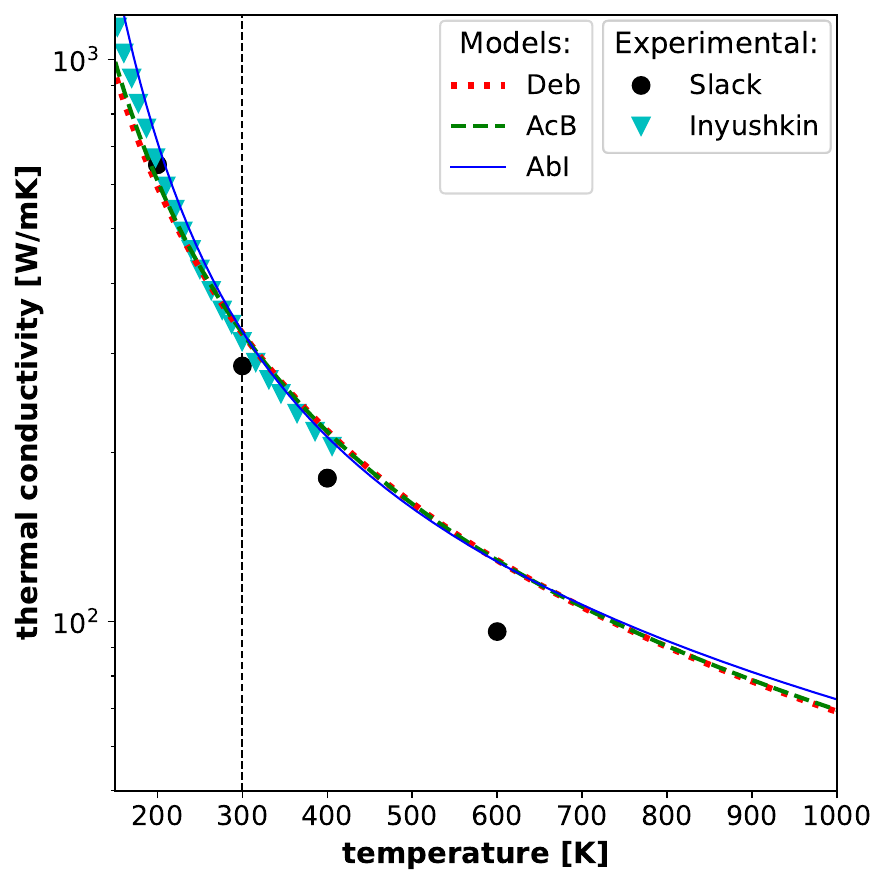}
\end{subfigure}
\hfill
\begin{subfigure}{0.329\textwidth}
\centering
\textbf{Al$_{0.5}$Ga$_{0.5}$N}
\includegraphics[width=\columnwidth,trim=0 0 0 0,clip]{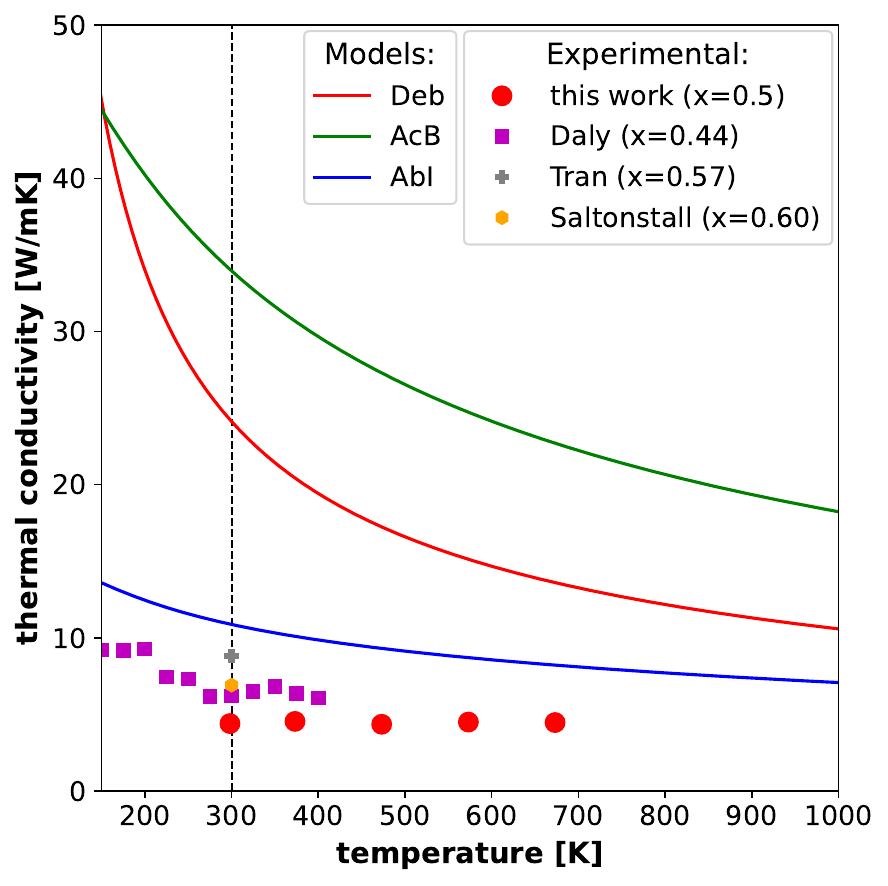}
\end{subfigure}
\hfill
\begin{subfigure}{0.329\textwidth}
\centering
\textbf{Al$_{0.15}$Ga$_{0.85}$N}
\includegraphics[width=\columnwidth,trim=0 0 0 0,clip]{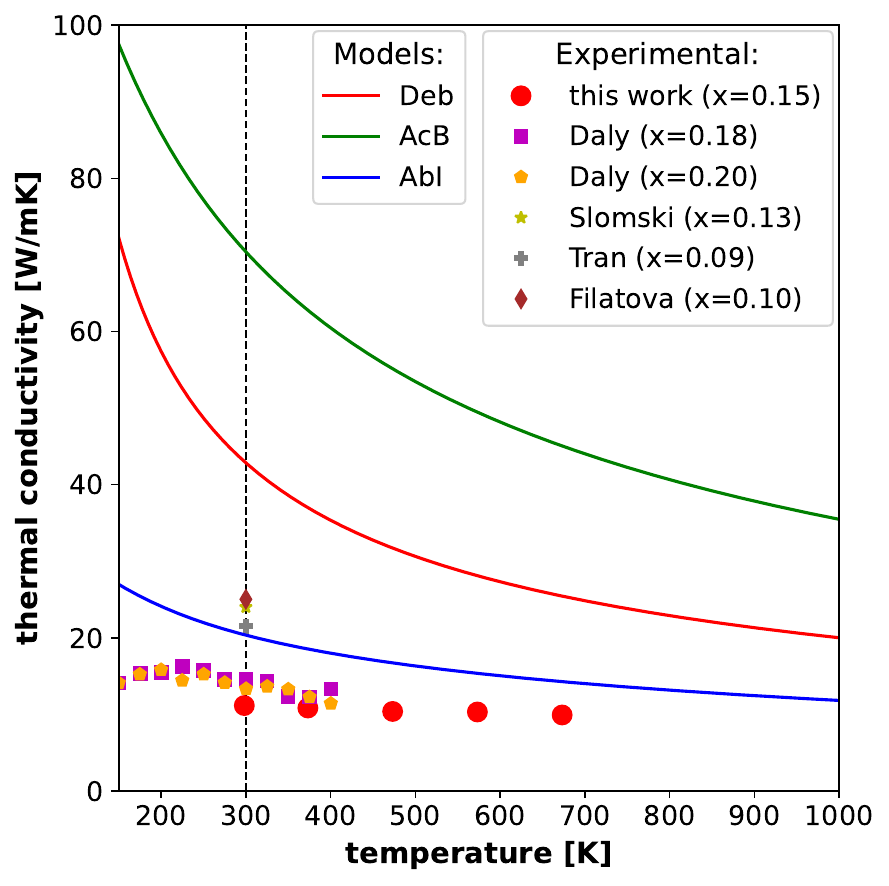}
\end{subfigure}
\caption{Comparison of the temperature-dependent thermal conductivity predicted by the Deb, AcB, and AbI models, obtained by fitting to first-principles calculations for single-crystalline bulk materials, with experimental data from the literature. The left, center, and right panels show results for AlN, Al$_{0.5}$Ga$_{0.5}$N, and Al$_{0.15}$Ga$_{0.85}$N, respectively. The experimental data for AlN reported by Slack et al. \cite{Slack1987} (black diamonds) and Inyushkin et al. \cite{Inyushkin2020} (cyan down triangles) were for bulk single crystals. In contrast, the experimental results for Al$_{x}$Ga$_{1\!-\!x}$N were obtained for single-crystalline thin films. The data were reported by Daly et al. \cite{Daly2002} (purple squares and orange pentagon symbols), Tran et al. \cite{DQTran2020} (grey plus signs), Saltonstall et al. \cite{Saltonstall2017} (orange hexagons), Slomski et al. \cite{Slomski2017} (yellow star), Filatova et al. \cite{Filtaova2020} (brown thin diamond symbol). The red circles represent the data measured in this work. More detailed descriptions of the experimental data can be found in the respective overview figures for AlN (\autoref{fig:AlN_ktemp_overview}) and Al$_{x}$Ga$_{1\!-\!x}$N (\autoref{fig:ktemp_AlGaN_overview}).}
\label{fig:single_crystal_overview}
\end{figure*}

Having established that the models reproduce the first-principles reference data when only Umklapp scattering is considered, we next assess their agreement with experimental measurements. The results are presented in \autoref{fig:single_crystal_overview}. Following the parametrization of the Umklap scattering, this comparison evaluates the models under more realistic conditions. For Al$_{x}$Ga$_{1\!-\!x}$N, the measured thermal conductivities additionally reflect the combined effects of multiple scattering mechanisms, including isotope and alloy scattering, which occur even in single-crystalline bulk samples. The present comparison, therefore, provides a stringent test of the models' predictive capabilities.

For AlN, the situation is relatively straightforward, as Umklapp scattering is the primary mechanism limiting the thermal conductivity of the single-crystalline bulk material. Consequently, the models shown in \autoref{fig:single_crystal_overview} are identical to those presented in \autoref{fig:res_single_crystal}. They largely agree with one another and, more importantly, all of them reproduce the experimental measurements well. The agreement is particularly good for the measurements by Inyushkin et al. \cite{Inyushkin2020}, who reported a thermal conductivity of 316 W/(m K) at 300 K for one of the purest AlN samples reported to date. Slack et al. \cite{Slack1987} reported lower thermal conductivities, particularly at elevated temperatures. The reduced thermal conductivity was attributed to small amounts of oxygen impurities experimentally detected in the sample. Such impurity-related effects are not accounted for in our models. A detailed comparison reveals noticeable differences between our models primarily emerge below approximately 300 K (mind the logarithmic scale), where the models need to extrapolate (the fitting of Umklapp scattering depicted in Fig. 2 only included data starting at 300K). Moreover, at very low temperatures, the treatment of Normal scattering within the relaxation-time approximation (RTA) limits the accuracy.

For AlGaN, comparison between the simulations and experimental data is more challenging for three reasons. First, accurately describing alloy scattering is more difficult than modelling intrinsic phonon–phonon scattering, such as Umklapp scattering. Second, the strong influence of alloy composition on thermal conductivity, combined with variations in composition among the experimental samples, complicates direct comparisons. Third, to the best of our knowledge, experimental data for bulk single-crystalline AlGaN are not available. The latter is particularly relevant for alloys, which can exhibit pronounced size effects \cite{Saltonstall2019,Cheaito2012}. We, therefore, compare our simulations to measurements on the thickest available samples with comparable alloy compositions. The ab-initio-derived model, which accounts for alloy scattering using the Tamura model (see above), reproduces the experimental data reasonably well for both target compositions. As expected, alloy scattering substantially reduces thermal conductivity compared to AlN. At 300 K, the ab-initio-derived model predicts a reduction in thermal conductivity by a factor of about 30 when going from AlN to Al$_{0.5}$Ga$_{0.5}$N. Furthermore, the thermal conductivity becomes only weakly temperature-dependent (mind the logarithmic scale in the leftmost panel of \autoref{fig:single_crystal_overview} in contrast to the linear scales used in the remaining panels). The lower experimental thermal conductivities compared with the ab-initio-derived model predictions can be attributed primarily to boundary scattering from finite sample thickness. In contrast to the ab-initio model's good overall performance, both the Debye-approximation and the acoustic band model substantially overestimate the experimentally measured thermal conductivities when combined with the Klemens expression for alloy scattering. Since the models agree well with the reference data when alloy scattering is neglected (see \autoref{fig:res_single_crystal}), we attribute these discrepancies primarily to the Klemens expression for point-defect (alloy) scattering (\autoref{eq:klemens_model}). The differences between the Debye-approximation and acoustic band models can be explained by their different representations of the phonon dispersion. The Klemens model is particularly sensitive to these variations because the scattering rates depend quartically on frequency. Furthermore, the Debye-approximation model neglects phonon-dispersion anisotropy, another factor contributing to the differences between the two models.

After evaluating our models’ ability to replicate thermal transport in both idealized and real bulk systems, we next proceed to discussing results for the thin-film samples studied in this work.


\subsection{Aluminium Nitride transition layer}\label{sec:results_id762}


\begin{figure*}[htbp]
\centering
\includegraphics[width=0.75\textwidth,trim=0 0 0 0,clip]{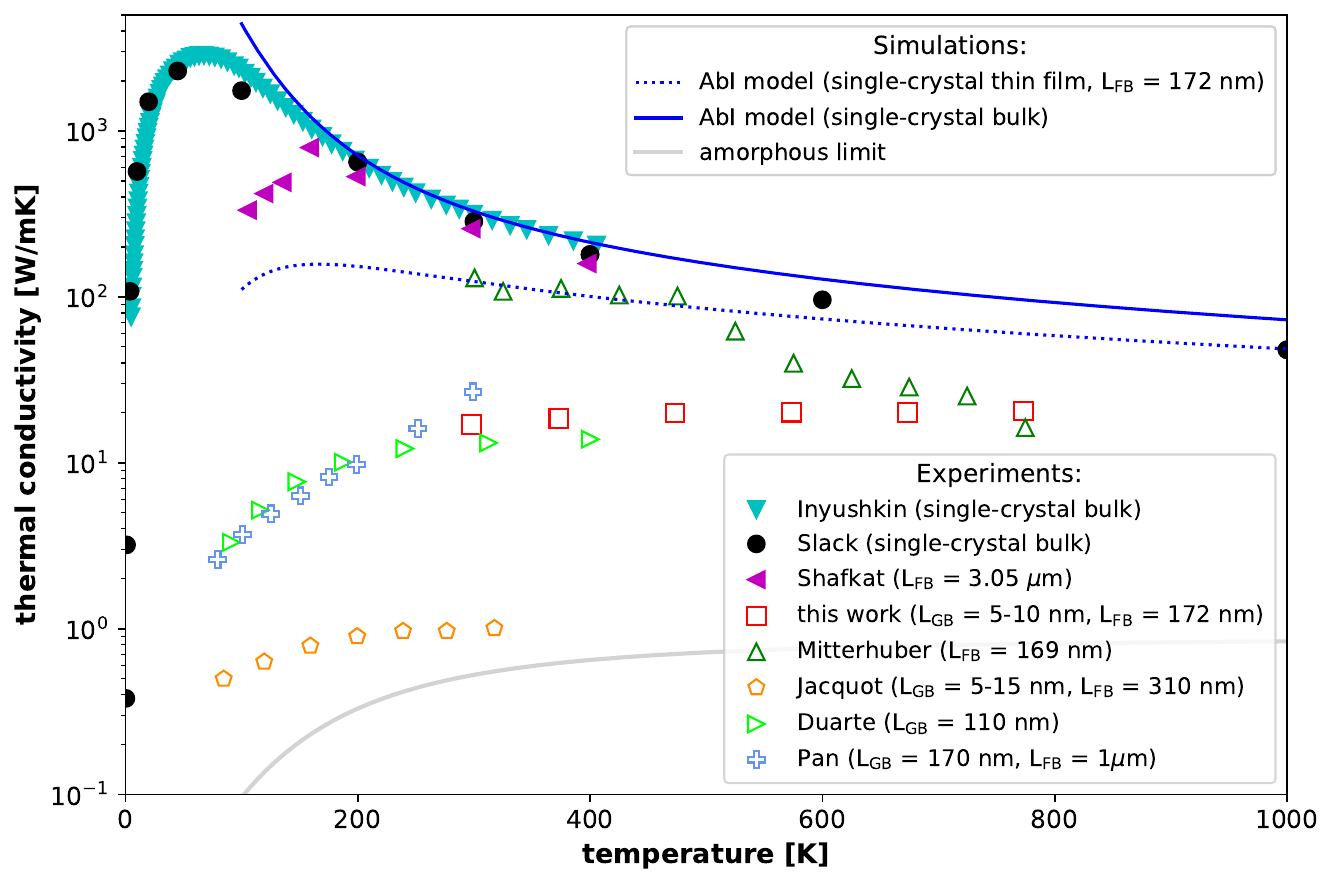}\vspace{5pt}
\caption{\label{fig:AlN_ktemp_overview} Overview of thermal conductivity data for AlN, including both theoretical predictions and experimental measurements. The theoretical results include calculations for ideal single-crystalline bulk (blue solid) and thin-film (blue dashed) systems computed with the ab-initio-derived model (\autoref{sec:AbI}). In the simulations of the thin film, the thickness corresponds to the sample thickness of 172 nm, and the boundary specularity is set to $p_{\text{FB}} = 0$. The amorphous limit (grey solid line) was computed according to \cite{Cahill1992}, and provides a lower bound for the thermal conductivity of AlN. The experimental data for bulk samples are depicted with filled symbols, whereas thin film samples which are shown by open symbols. L$_{\mathrm{FB}}$ refers to the thin film thickness reported in the respective papers and L$_{GB}$ to the average grain sizes in the case of polycrystalline samples whenever they are reported. The measurements by Slack \cite{Slack1987} (black spheres) and Inyushkin \cite{Inyushkin2020} (cyan downward triangles) correspond to single-crystalline bulk samples. Shafkat \cite{Shafkat2021} (magenta left-pointing triangle) reports measurements of a thick (L$_{\text{BD}}$ = 3.05 $\mu$m) single-crystalline thin film. Our measurements (red squares) were performed on a 172 nm thick polycrystalline film. The sample reported by Duarte \cite{Duarte2021} (lime green right-pointing triangle) is a porous polycrystalline bulk sample (L$_{\text{GB}}$ = 110 nm), thus the impact of grain boundary scattering is amplified. Jacquot \cite{Jacquot2002} (orange pentagons) measured a polycrystalline thin film (L$_{\text{GB}}$ = 5-15 nm, L$_{\text{BD}}$ = 310 nm). Measurements by Pan \cite{Pan2012} (light blue plus signs) were performed on a polycrystalline thin film (L$_{\text{GB}}$ = 170 nm, L$_{\text{BD}}$ = 1 $\mu$m) containing a high concentration of oxygen defects (x$_{\text{O}} \approx 25\%$). The measurements by Mitterhuber \cite{Mitterhuber2020} (green upward triangles) are from a thin film (L$_{\text{BD}}$ = 169 nm) of unknown, but apparently relatively high crystallinity, as inferred from the high thermal conductivity values.}
\end{figure*}



\begin{figure*}[ht]
\centering
\begin{subfigure}{0.326\textwidth}
\centering
\textbf{Debye-approximation}
\includegraphics[width=\textwidth,trim=0 0 0 0,clip]{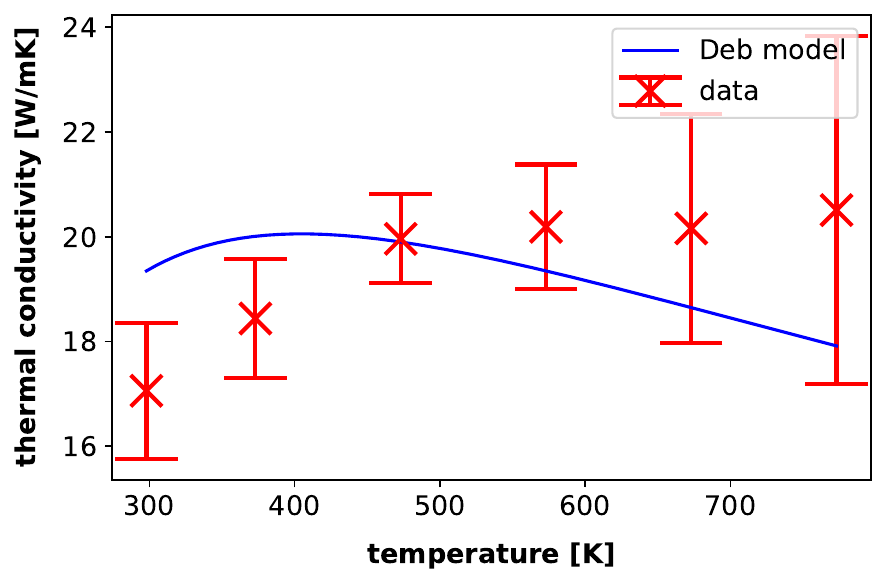}
\includegraphics[width=\textwidth,trim=0 0 0 0,clip]{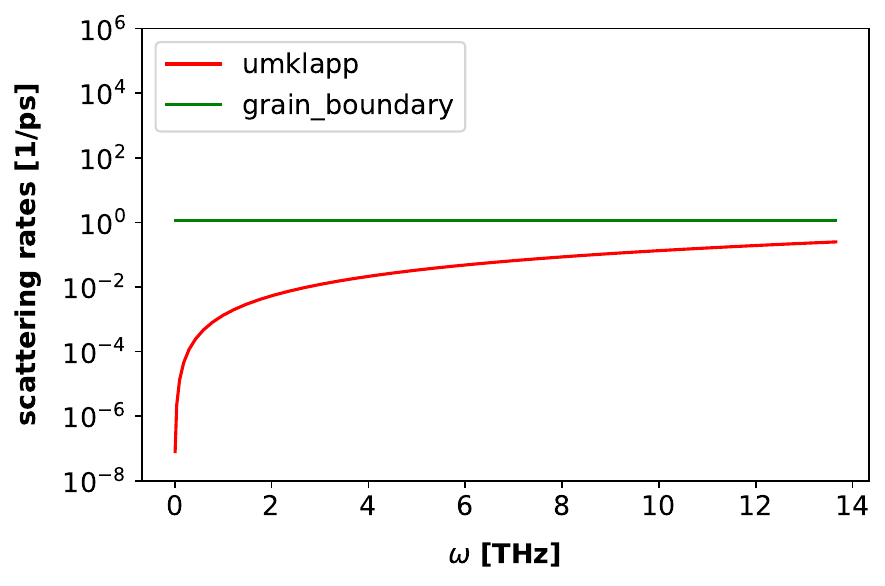}
\end{subfigure}
\hfill
\begin{subfigure}{0.326\textwidth}
\centering
\textbf{acoustic band}
\includegraphics[width=\textwidth,trim=0 0 0 0,clip]{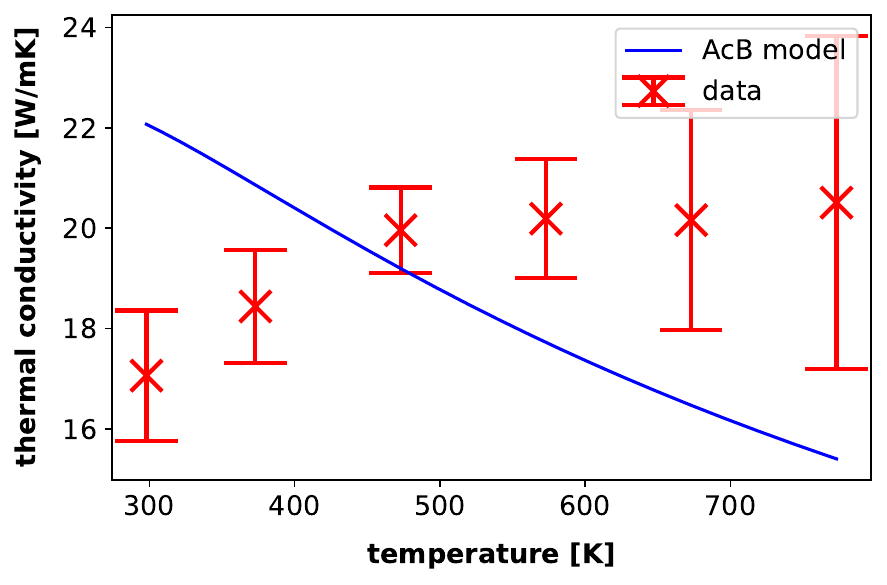}
\includegraphics[width=\textwidth,trim=0 0 0 0,clip]{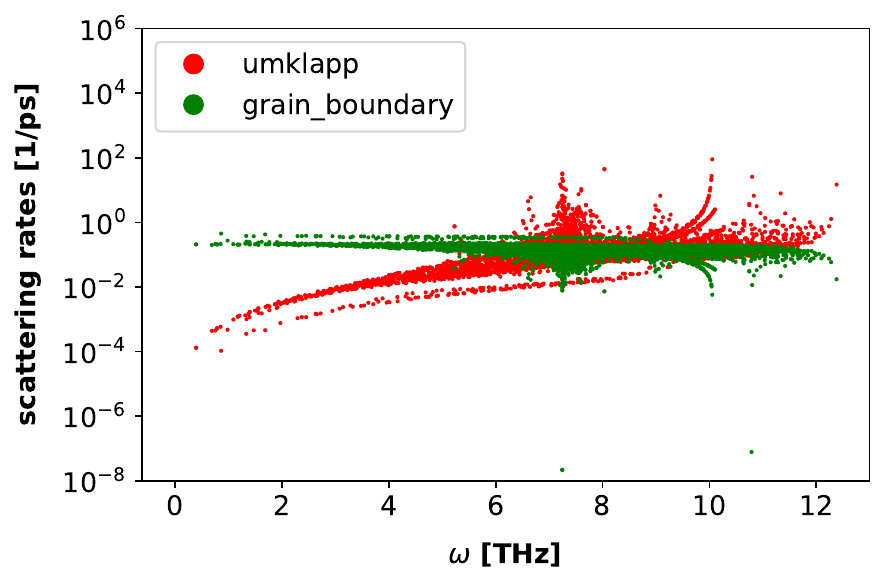}
\end{subfigure}
\hfill
\begin{subfigure}{0.326\textwidth}
\centering
\textbf{ab-initio-derived}
\includegraphics[width=\textwidth,trim=0 0 0 0,clip]{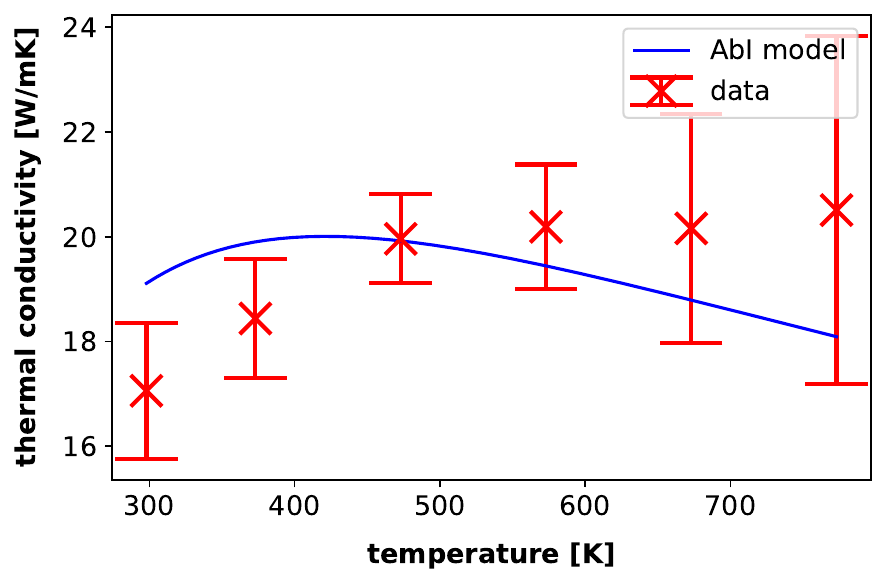}
\includegraphics[width=\textwidth,trim=0 0 0 0,clip]{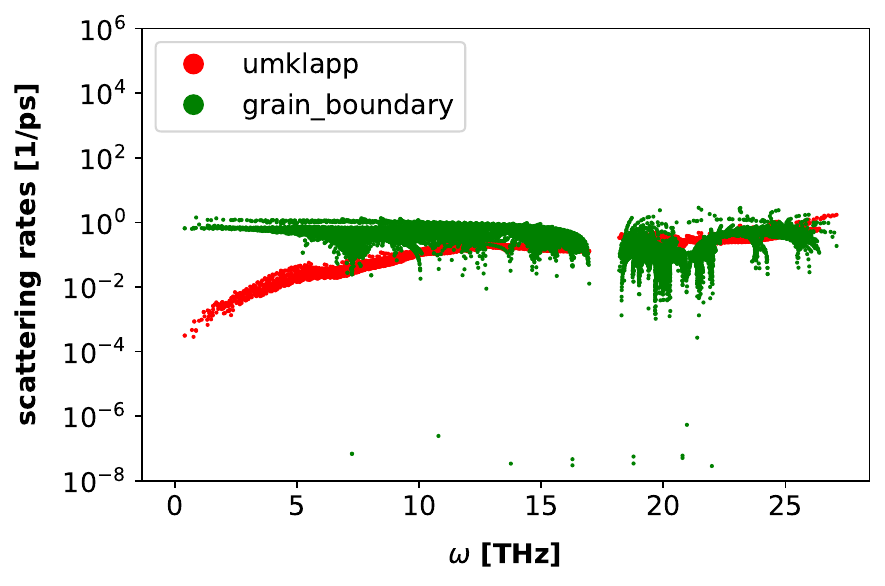}
\end{subfigure}
\caption{The top row shows the measured temperature-dependent thermal conductivity of the aluminium nitride transition layer (red error bars), together with the fitted models (blue lines). The models consider both Umklapp and grain boundary scattering in this case. The bottom row shows the corresponding scattering rates as a function of phonon frequency and at a temperature of 300 K. From left to right, the panels correspond to the Debye-approximation model (\autoref{sec:Deb}), the acoustic band model (\autoref{sec:AcB}), and the ab-initio-derived model (\autoref{sec:AbI}).}
\label{fig:id762_res_umklapp}
\end{figure*}


The first sample is the 172 nm-thick AlN transition layer (see \autoref{fig:samples_schematics}). To put our measurements into context, \autoref{fig:AlN_ktemp_overview} compares our data (red squares) to literature data for AlN samples spanning a wide range of structural configurations. These include single-crystalline bulk materials (Inyushkin \cite{Inyushkin2020} and Slack \cite{Slack1987}), single-crystalline thin films (Shafkat \cite{Shafkat2021} and Mitterhuber \cite{Mitterhuber2020}), and polycrystalline and amorphous thin films (Pan \cite{Pan2012}, Duarte \cite{Duarte2021} and Jacquot \cite{Jacquot2002}). The single-crystal simulation results for bulk AlN obtained with the ab-initio-derived model, which were discussed in the preceding section, are included as solid blue lines. Additionally, the data simulated for a thin film with a thickness of 172 nm corresponding to that of our sample are shown as dotted blue line. For the thin-film calculation, the boundaries are assumed to be fully diffusive, with a specularity of p$_{\text{\tiny FB}}$=0 (see \autoref{sec:boundary_scattering}). The results of this calculation demonstrate that boundary scattering alone is insufficient for explaining the measured thermal conductivity. To establish a lower bound for the thermal conductivity of the sample, also the amorphous-limit values calculated using the model of Cahill et al. \cite{Cahill1992} are shown.

\begin{figure*}[ht]
\centering
\begin{subfigure}{0.329\textwidth}
\vspace{10pt}
\centering
\textbf{Debye-approximation}
\includegraphics[width=\textwidth,trim=0 0 0 0,clip]{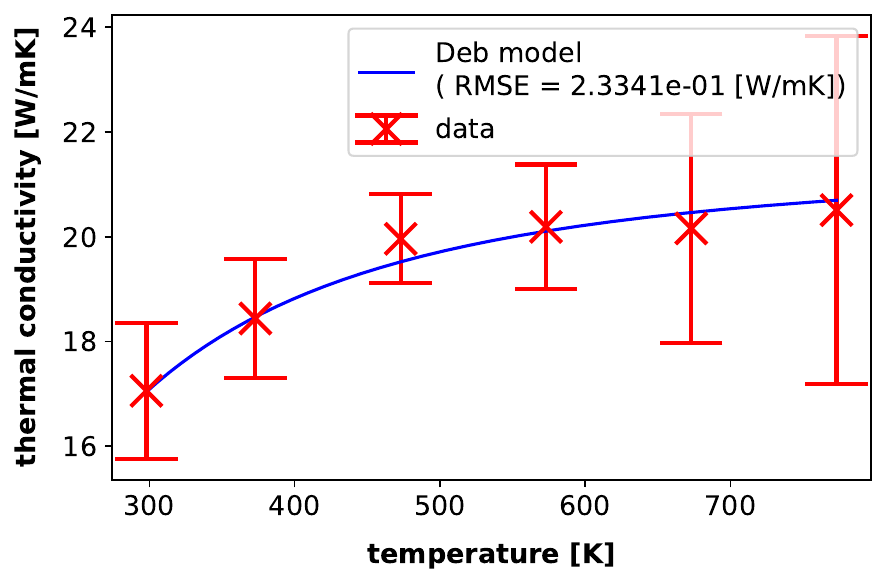}
\includegraphics[width=\textwidth,trim=0 0 0 0,clip]{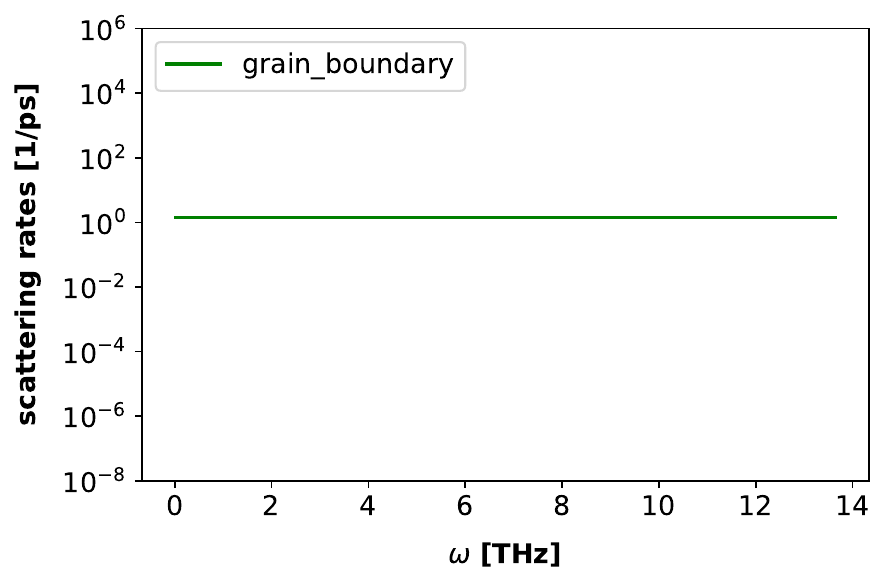}
\end{subfigure}
\hfill
\begin{subfigure}{0.329\textwidth}
\centering
\textbf{acoustic band}
\includegraphics[width=\textwidth,trim=0 0 0 0,clip]{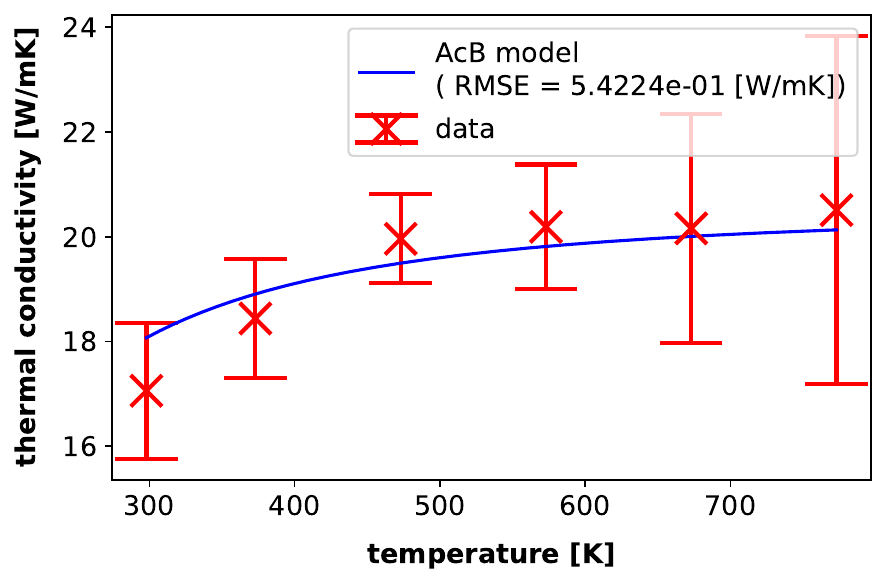}
\includegraphics[width=\textwidth,trim=0 0 0 0,clip]{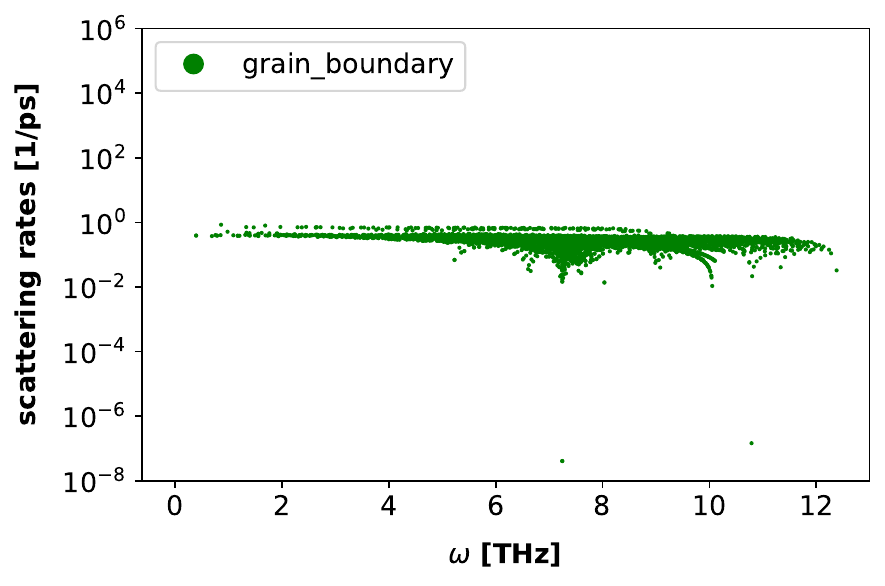}
\end{subfigure}
\hfill
\begin{subfigure}{0.329\textwidth}
\centering
\textbf{ab-initio-derived}
\includegraphics[width=\textwidth,trim=0 0 0 0,clip]{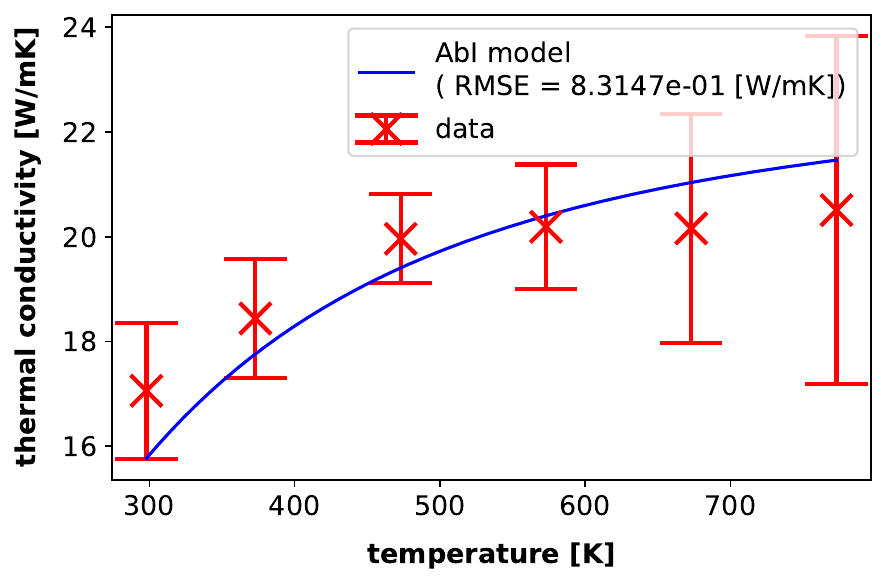}
\includegraphics[width=\textwidth,trim=0 0 0 0,clip]{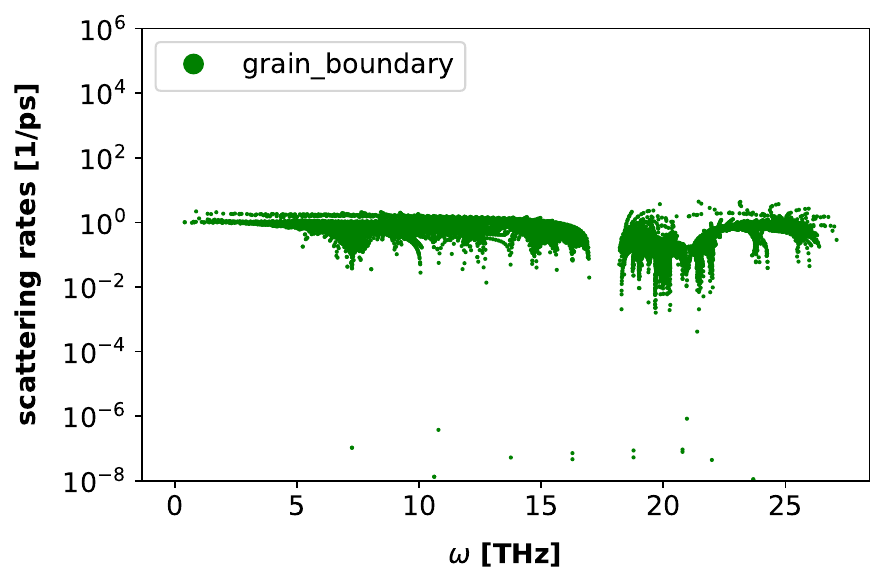}
\end{subfigure}    
\caption{The top row shows the measured temperature-dependent thermal conductivity of the aluminium nitride transition layer (red error bars), together with the fitted models (blue lines). The bottom row shows the corresponding scattering rates as a function of phonon frequency and at a temperature of 300 K. From left to right, the panels correspond to the Debye-approximation model (\autoref{sec:Deb}), the acoustic band model (\autoref{sec:AcB}), and the ab-initio-derived model (\autoref{sec:AbI}). As evident from the scattering rates, the results shown are obtained without including Umklapp scattering and nevertheless reproduce the experimentally observed thermal conductivity.}
\label{fig:id762_res}
\end{figure*}

The data in \autoref{fig:AlN_ktemp_overview} allow us to draw several conclusions: (i) The thermal conductivity of AlN varies by orders of magnitude depending on the quality and structural characteristics of the samples, as evident by the need for a logarithmic scale in \autoref{fig:AlN_ktemp_overview}. (ii) Accounting for film-boundary scattering in a 172 nm-thick film, reduces the simulated thermal conductivity by a factor ranging from approximately $\sim 4.0$ at 100 K to $\sim 1.5$ at 1000 K. The resulting predictions agree well with the measurements reported by Mitterhuber et al. \cite{Mitterhuber2020} for a film of comparable thickness but unspecified crystal quality, particularly between 300 and 500 K. This agreement suggests that the sample used in the respective measurements was highly crystalline despite its finite thickness. (iii) Overall, the simulated thermal conductivities remain substantially higher than those measured for the AlN transition layer studied here, indicating that in our sample additional scattering mechanisms significantly limit heat transport. (iv) Despite its comparatively low thermal conductivity, the measured values remains well above the amorphous limit and also above the values reported by Jacquot et al. \cite{Jacquot2002}. (v) Finally, our measurements exhibit a glass-like temperature dependence, with the thermal conductivity increasing up to approximately 500 K before approaching saturation at higher temperatures (see \autoref{fig:id762_res}).


This behavior suggests that neither Umklapp scattering nor film-boundary scattering alone dominates thermal transport in our sample. Instead, other scattering mechanisms associated with the supposedly polycrystalline microstructure of the studied transition layer play a significant role. This interpretation is consistent with reports that AlN transition layers often exhibit poor crystal quality \cite{Seppaenen2019,Aggarwal2021,Pan2018,Pan2012,Die2019}, owing in part to the substantial mismatches in lattice constants and thermal expansion coefficients between AlN and common substrates such as silicon and sapphire (with the former used also in our sample). Furthermore, the short surface migration lengths of aluminum atoms can lead to a poor surface morphology and to island growth, thereby promoting the formation of polycrystalline structures \cite{Pan2018,Simsek2021}. Defects particularly relevant for thermal transport in such layers include grain boundaries, threading dislocations, and point defects such as oxygen impurities, vacancies, and interstitials \cite{Shafkat2021,Cho2014,Xu2019,Mitterhuber2024}.

However, treating all of these defects as independent scattering sources and fitting their contributions individually to the experimental data is highly problematic. Their impacts are difficult to disentangle and, thus, the resulting parameter estimation would be highly ill-conditioned. Indeed, our attempts to fit the experimental data employing such an approach yielded unrealistically high concentrations of threading dislocations and point defects, whereas imposing lower, more realistic concentrations  of these defect types resulted in only minor changes in the predicted thermal conductivity. Thus, given the limited availability of experimental data, we restricted our modeling to the scattering mechanism considered most significant for the sample under investigation: grain-boundary scattering (with the corresponding modeling approach described in \autoref{sec:boundary_scattering}). We further simplified the model by assuming a single, effective grain size rather than modelling a size-distribution. Consequently, the grain size, (L$_{\mathrm{GB}}$), obtained by fitting the models to the experimental data should be interpreted as a qualitative estimate rather than as quantitatively reliable measure of the actual average grain size. 

When including Umklapp- and grain-boundary scattering (and fitting the average grain size), the "order of magnitude" of the thermal conductivity can be reproduced, but the obtained temperature-dependence even for the ab-initio-derived model does not match the experimental data very well. This is illustrated in \autoref{fig:id762_res_umklapp} together with the respective scattering rates. In fact, all fits incorporating Umklapp scattering exhibit a pronounced decrease in thermal conductivity at elevated temperatures. Although this behavior is typical of crystalline semiconductors, it does not match the experimental observations for the AlN thin film investigated here. Given the extremely small predicted grain sizes (with an effective grain size of 4.5 nm extracted from the fit of the AbI model), phonon mean free paths are expected to be predominantly limited by grain-boundary scattering. This is supported by the scattering rates shown in the bottom row of \autoref{fig:id762_res_umklapp}, where grain-boundary scattering dominates especially at low frequencies. At higher frequencies, grain-boundary scattering rates (at 300 K) fall below those associated with Umklapp scattering for some of the optical phonon modes, indicating that intrinsic phonon–phonon scattering becomes comparatively more important in these spectral regions. Interestingly, suppressing the Umklapp scattering in optical modes substantially improves the agreement with the experimental data as discussed in Sec. S6 of the SI. This suggests that treating both mechanisms as independent scattering contributions may overestimate their combined effect on thermal transport.

To rationalize this observation, one has to keep in mind that we here describe thermal transport employing the Boltzmann transport equation in the relaxation time approximation (RTA) combined with Matthiessen's rule. This approach assumes that distinct scattering mechanisms contribute independently and additively to the total phonon scattering rate, with intrinsic Umklapp scattering remaining unchanged by additional scattering sources. In the present case, Umklapp scattering reproduces the temperature-dependent thermal conductivity of bulk single-crystalline AlN (see left figure in \autoref{fig:single_crystal_overview}), while grain-boundary scattering alone can provide a good description of the experimental data for the current sample, as demonstrated below when discussing \autoref{fig:id762_res}. We attribute the comparatively poor performance of a model comprising both scattering mechanisms primarily to the following four factors: 

\begin{itemize}
    \item Limitations of Matthiessen's rule and the RTA: The assumption of independent and additive scattering rates may become inadequate in the region of similarly strong grain-boundary and Umklapp scattering. In that case, the combined scattering processes may affect the phonon population and its deviation from equilibrium, such that their contributions to thermal transport potentially cannot be described by simply adding the individual scattering rates.
    \item Shortcomings in the description of Umklapp scattering: The rates calculated for bulk single-crystalline AlN may not adequately represent those in highly polycrystalline or disordered materials. This would be a consequence of structural disorder modifying the phonon states and anharmonic interactions, altering the intrinsic scattering rates.
    \item Limitations of the phonon-gas picture: The estimated grain sizes are exceptionally small. For example, the grain size in the ab-initio-derived model amounts to 4.50 nm. This corresponds to fewer than ten AlN unit-cell lengths along the \(c\)-direction. At such length scales, strong structural disorder and frequent grain-boundary interactions may challenge the description of thermal transport in terms of well-defined phonon modes with identifiable scattering rates and mean free paths.
    \item The description of grain-boundary scattering may not fully capture the relevant interactions between phonons and grain boundaries. To investigate this possibility, we also tested a mode-dependent specularity parameter for grain-boundary scattering. However, this refinement did not improve the agreement with the experimental data, as demonstrated in Sec. S7 of the SI.
\end{itemize}

At this stage, it is not clear, which of the above aspects predominantly causes the discrepancies. A more in-depth study of testing (technically extremely challenging) approaches for overcoming the said limitations is, however, beyond the scope of the present paper. Still, as a final test and given the expected dominance of grain-boundary scattering, we performed simulations in which Umklapp scattering was neglected. In this case the focus is exclusively on grain-boundary scattering. This approach yields a good overall agreement with the experimental results and, in particular, reproduces the observed high-temperature behavior. The corresponding results are shown in \autoref{fig:id762_res}, where the fitted models (blue lines) and experimental data (red error bars) are shown in the top row, while the corresponding phonon scattering rates as a function of frequency are shown in the bottom row. The fitted grain sizes are 1.69 nm, 7.50 nm, and 2.90 nm for the Debye-approximation model, the acoustic band model, and the ab-initio-derived model, respectively, as summarized in \autoref{tab:id762_parameters}. Relative variations between the different models to describe the phonon bands are attributed to a compensation of model shortcomings in the fitting process especially for the Deb and AcB models. 

Notably, the observation that the fitted grain sizes are much smaller than the film thickness, provides an a posteriori justification for neglecting scattering associated with the finite film thickness.

\begin{table}[htp!]
\centering
\caption{Fitted parameters and root-mean-square errors (RMSE) of the fits for the different models applied to the AlN transition layer. To reproduce the measured thermal conductivity, only grain-boundary scattering was considered for the Deb and Acb models and for the first entry of the AbI model. The second and third values in the AbI entry, correspond to cases where Umklapp scattering is included only for acoustic phonons and for all phonon modes, respectively. The corresponding fit parameter is the average grain size L$_{\text{\tiny GB}}$. The RMSE quantifies the deviation between the predicted and measured thermal conductivties.}
\begin{tabular}{|c|c|c|}
\hline
Model & L$_{\text{\tiny GB}}$ [nm] & RMSE [W/mK] \\
\hline
\makecell{Deb \\(Sec. \ref{sec:Deb})} &  1.69 & 0.23  \\
\hline
\makecell{AcB \\(Sec. \ref{sec:AcB})} & 7.50 & 0.54 \\
\hline
\makecell{AbI \\(Sec. \ref{sec:AbI})} & 2.90, 3.09, 4.48 & 0.83, 0.53, 1.56 \\
\hline
\end{tabular}\vspace{10pt}
\label{tab:id762_parameters}
\end{table}

Not unexpectedly, completely neglecting Umklapp scattering also introduces some limitations: especially, the ab-initio-derived model predicts a somewhat exaggerated rise of the thermal conductivity with temperature, as shown in the top, right panel of \autoref{fig:id762_res}. We attribute this observation to an overestimation of phonon lifetimes, when anharmonic phonon scattering (as an effect that increases with temperature) is largely neglected. Despite these limitations, the modeling results, in combination with the experimental data for polycrystalline samples reported in the literature, provide strong evidence that the aluminium nitride transition layer is highly polycrystalline.


\subsection{Al$_{0.5}$Ga$_{0.5}$N buffer layer}\label{sec:results_id744}

\begin{figure*}[htp]
\centering
\includegraphics[width=0.75\textwidth,trim=0 0 0 0,clip]{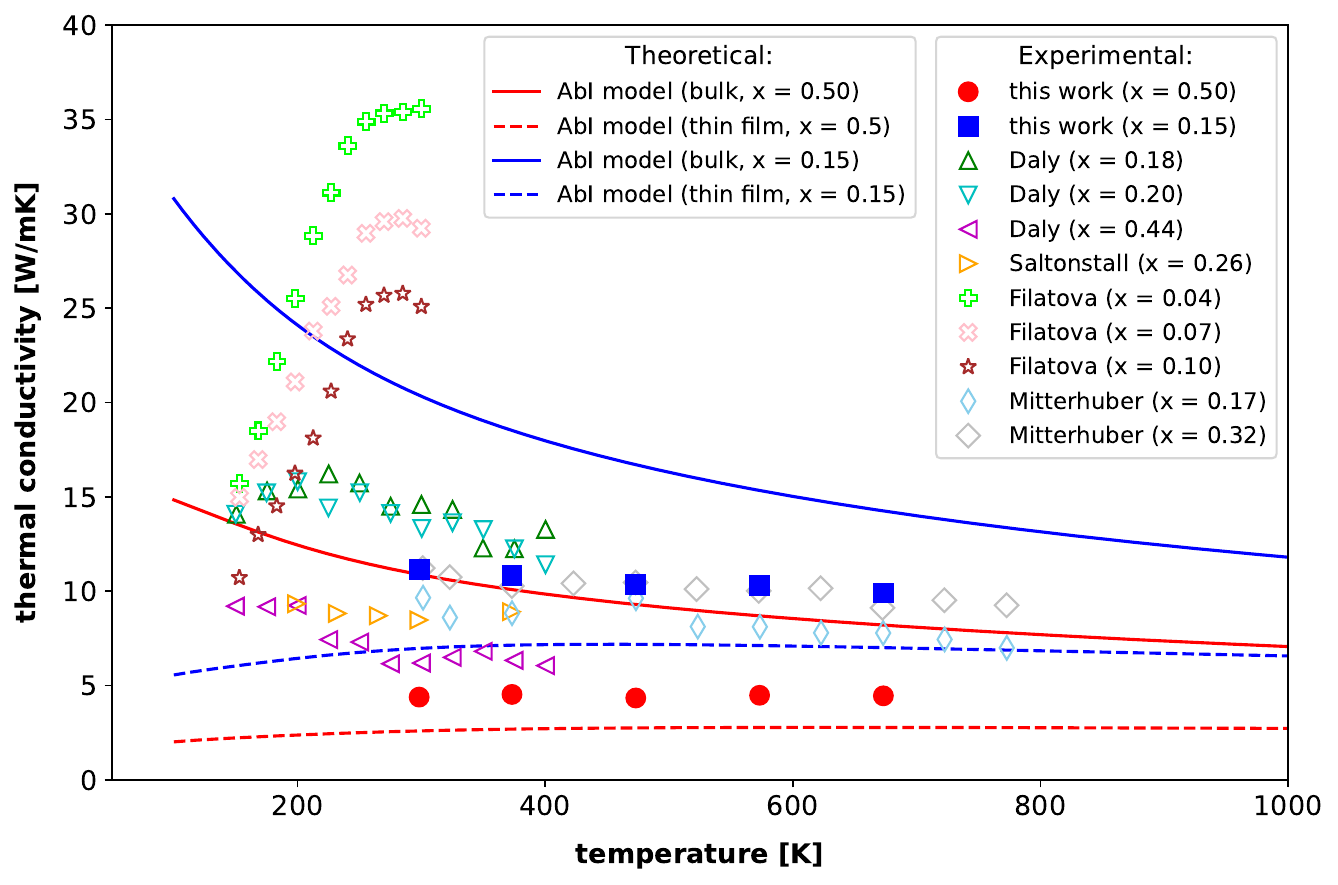}
\vspace{5pt}
\caption{Overview of thermal conductivity data for Al$_{x}$Ga$_{1-x}$N. Theoretical results for the single-crystalline bulk and thin-film systems are computed with our ab-initio-derived model (described in \autoref{sec:AbI}). Blue lines correspond to Al$_{0.15}$Ga$_{0.85}$N, and red lines correspond to Al$_{0.5}$Ga$_{0.5}$N. Results for the bulk are shown as solid lines, and thin-film results are plotted as dotted lines. For the thin-film results, we use the experimental sample thicknesses and assume the limiting case of zero boundary specularity $p_{\text{BD}} = 0$. Our measurements are shown as blue squares for the 407-nm-thick Al$_{0.15}$Ga$_{0.85}$N sample (id632) and as red circles for the 103-nm-thick Al$_{0.5}$Ga$_{0.5}$N sample (id744). The data from Daly et al. \cite{Daly2002} pertain to single-crystalline Al$_{0.18}$Ga$_{0.82}$N (green, upward triangles), Al$_{0.2}$Ga$_{0.8}$N (cyan, downward triangles), and Al$_{0.44}$Ga$_{0.56}$N (magenta, leftward triangles) thin films with unspecified thicknesses. Mitterhuber et al. \cite{Mitterhuber2020} reported data for a 65-nm-thick Al$_{0.17}$Ga$_{0.83}$N sample (light blue, thin diamond symbols) and a 423-nm-thick Al$_{0.32}$Ga$_{0.68}$N sample (grey, wide diamond symbols). Saltonstall et al. \cite{Saltonstall2017} report data for a 1-$\mu$m-thick Al$_{0.26}$Ga$_{0.74}$N film (orange, right-sided triangles). All of these measurements were performed using the optical pump-probe technique. The results from Filatova et al. \cite{Filtaova2020} were obtained using the $3\omega$ technique. These data correspond to 500-nm-thick films of Al$_{0.04}$Ga$_{0.96}$N (lime, plus symbols), Al$_{0.07}$Ga$_{0.93}$N (pink, × symbols), and Al$_{0.1}$Ga$_{0.9}$N (brown, star symbols).}
\label{fig:ktemp_AlGaN_overview}
\end{figure*}

\begin{figure*}[ht]
\centering
\begin{subfigure}{0.329\textwidth}
\centering
\textbf{Debye-approximation}
\includegraphics[width=\textwidth,trim=0 0 0 0,clip]{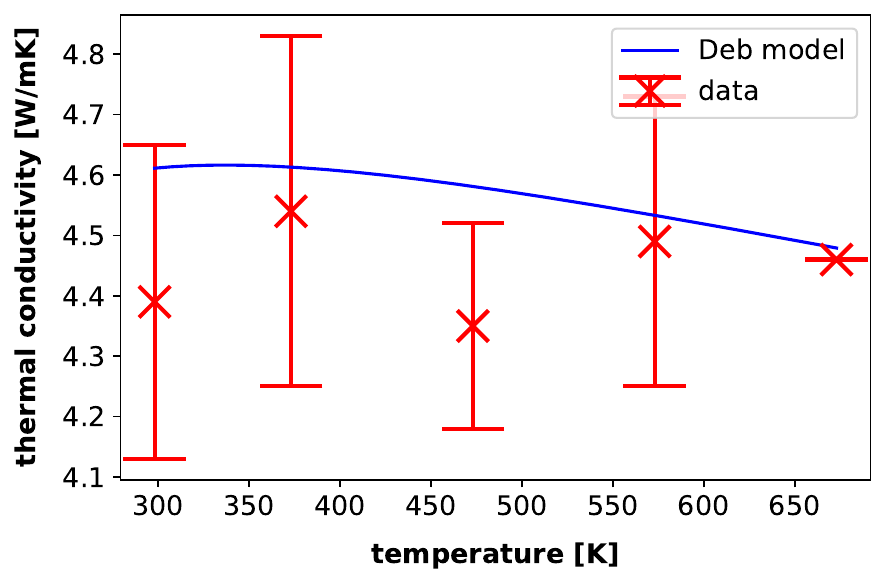} \\[5pt]
\includegraphics[width=\textwidth,trim=0 0 0 0,clip]{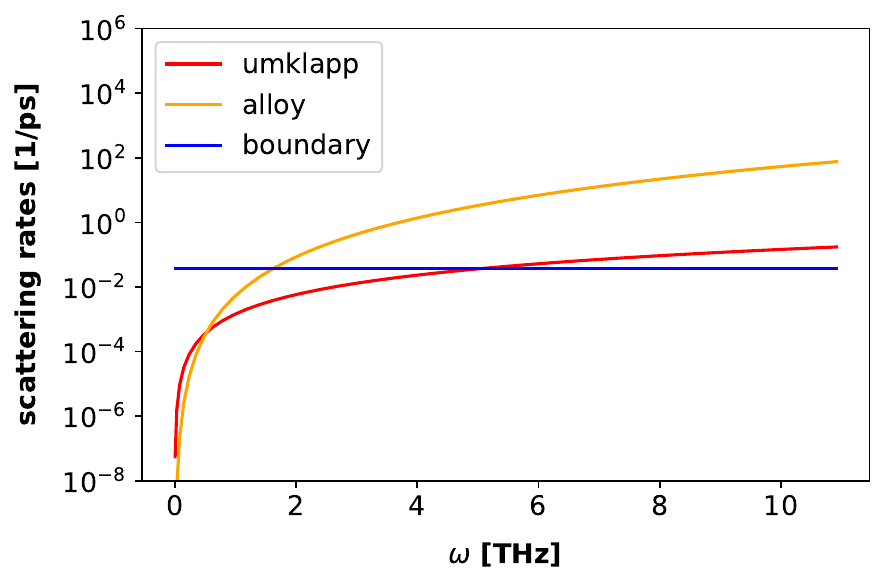}
\label{fig:id744_res_callaway}
\end{subfigure}
\hfill
\begin{subfigure}{0.329\textwidth}
\centering
\textbf{acoustic band}
\includegraphics[width=\textwidth,trim=0 0 0 0,clip]{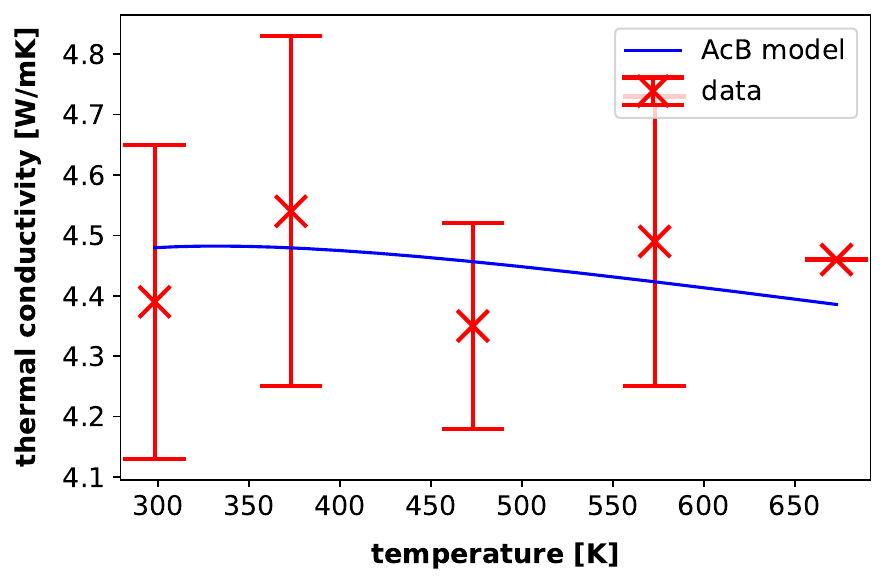} \\[5pt]
\includegraphics[width=\textwidth,trim=0 0 0 0,clip]{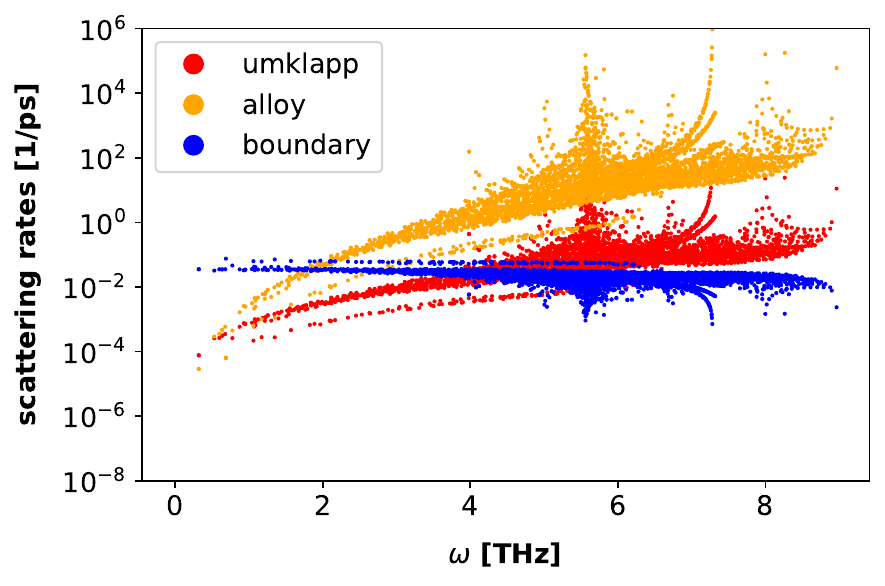}
\label{fig:id744_res_band}
\end{subfigure}
\hfill
\begin{subfigure}{0.329\textwidth}
\centering
\textbf{ab-initio-derived}
\includegraphics[width=\textwidth,trim=0 0 0 0,clip]{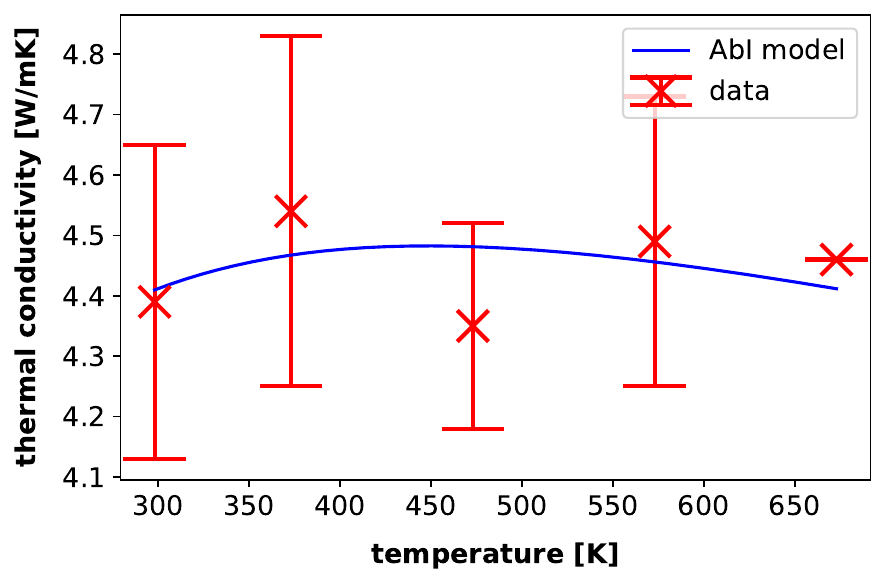} \\[5pt]
\includegraphics[width=\textwidth,trim=0 0 0 0,clip]{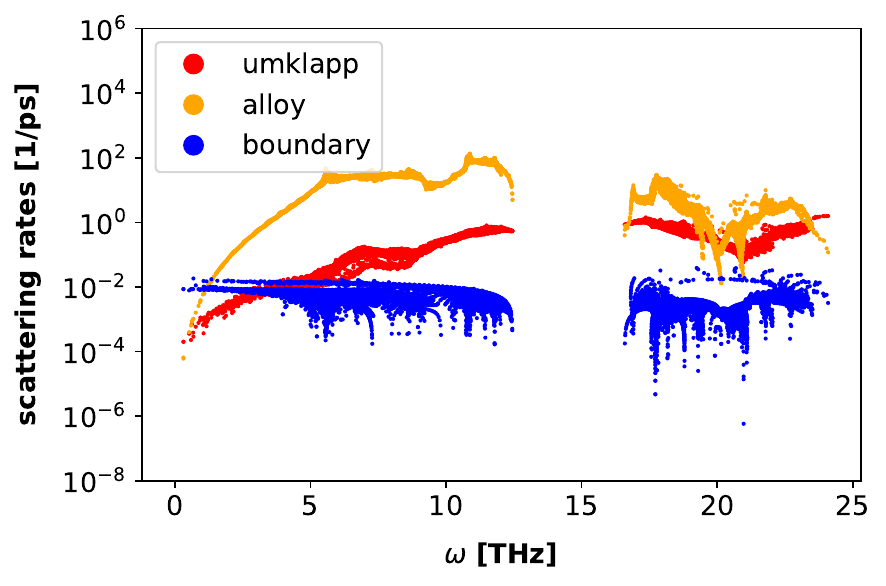}
\label{fig:id744_res_ab_initio}
\end{subfigure}
\caption{The top row shows the measured temperature-dependent thermal conductivity of the Al$_{0.5}$Ga$_{0.5}$N buffer layer (red error bars), together with the fitted models (blue lines). The bottom row shows the corresponding scattering rates as a function of phonon frequency and at a temperature of 300 K. From left to right, the panels correspond to the Debye-approximation model (\autoref{sec:Deb}), the acoustic band model (\autoref{sec:AcB}), and the ab-initio-derived model (\autoref{sec:AbI}).}
\label{fig:id744_res}
\end{figure*}

The second considered sample is a 103 nm-thick Al$_{0.5}$Ga$_{0.5}$N layer used as a buffer layer in the HEMT stack designs. \autoref{fig:ktemp_AlGaN_overview} provides an overview of the available thermal conductivity data for Al$_{x}$Ga$_{1\!-\!x}$N samples, including our measurements, predictions from our ab-initio-derived model for the bulk material, predictions for a thin film with the same thickness as our experimentally investigated sample, and experimental results from the literature. The latter cover thin films with various thicknesses and alloy compositions. Interestingly, our measurements display a glass-like temperature dependence, similar to that observed for the previously discussed AlN transition layer. This raises the question of whether such a behavior necessarily requires strong grain-boundary scattering or whether one could also obtain it relying on other scattering mechanisms.

To address this question, we compare the experimental results (red circles) with predictions from our ab-initio-derived model for the bulk material, without film-boundary scattering (\autoref{fig:ktemp_AlGaN_overview} red solid line), and with a model comprising fully diffusive boundaries (p$_{\text{\tiny FB}}=0$) and a thickness matching that of the sample (L$_{\text{\tiny FB}}=103$ nm; \autoref{fig:ktemp_AlGaN_overview} red dashed line). The experimental data lie within the range spanned by these two predictions, suggesting that additional scattering mechanisms, such as grain-boundary scattering, are not necessary to reproduce the measured thermal conductivity. This implies an improved crystalline quality in the buffer layer compared to the previously discussed transition layer, an interpretation that is plausible, as the underlying AlN layer on which the Al$_{0.5}$Ga$_{0.5}$N layer is grown already accommodates the structural mismatch with respect to the Si substrate \cite{Pan2018,Mansor2023}. 

To investigate whether our models, emplyoing Umklapp, alloy and film-boundary scattering, can reproduce the measured temperature dependence also quantitatively, they are fitted to the experimental data. The only fitting parameter in this case is the specularity of the film boundaries $\text{p}_{\text{FB}}$. The resulting $\text{p}_{\text{FB}}$ values, together with the RMSE fit error for each model, are reported in \autoref{tab:id744_parameters}. The fitted thermal conductivities are compared to the measurements in the top row of \autoref{fig:id744_res}, while the corresponding scattering rates required to reproduce the measured thermal conductivities are shown as a function of frequency for a temperature of 300 K in the bottom row. All three approaches yield satisfactory fits, despite the nearly temperature-independent thermal conductivity observed experimentally, which is more characteristic of glass-like or amorphous materials than of crystalline semiconductors. Nevertheless, the measured thermal conductivity remains significantly above the amorphous limit.

\begin{table}[t]
\centering
\caption{Fitted parameters and root-mean-square errors (RMSE) of the fits for the different models applied to the Al$_{0.5}$Ga$_{0.5}$N buffer layer. To reproduce the measured thermal conductivity, Umklapp, alloy, and film-boundary scattering were included, with the film-boundary specularity $p_{\text{\tiny FB}}$ serving as the sole fitting parameter. The RMSE quantifies the deviation between the predicted and measured thermal conductivities.}
\resizebox{0.6\columnwidth}{!}{
\begin{tabular}{|c|c|c|}
\hline
\multicolumn{3}{|c|}{Al$_{0.5}$Ga$_{0.5}$N buffer layer} \\
\hline
Model & p$_{\text{\tiny FB}}$  & RMSE [W/mK] \\
\hline
Deb (\ref{sec:Deb}) & 0.0 & 0.15  \\
\hline
AcB (\ref{sec:AcB}) &  0.25 & 0.08 \\
\hline
AbI (\ref{sec:AbI}) & 0.82 & 0.07  \\
\hline
\end{tabular}
}
\vspace{10pt}
\label{tab:id744_parameters}
\end{table}

The fitted specularities, p$_{\text{\tiny FB}}$, differ considerably between the models. This variation arises primarily from the limitations of the Debye-approximation and acoustic band models in describing the bulk thermal conductivity of single-crystalline AlGaN, as discussed in \autoref{sec:results_single_crystal} (see \autoref{fig:single_crystal_overview}). During fitting, the specularity in these models is driven to lower values to compensate for the overestimated bulk thermal conductivity. Consequently, only for the ab-initio-derived model with its much more accurate description of bulk transport, the specularity can be interpreted as a physically meaningful effective sample parameter.

To put our results in the context of previous experimental studies, in \autoref{fig:ktemp_AlGaN_overview} we compare them to available literature data. To the best of our knowledge, no measurements are available for samples with exactly the same film thickness and alloy composition as those investigated here. The closest match in terms of alloy composition is the $x=0.44$ sample studied by Daly et al. \cite{Daly2002} (\autoref{fig:ktemp_AlGaN_overview}, purple left-pointing triangles). However, this sample is a single-crystalline thin film of unknown thickness, which complicates a direct comparison. Nevertheless, their measurements agree reasonably well with ours, although they observe slightly higher thermal conductivities. Moreover, their thermal conductivities decrease with the temperature (albeit more weakly than is generally expected for crystalline semiconductors). In contrast, our measurements for this sample (red circles) show a roughly temperature-independent thermal conductivity. The peculiar temperature dependence observed in different AlGaN thin films and its sample-to-sample variation can be traced back to the strong influence of boundary scattering on low-frequency phonons, which provide a substantial contribution to the thermal conductivity in alloys  (see also the bottom panels in \autoref{fig:id744_res}). This is because alloy and Umklapp scattering strongly suppress the contributions of high-frequency phonons, with point-defect scattering rates scaling approximately as $\propto\omega^4$ in the Rayleigh-scattering regime. Consequently, low-frequency phonons with comparatively long mean free paths dominate heat transport. At the same time, they are particularly susceptible to scattering at the film boundaries. As temperature decreases, these phonons contribute more to heat transport, making thermal conductivity increasingly sensitive to boundary scattering. This suppresses the characteristic low-temperature peak in the thermal conductivity of crystalline semiconductors and produces a flatter temperature dependence. This behavior becomes apparent when comparing our ab-initio-derived model for the bulk material with that for the thin film (\autoref{fig:ktemp_AlGaN_overview}, red solid and dashed lines, respectively). It is also consistent with our measurements: the thinner, 103 nm Al$_{0.5}$Ga$_{0.5}$N layer exhibits a slightly flatter temperature dependence than the thicker, 407 nm Al$_{0.15}$Ga$_{0.85}$N layer. These results demonstrate that a glass-like temperature dependence does not necessarily require a highly polycrystalline film but can also arise from strong film-boundary scattering.

\begin{figure}[t!]
\centering
\includegraphics[width=\columnwidth,trim=0 0 0 0,clip]{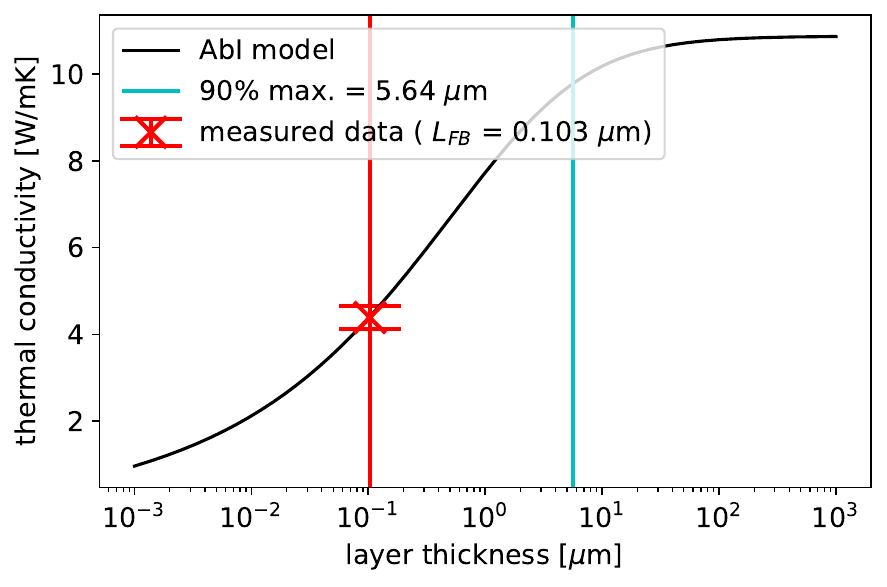}
\caption{Thermal conductivity as a function of layer thickness for the Al$_{0.5}$Ga$_{0.5}$N thin film (sample id744), calculated using the ab-initio-derived model (\autoref{sec:AbI}). The film-boundary specularity p$_{\text{\tiny FB}}$ was determined from fitting the measured temperature-dependent data shown in \autoref{fig:id744_res}.}
\label{fig:id744_kthick_ab_initio}
\end{figure}

Consistent with the discussion above, a significant thickness-dependent thermal conductivity (size-effects) has been reported for AlGaN by Saltonstall et al. \cite{Saltonstall2019} and for SiGe by Cheaito et al. \cite{Cheaito2012}. To quantify thickness effects, \autoref{fig:id744_kthick_ab_initio} shows the thermal conductivity predicted by our ab-initio-derived model (with the fitted value of p$_{\text{\tiny FB}}$) as a function of film thickness. The thermal conductivity decreases to less than half of its bulk value at a thickness of 100 nm and to approximately one-quarter at 10 nm. Recovering 90\% of the bulk thermal conductivity requires a film thickness greater than 5.64 $\mu$m.

\begin{figure*}[ht]
\begin{subfigure}{0.329\textwidth}
\centering
\hspace{15pt}\textbf{Debye-approximation}
\includegraphics[width=\textwidth,trim=0 0 0 0,clip]{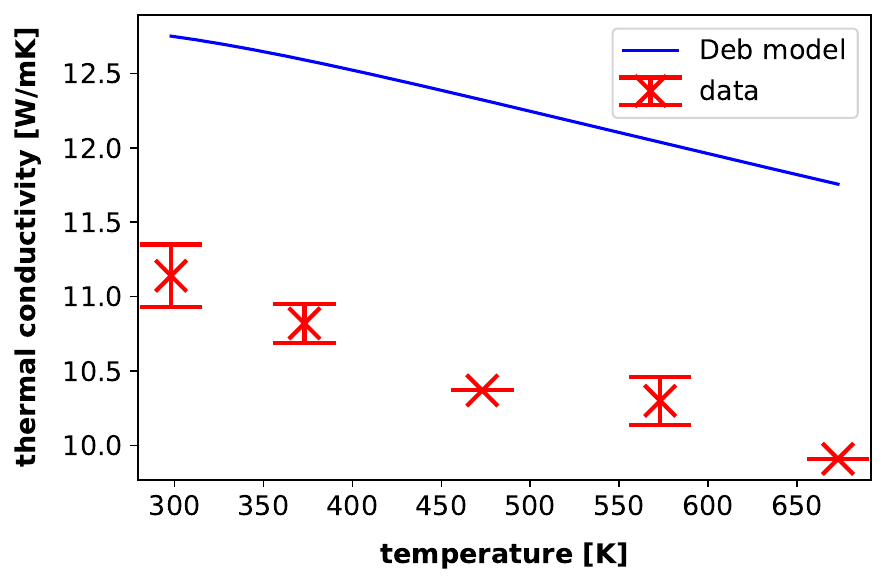} \\[5pt]
\includegraphics[width=\textwidth,trim=0 0 0 0,clip]{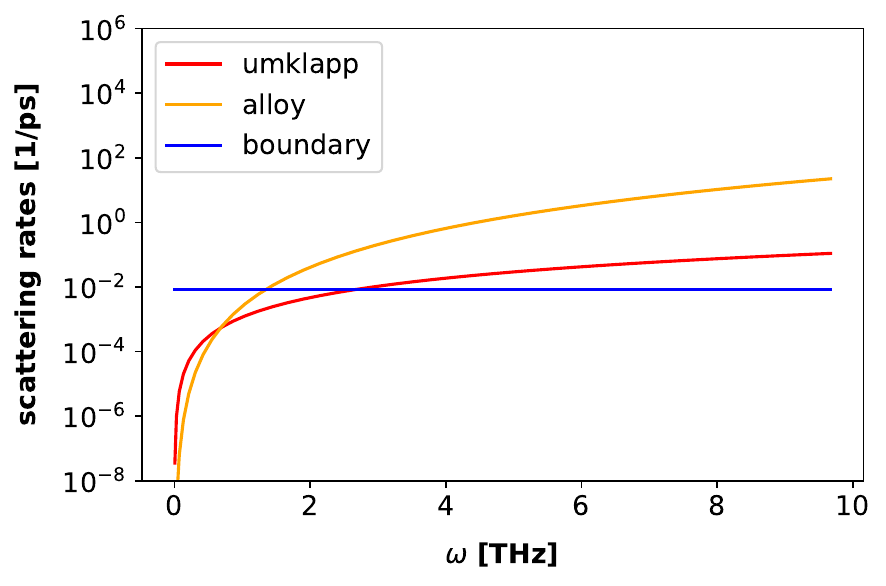}
\end{subfigure}
\hfill
\begin{subfigure}{0.329\textwidth}
\centering
\hspace{15pt}\textbf{acoustic band}
\includegraphics[width=\textwidth,trim=0 0 0 0,clip]{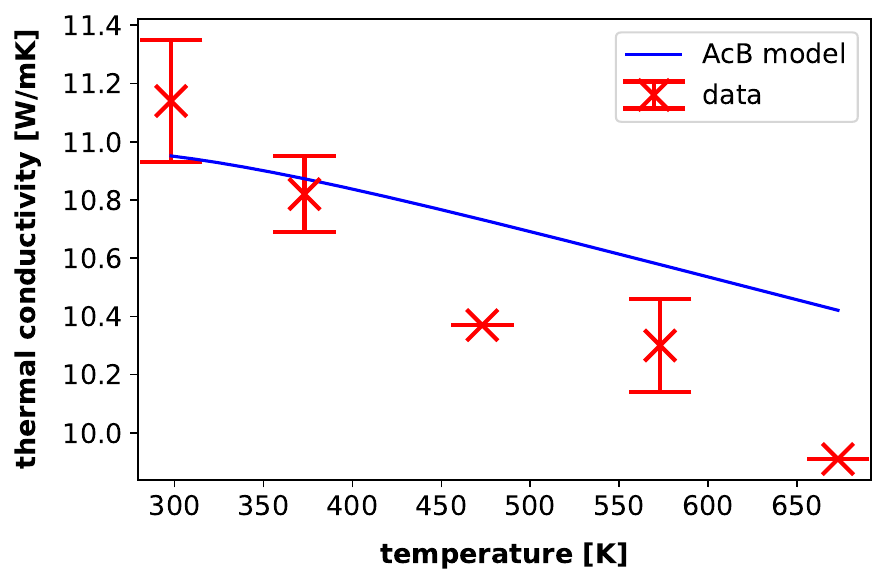} \\[5pt]
\includegraphics[width=\textwidth,trim=0 0 0 0,clip]{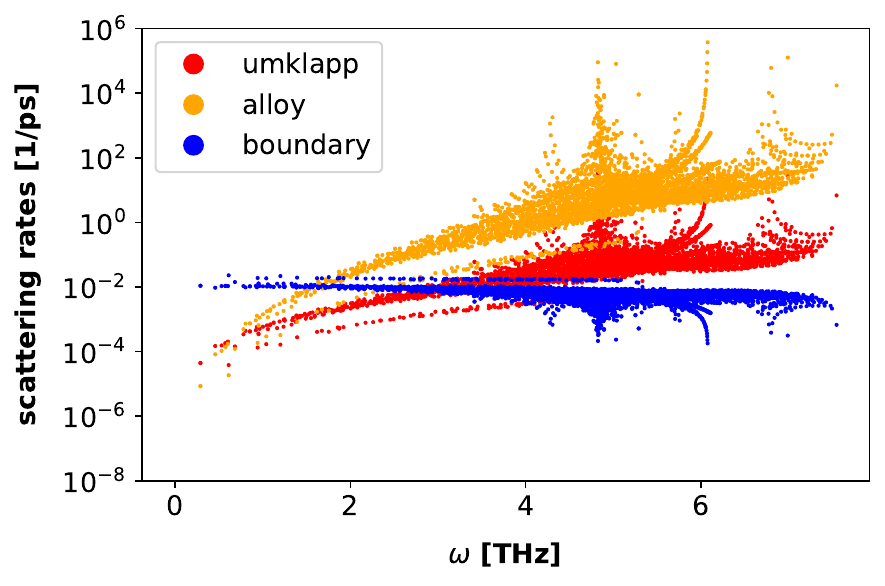}
\end{subfigure}
\hfill
\begin{subfigure}{0.329\textwidth}
\centering
\hspace{15pt}\textbf{ab-initio-derived}
\includegraphics[width=\textwidth,trim=0 0 0 0,clip]{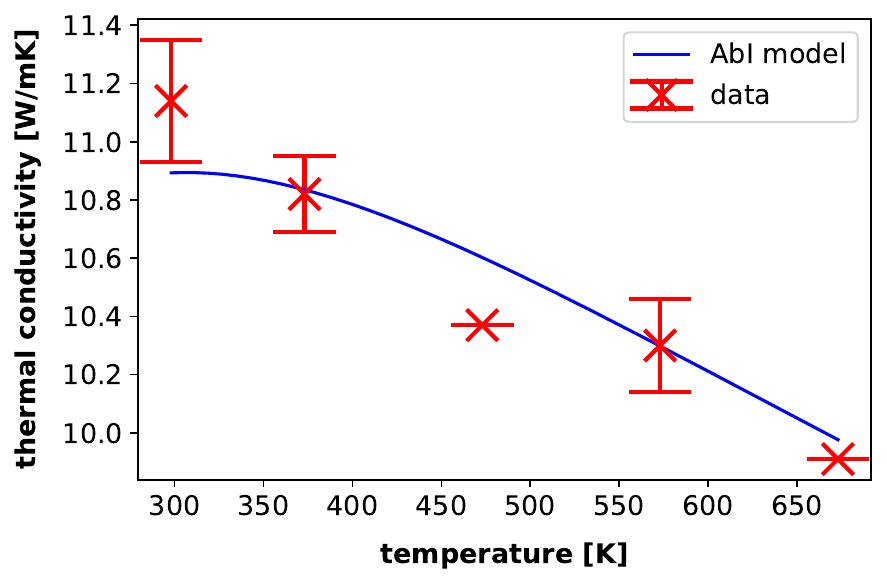} \\[5pt]
\includegraphics[width=\textwidth,trim=0 0 0 0,clip]{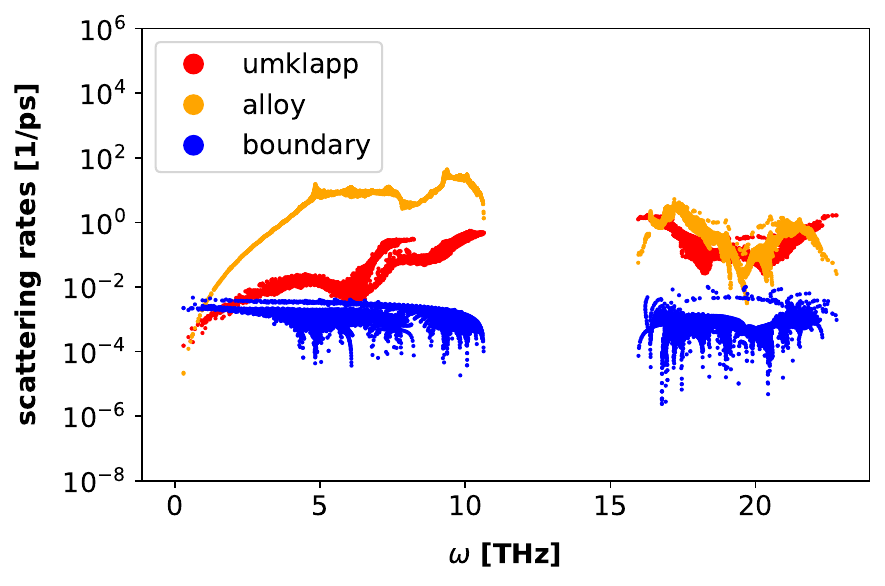}
\end{subfigure}
\caption{The top row shows the measured temperature-dependent thermal conductivity of the Al$_{0.15}$Ga$_{0.85}$N layer (red error bars), together with the fitted models (blue lines). The bottom row shows the corresponding scattering rates as a function of phonon frequency and at a temperature of 300 K. From left to right, the panels correspond to the Debye-approximation model (\autoref{sec:Deb}), the acoustic band model (\autoref{sec:AcB}), and the ab-initio-derived model (\autoref{sec:AbI}).}
\label{fig:id632_res}
\end{figure*}


\subsection{Al$_{0.15}$Ga$_{0.85}$N - isolated superlattice layer}\label{sec:results:id632}

\begin{table}[htbp!]
\centering
\caption{Fitted parameters and root-mean-square errors (RMSE) of the fits for the different models applied to the Al$_{0.15}$Ga$_{0.85}$N layer. To reproduce the measured thermal conductivity, Umklapp, alloy, and film-boundary scattering were included, with the film-boundary specularity $p_{\text{\tiny FB}}$ serving as the sole fitting parameter. The RMSE quantifies the deviation between the predicted and measured thermal conductivities.}
\resizebox{0.6\columnwidth}{!}{
\begin{tabular}{|c|c|c|}
\hline
\multicolumn{3}{|c|}{Al$_{0.15}$Ga$_{0.85}$N thin film} \\
\hline
Model  & p$_{\text{\tiny FB}}$  & RMSE [W/mK] \\
\hline
Deb (\ref{sec:Deb}) & 0.0  & 1.79 \\
\hline
AcB (\ref{sec:AcB})  & 0.0 & 0.32  \\
\hline
AbI (\ref{sec:AbI}) & 0.79 & 0.15 \\
\hline
\end{tabular}
}
\label{tab:id632_parameters}
\end{table}

\begin{figure}[htp!]
\centering
\includegraphics[width=\columnwidth,trim=0 0 0 0,clip]{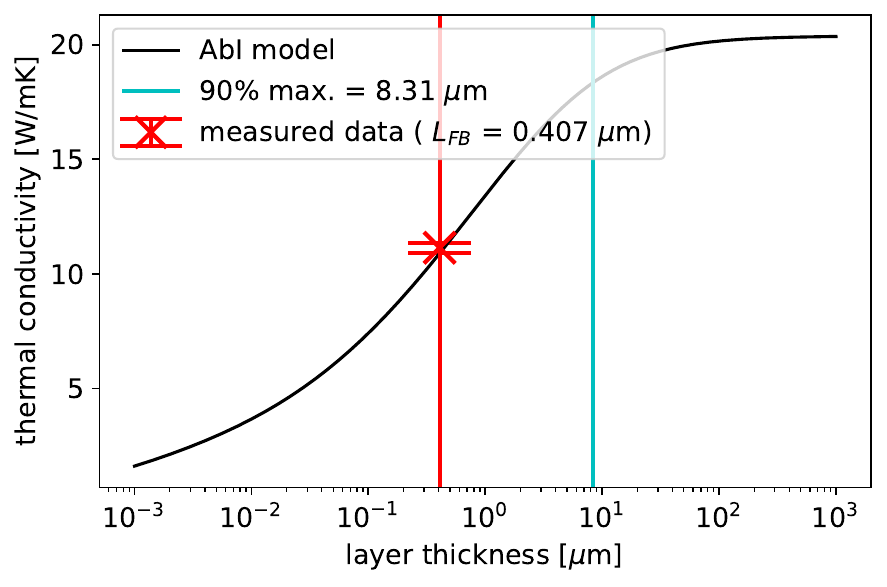} 
\caption{Thermal conductivity as a function of layer thickness for the Al$_{0.15}$Ga$_{0.85}$N thin film, calculated using the ab-initio-derived model (\autoref{sec:AbI}). The film-boundary specularity p$_{\text{\tiny FB}}$ was determined from fitting the measured temperature-dependent data shown in \autoref{fig:id632_res}.}
\label{fig:id632_kthick_ab_initio}
\end{figure}

The third sample, is a 407 nm-thick AlGaN layer that is used in AlGaN/AlN superlattice configurations. These have been proposed as an alternative to the conventional AlGaN buffer layer in HEMT stack designs. As shown in \autoref{fig:ktemp_AlGaN_overview}, the measured thermal conductivity of this layer (blue squares) again lies between the predictions of our ab-initio-derived model for the bulk-material and the fully diffusive thin-film limit, described by zero boundary specularity (p$_{\text{\tiny FB}}$ = 0; see discussion in \autoref{sec:boundary_scattering}). This indicates that, as in the previously discussed Al$_{0.5}$Ga$_{0.5}$N layer (see \autoref{sec:results_id744}), the film has high crystalline quality, so grain-boundary scattering need not be included in the model. In contrast to the essentially temperature-independent thermal conductivity of the buffer layer discussed in \autoref{sec:results_id744}, we here observe a weak decrease of the thermal transport with the sample temperature. In a sense, this is reminiscent of what one would expect for crystalline samples, but the decrease is comparably small. We attribute this difference to the larger sample thickness, which substantially reduces film-boundary scattering, as discussed in detail in the previous section.

Although some variation exists among the published datasets, our measurements again consistently fall within the range spanned by previous studies, providing independent support for the reliability of our results. Overall, we find close agreement with the results reported by Mitterhuber et al. \cite{Mitterhuber2020} for their 423 nm-thick Al$_{0.32}$Ga$_{0.68}$N sample. Interestingly, this agreement is observed despite the difference in alloy composition. This can be explained by the relatively weak dependence of thermal conductivity on alloy composition in the intermediate composition range ($\sim$15–85\%), where alloy scattering approaches saturation \cite{Abeles1963,Balandin2005,Ma2016}. Consequently, the measurements by Mitterhuber et al. agree more closely with our results for the present 407 nm-thick Al$_{0.15}$Ga$_{0.85}$N sample than with those for the previously discussed 103 nm-thick Al$_{0.5}$Ga$_{0.5}$N sample. Similarly, our measurements are in reasonable agreement with the second sample investigated by Mitterhuber et al., a substantially thinner 65 nm-thick Al$_{0.17}$Ga$_{0.83}$N film. The clearly reduced thermal conductivity of that film is attributed to the stronger film-boundary scattering in the thinner sample. In general, the thermal conductivity of our Al$_{0.15}$Ga$_{0.85}$N sample is also comparable to values reported for other AlGaN alloys with similar compositions ($x = 0.17$–$0.20$), provided that the reduced thermal conductivity due to the finite film thicknesses is accounted for.

As a next step, our models are fitted to the measurements following the same procedure as for the previous sample. This means that Umklapp, alloy, and film-boundary scattering are accounted for, again requiring only a single adjustable parameter: the film-boundary specularity, p$_{\text{\tiny FB}}$. The fitted results, are shown in the top row of \autoref{fig:id632_res}, together with the measured data. The corresponding scattering rates are presented in the bottom row. In contrast to the previous sample, where all the models could reproduce the experimental data reasonably well upon adjusting p$_{\text{\tiny FB}}$ (see \autoref{sec:results_id744}), now the ab-initio-derived model clearly provides the best match between theory and experiment. As discussed for the previous sample (see \autoref{sec:results_id744}), the overestimation of the bulk thermal conductivity of the Al$_{x}$Ga$_{1\!-\!x}$N alloy by the Deb and AcB models needs to be compensated by an artificially increased film-boundary scattering (i.e., a lower $p_{\text{\tiny FB}}$). For the previous sample, this compensation by diminishing the value of $p_{\text{\tiny FB}}$ was possible due to the substantially smaller thickness. However, for the thicker layer considered here, even fully diffuse film-boundary scattering ($p_{\text{\tiny FB}} = 0$) is insufficient to reproduce the measured thermal conductivity. Encouragingly, and in sharp contrast to the failure of the Debye-approximation and acoustic band models, for the ab-initio-derived model the fitted film-boundary specularity $p_{\text{\tiny FB}} = 0.79$ is very close to that obtained for the previous sample. The consistency of this parameter across samples with different thicknesses suggests that the model provides a robust description of the intrinsic thermal conductivity, further supporting the reliability and consistency of the ab-initio-derived model. 

As in the case of the Al$_{0.5}$Ga$_{0.5}$N buffer layer discussed in the previous section, significant size effects are also observed for the film considered here (see \autoref{fig:id632_kthick_ab_initio}). According to the ab-initio-derived model, the film thickness must reach 8.31 $\mu$m for the thermal conductivity to attain 90\% of its bulk value. This is comparable in magnitude to the corresponding thickness of 5.61 $\mu$m obtained for the previous sample.


\section{Summary and Conclusions}\label{sec:conclusion}

The thermal transport properties of thin film layers typically used in AlGaN HEMT designs were analyzed. To determine them experimentally, the Time-Domain Thermoreflectance method was applied, and three computational models of increasing sophistication were fitted to the data. They consist of a model relying solely on isotropic acoustic phonons with a linear dispersion in the spirit of the Debye-approximation (Deb), a model fully accounting for the acoustic bands only (AcB), and a full ab-initio-based model considering the detailed properties of all phonons (AbI). These models allow us to identify the main scattering sources, some of which are intimately linked to underlying structural properties, such as alloy disorder, grain boundaries and finite film thicknesses. In this work, Umklapp scattering contributions in the Deb and AcB models are determined by fitting to first-principles data for ideal bulk single crystals. This avoids potential artefacts from fitting an expression intended to describe only Umklapp scattering to experimental data, which inevitably include other scattering mechanisms. In addition, fixing the Umklapp contributions prevents errors in the model from being compensated by adjusting the temperature dependence of the scattering. Importantly, following this procedure, all models accurately reproduced the experimental temperature-dependent thermal conductivity of bulk single-crystalline AlN. 

Nevertheless, our measurements of the 172 nm-thick AlN thin film, show a thermal conductivity significantly differing from that of a single crystal, with thermal conductivity values roughly an order of magnitude smaller and with a massively reduced temperature dependence. The latter is unusal for a highly crystalline sample and resembles the situation in amorphous or fine-grained, polycrystalline materials. Consistently, reproducing the experimental data by the models requires the inclusion of grain-boundary scattering with effective grain sizes well below 10 nm. In sharp contrast, both the 103 nm-thick Al$_{0.5}$Ga$_{0.5}$N and the 407 nm-thick Al$_{0.15}$Ga${_0.85}$N samples were identified as being highly crystalline. Despite the particular relevance of alloy scattering in these samples, also film-boundary scattering plays a decisive role, especially in the low-frequency, long mean-free path regime. This becomes particularly apparent from the observation that in the thinner film, the thermal conductivity weakly increases with temperature, while for the thicker film it decreases with temperature. In fact, for Al$_{0.5}$Ga$_{0.5}$N and Al$_{0.15}$Ga$_{0.85}$N, the developed ab-initio-derived model predicts that thicknesses of 5.6 $\mu$m and 8.3 $\mu$m are required to achieve 90\% of the bulk thermal conductivity. 

In conclusion, this work highlights the limitations of bulk-based descriptions of thermal transport when applied to real-world thin-film structures. The thermal conductivity of the investigated layers is significantly reduced compared to bulk single crystals, reflecting the importance of additional scattering mechanisms and the specific characteristics of the materials and structures involved. An accurate description of their thermal transport, therefore, requires a detailed treatment of the full phonon dispersion including optical modes, as well as the consideration of anharmonic scattering rates. Simplified analytical models, such as the Debye-approximation and acoustic band models, can provide qualitative insights but cannot fully capture the relevant physics across the investigated systems. Especially interfacial layers can be highly defective resulting in small crystallite sizes leading to a dominance of grain-boundary scattering. Unexpectedly, this renders the AlN buffer layer the heat-transport bottleneck in the considered multi-layer structure, illustrating the importance of considering the thermal properties of individual layers rather than focusing exclusively on the nominally dominant materials. The pronounced dependence of thermal conductivity on film thickness is associated with the intricate interplay between Umklapp, alloy, and film-boundary scattering, which also helps explaining the large variations in experimental results reported in the literature. In this context, reliable experimental data for single-crystalline bulk Al$_{x}$Ga$_{1\!-\!x}$N would be highly desirable as a reference for further modelling efforts.

\vspace{10pt}
\textbf{Funding:} This work was conducted as part of the iRel4.0 project which is a European co-funded innovation project that has been granted by the ECSEL Joint Undertaking (JU) under grant agreement No 876659. The funding of the project comes from the Horizon 2020 research program and participating countries. National funding is provided by Germany, including the Free States of Saxony and Thuringia, Austria, Belgium, Finland, France, Italy, the Netherlands, Slovakia, Spain, Sweden, and Turkey. Additional funding by the Austrian Federal Government (in particular from Federal Ministry for Climate Action, Environment, Energy, Mobility, Innovation and Technology) represented by the Austrian Research Promotion Agency (FFG), within the framework of the “IKT der Zukunft” Programme (project number: 877534) is gratefully acknowledged.

\vspace{10pt}
\textbf{Acknowledgments:} 
The authors acknowledge Infineon Technologies AG for the samples used for this study.
The author acknowledges the use of ChatGPT 5 (OpenAI) for assistance with language editing and improving the readability of the manuscript. All content was reviewed and edited by the authors, who assume full responsibility for the final text.\\
The authors would also like to thank Markus Krainz for providing the software that was used in the evaluation of TDTR measurement, as well as, his assistance with measurements.


\bibliographystyle{IEEEtran}
\bibliography{references}

\onecolumn

\title{\textbf{\Huge Supporting Information}\\[20pt]{\Huge Thermal investigation of buffer and transition layers in AlGaN HEMT stacks}\vspace{20pt}}

\date{\today}

\author{%
\begin{minipage}[t]{0.45\textwidth}
\centering
\textbf{Simon Fernbach\thanks{Corresponding author: simon.fernbach@live.at}}\\
\textit{Microelectronics}\\
\textit{Materials Center Leoben}\\
Leoben, Austria\\
ORCID: 0009-0008-2169-3869
\end{minipage}
\hfill
\begin{minipage}[t]{0.45\textwidth}
\centering
\textbf{Lisa Mitterhuber}\\
\textit{Microelectronics}\\
\textit{Materials Center Leoben}\\
Leoben, Austria\\
ORCID: 0000-0001-7734-5603
\end{minipage}

\par\vspace{1.5em}

\begin{minipage}[t]{0.45\textwidth}
\centering
\textbf{Egbert Zojer}\\
\textit{Institute of Solid State Physics}\\
\textit{Technische Universität Graz}\\
Graz, Austria\\
ORCID: 0000-0002-6502-1721
\end{minipage}
\hfill
\begin{minipage}[t]{0.45\textwidth}
\centering
\textbf{Natalia Bedoya-Martínez}\\
\textit{Simulation}\\
\textit{Materials Center Leoben}\\
Leoben, Austria\\
ORCID: 0000-0001-9824-6129
\end{minipage}
}

\maketitle

\end{comment}

\newpage

\section*{\centering\textbf{\Huge Supporting Information}\\[20pt]}
\section{SEM Images}\label{sec:SEM}

\vspace{10pt}

The layer thicknesses were characterized by cross-sectional scanning electron microscopy (SEM) using a Crossbeam 550 (Zeiss, Oberkochen, Germany). Cross-sectional specimens were prepared by focused ion beam (FIB) milling to obtain high-quality cross-sections suitable for thickness measurements. SEM imaging was performed at an accelerating voltage of 2 kV, and all micrographs presented in this work were acquired at a magnification of 50,000 times. Layer thicknesses were determined directly from the cross-sectional SEM images.\vspace{20pt}

\begin{figure*}[htp!]
\begin{subfigure}{0.326\textwidth}
\includegraphics[width=\textwidth,trim=0 0 0 0,clip]{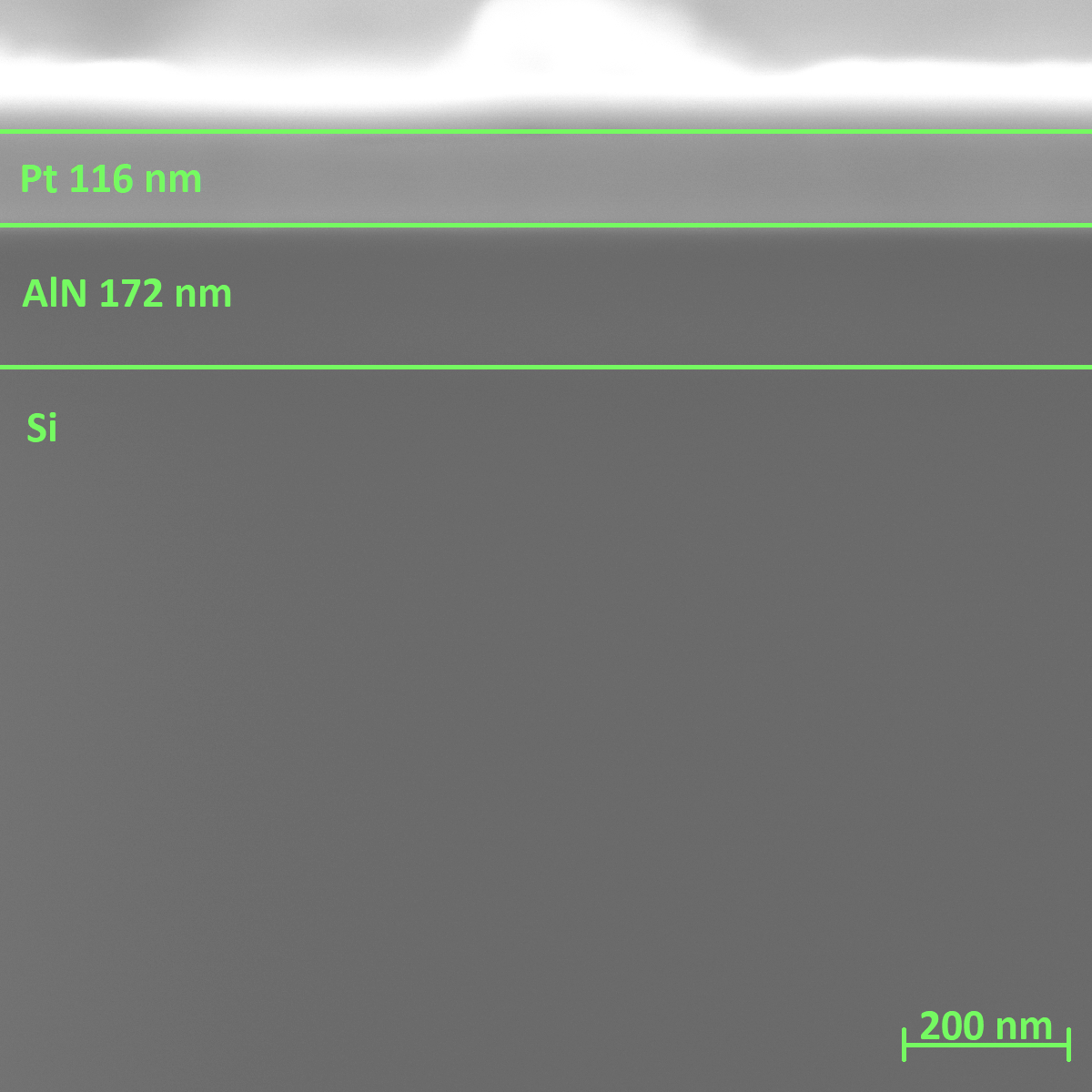}
\end{subfigure}
\begin{subfigure}{0.326\textwidth}
\includegraphics[width=\textwidth,trim=0 0 0 0,clip]{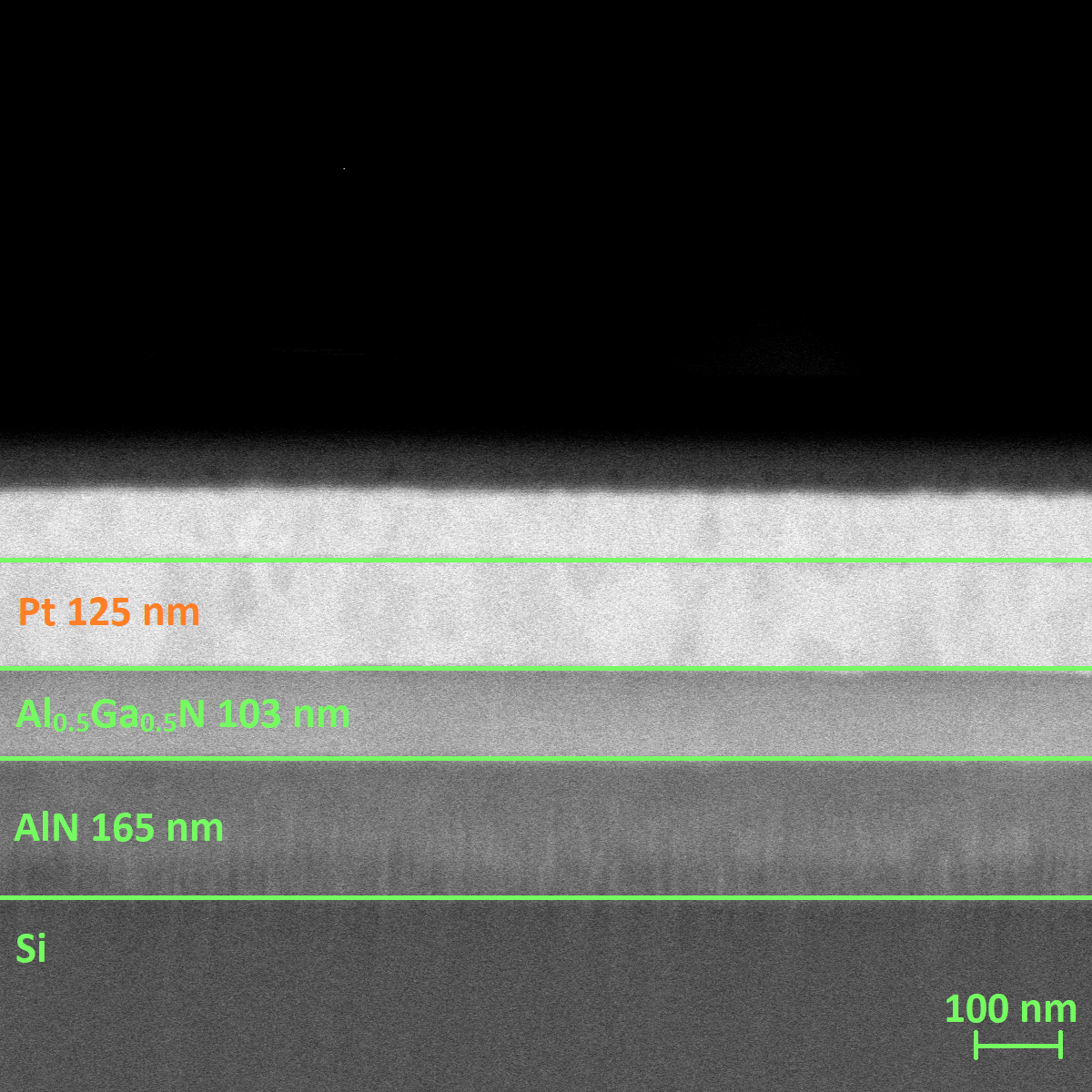}
\end{subfigure}
\begin{subfigure}{0.326\textwidth}
\includegraphics[width=\textwidth,trim=0 0 0 0,clip]{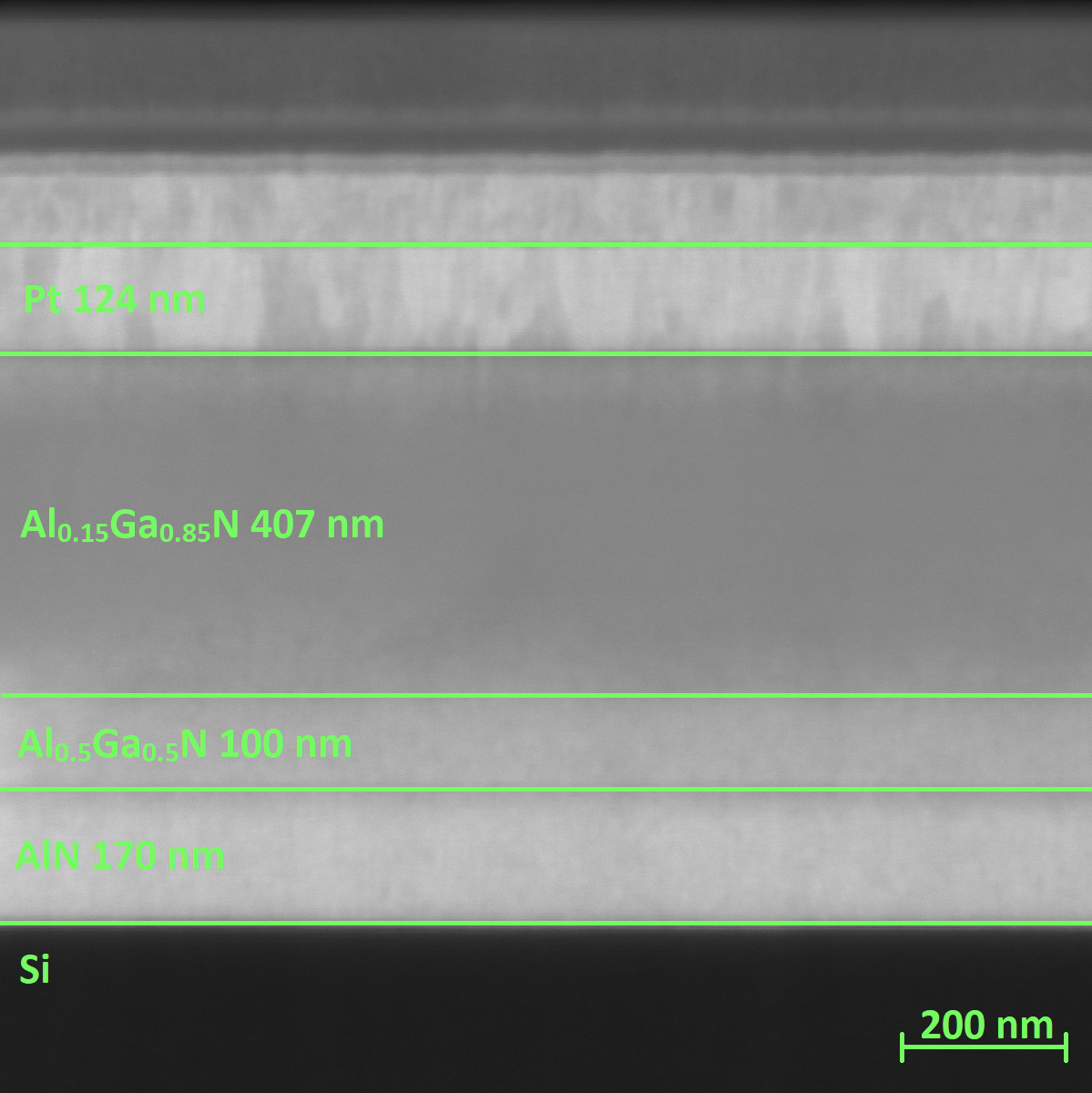}
\end{subfigure}
\caption{SEM images of the three samples investigated in this work.}
\label{fig:sample_REM}
\end{figure*}


\newpage
\section{Relaxation time approximation and normal scattering}\label{sec:RTA_test}

\vspace{10pt}

In the relaxation time approximation, both Umklapp and Normal processes are treated as resistive, while the actual impact of normal scattering is primarily indirect via modifying the phonon distribution without changing the overall phonon momentum.

To assess the impact of this incorrect treatment of phonon scattering, we compare RTA-level results with results obtained in an iterative solution of the BTE for the materials considered in this work at 300 K and 1000 K. The results, shown in \autoref{tab:normal_scattering}, demonstrate that for AlN the impact is relatively small, albeit non-vanishing; for the alloys it is negligible.

\vspace{20pt}

\begin{table}[htbp]
\centering
\caption{Scalar thermal conductivity ($\kappa = \frac{1}{3}\sum_{\alpha}\kappa_{\alpha\alpha}$) calculated within the relaxation-time approximation (RTA) and by iteratively solving the BTE for the materials considered in this work (AlN, Al$_{0.5}$Ga$_{0.5}$N, and Al$_{0.15}$Ga$_{0.85}$N) at 300 and 1000 K.}
\resizebox{0.7\columnwidth}{!}{
\begin{tabular}{|c|c|c|c|c|}
\cline{3-5}
\multicolumn{2}{c|}{} & AlN & Al$_{0.15}$Ga$_{0.85}$N & Al$_{0.5}$Ga$_{0.5}$N \\
\hline
\multirow{2}{*}{300 K} & RTA & 308.13 & 22.82 & 12.25  \\
& full & 332.69 & 22.80 & 12.25 \\
\hline
\multirow{2}{*}{1000 K} & RTA & 68.81 & 12.43 & 7.40  \\
& full & 73.80 & 12.40 & 7.39 \\
\hline
\end{tabular}}
\label{tab:normal_scattering}
\end{table}


\newpage
\section{Derivation of the isotropic thermal conductivity expression}\label{sec:callaway_derivation}

\vspace{10pt}

Models that utilize analytically descriptions of the phonon dispersion, such as the Callaway, Holland or Born-von-Karman models, typically employ an isotropic variant of the lattice thermal conductivity expression, which is given by Eq. 4 in the main text. In the following we will derive this formula from the lattice thermal conductivity expression given by Eq. 1 of the main text.\\

Taking the thermodynamic limit $\frac{1}{V}\sum_{\vec{q}}\rightarrow \frac{1}{\left(2\pi\right)^3}\int_{q}d\vec{q}$, i.e., assuming large systems, one can replace the sum over reciprocal lattice vectors by an integral:

\begin{equation*}
\begin{split}
     \kappa_{\alpha\beta}&=\frac{1}{V}\sum_{\vec{q}}\sum_\nu^{3\,N} C_\text{ph}(\vec{q},\nu)\,{v}_\text{g}^{\alpha}(\vec{q},\nu)\,{v}_\text{g}^{\beta}(\vec{q},\nu)\,\tau(\vec{q},\nu)\\
    &= \frac{1}{\left(2\pi\right)^3}\sum_{\nu}\int C_\text{ph}(\vec{q},\nu)\,{v}_\text{g}^{\alpha}(\vec{q},\nu)\,{v}_\text{g}^{\beta}(\vec{q},\nu)\,\tau(\vec{q},\nu) d\vec{q}    
    \end{split}
\end{equation*}

Optical phonons are neglected, since it is assumed they do not significantly contribute to the thermal conductivity. The phonon band structure of the acoustic branches is approximated by a linear and three-fold degenerate Debye-like dispersion $\omega(\vec{q},\nu) = \omega(\vec{q}) = \text{v}_s \,\vert\vec{q}\vert$, with v$_s$ being the speed of sound. Because the dispersion is degenerate, the sum over the bands can be replaced by a factor of 3. Moreover, due to the isotropic nature of the dispersion, the group velocity can be written as $v_{\text{g}} = \frac{d\omega}{d\vec{q}} = v_{s}\frac{\vec{q}}{\vert\vec{q}\vert} = v_s\,\vec{e}_{q}$, with $\vec{e}_q$ being the radial unit vector in spherical coordinates. Changing to spherical coordinates and integrating over the angular components, the integral in the expression above can then be written as:

\begin{equation*}
\begin{gathered}
    \kappa_{\alpha\beta} = \frac{3}{\left(2\pi\right)^3}\int_{q}C_\text{ph}\left(q, \nu\right)\tau\left(q,\nu\right)q^2v_s^2 dq \int_{\theta}\int_{\phi}\vec{e}_{q,\alpha}\vec{e}_{q,\beta}\sin\left(\theta\right)d\theta d\phi 
\end{gathered}
\end{equation*}

The angular integrals over the components of the group velocities yield a diagonal tensor, where all the trace elements are $\frac{4\pi}{3}$. Therefore, the tensor $\kappa_{\alpha\beta}$ can be replaced by a scalar 

\begin{equation} 
\begin{gathered}
    \kappa = \frac{4\pi}{3}\frac{3}{\left(2\pi\right)^3}\int_{q}C_{\text{ph}}\left(q,\nu\right)\tau\left(q,\nu\right)v_{s}^2q^2dq \\
\end{gathered}
\end{equation}

Then changing variables of integration from $q$ to $\omega$ and introducing a maximum frequency cutoff $\omega_D$ for the integration, finally yields

\begin{equation*}
\begin{gathered}
    \kappa = \frac{1}{3}\int_{\omega=0}^{\omega_D}\frac{3\omega^2}{2\pi^2v_s^3}C_{\text{ph}}\left(\omega,T\right)v_s^2\tau\left(\omega,T\right)d\omega \\
    = \frac{1}{3}\int_{\omega=0}^{\omega_D}D\left(\omega\right)C_{\text{ph}}\left(\omega,\text{\small T}\right)v_s^2\tau\left(\omega,T\right)d\omega
\end{gathered}
\end{equation*}

with the Debye frequency $\omega_D$ and the Debye density of states (DOS) $D\left(\omega\right) = \frac{3\omega^2}{2\pi^2v_s^3}$. The Debye frequency is determined from $\int_{0}^{\omega_D} D\left(\omega\right)d\omega = 3n$. This equation states that the number of states per volume, which is given by the integral over the DOS, equals 3 (for the three acoustic modes) times the atom density $n$.


\newpage
\section{Convergence tests}\label{sec:convergence_test}

\begin{figure}[htbp]
\begin{subfigure}{0.49\textwidth}
\centering
\textbf{AlN}
\includegraphics[width=\textwidth,trim=0 0 0 0,clip]{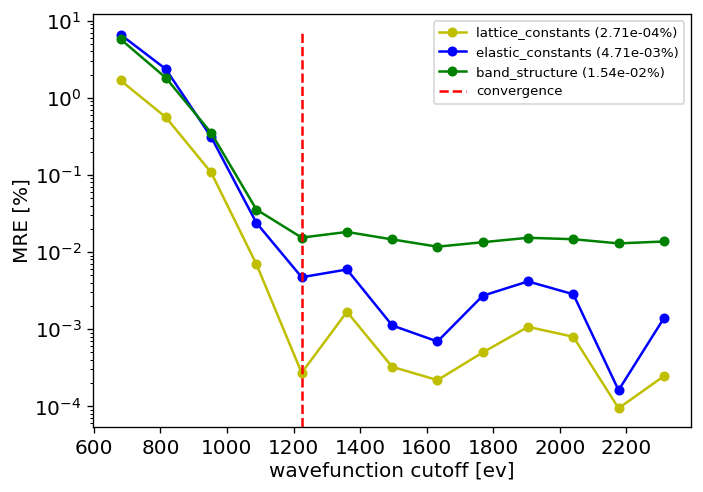} \\[5pt]
\includegraphics[width=\textwidth,trim=0 0 0 0,clip]{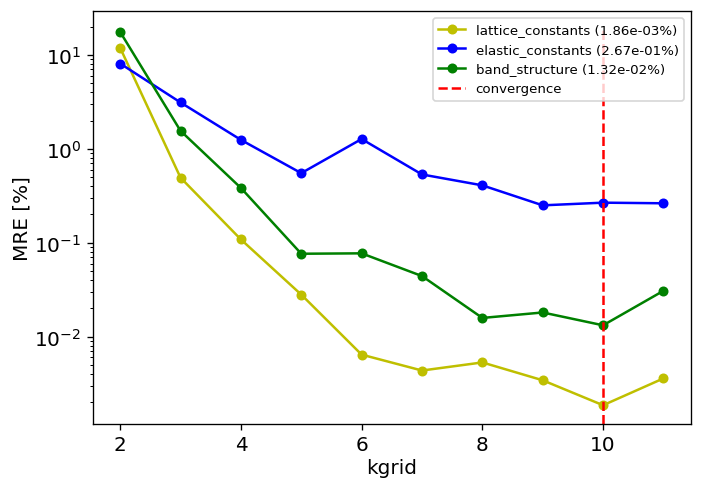}
\end{subfigure}
\begin{subfigure}{0.49\textwidth}
\centering
\textbf{GaN}
\includegraphics[width=\textwidth,trim=0 0 0 0,clip]{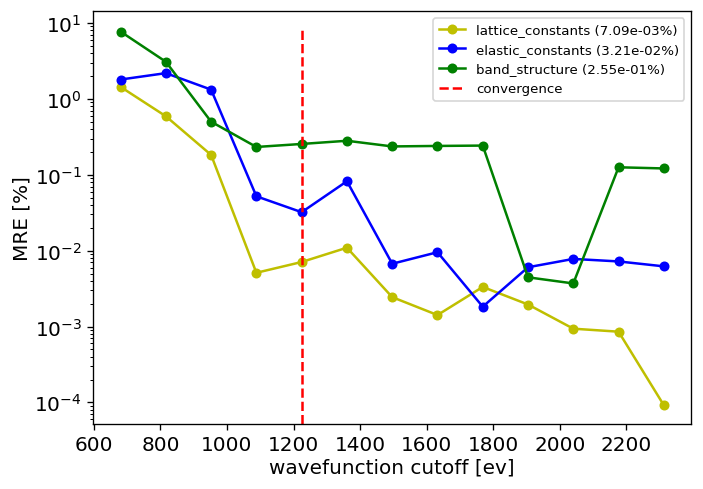} \\[5pt]
\includegraphics[width=\textwidth,trim=0 0 0 0,clip]{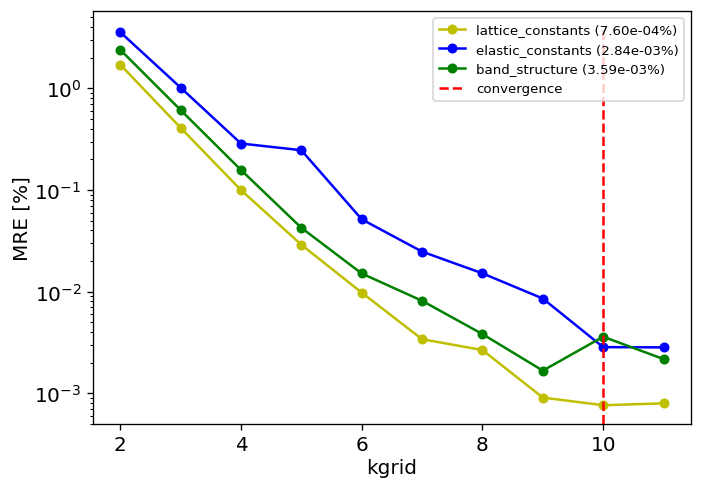}
\end{subfigure}
\caption{Mean relative errors of different material properties as a function of the DFT parameter (wavefunction cutoff or electronic reciprocal-space grid size), calculated with respect to the results obtained using the highest parameter values. The top and bottom rows show the results for the wavefunction cutoff and electronic reciprocal-space grid size (k-grid), respectively. The left and right columns show the results for AlN and GaN, respectively. The properties considered are the lattice constants, elastic constants, and $\Gamma$-point phonon frequencies, the latter denoted as "band\_structure".}
\label{fig:convergence_DFT}
\end{figure}

\begin{figure}[htbp]
\begin{subfigure}{0.49\textwidth}
\centering
\textbf{AlN}
\includegraphics[width=\textwidth,trim=0 0 0 0,clip]{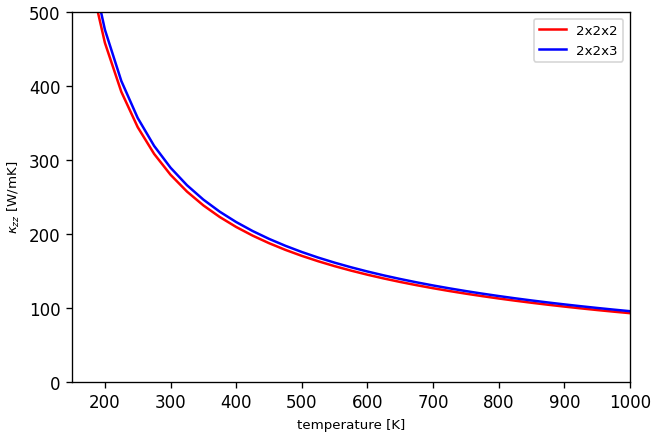}
\end{subfigure}
\begin{subfigure}{0.49\textwidth}
\centering
\textbf{GaN}
\includegraphics[width=\textwidth,trim=0 0 0 0,clip]{plots/convergence/convergence_GaN_FC3.png} 
\end{subfigure}
\caption{Comparison between the DFT calculated $\kappa_{zz}$ (RTA) with 2x2x2 (red) and 2x2x3 (blue) FC3 supercells. Results for AlN are shown on the left and for GaN on the right.}
\label{fig:convergence_FC3}
\end{figure}

\begin{figure}[htbp]
\centering
\includegraphics[width=0.5\linewidth]{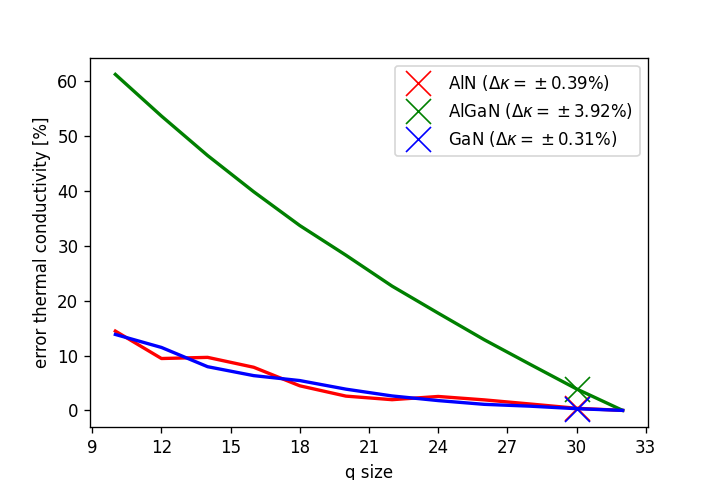}
\caption{Relative error in the scalar thermal conductivity ($\kappa = \frac{1}{3}\sum_{\alpha}\kappa_{\alpha\alpha}$) as a function of the reciprocal-space phonon grid size, with respect to the result obtained using the largest grid size considered ($32 \times 32 \times 32$). All calculations were performed using ShengBTE within the relaxation-time approximation (RTA).}
\label{fig:convergence_qgrid}
\end{figure}

We performed a number of convergence tests in the context of our ab initio simulations, which are summarized in the following. The convergence of our DFT calculations with respect to electronic reciprocal grid size (k-grid) and wave function cutoff was tested for AlN and GaN, capturing the impact of all pseudopotentials involved. For assessing the convergence, we consider the change in mean relative errors of the lattice constants, the elastic constants, and the Gamma-point phonons frequencies with respect to the highest values of the parameters used in the DFT calculations. The results are shown in \autoref{fig:convergence_DFT}. There we can see that for a wave function cutoff of 90 Ry (1225 eV), as used in our calculations, all properties are well converged. The errors for the lattice constants, elastic constants and Gamma-point phonon frequencies of AlN and GaN are $2.71\cdot10^{-4}\,\%$, $4.71\cdot10^{-3}\,\%$, $1.54\cdot10^{-2}\,\%$ and $7.09\cdot10^{-3}\,\%$, $3.21\cdot10^{-2}\,\%$, $2.55\cdot10^{-1}\,\%$. For the reciprocal grid size we use a 10x10x10 grid, which is also well converged. The corresponding errors are $1.86\cdot10^{-3}\,\%$, $2.67\cdot10^{-1}\,\%$, $1.32\cdot10^{-2}\,\%$  and $7.60\cdot10^{-4}\,\%$, $2.84\cdot10^{-3}\,\%$, $3.59\cdot^10{-3}\,\%$. The convergence of the thermal conductivity with respect to the reciprocal phonon grid size is shown in \autoref{fig:convergence_qgrid}. There we can see that the grid size of $30\times30\times30$ used in this work, is well converged with relative errors of 0.39\%, 3.92 and 0.31\%. The larger relative errors for Al$_{0.5}$Ga$_{0.5}$N are caused by alloy scattering in terms of the Tamura expression given in Eq. 7 of the main text. It samples the phonon DOS through the energy conserving delta function and is, therefore, much more sensitive to the grid size. Lastly, the impact of a larger FC3 supercell size ($2\times2\times3$ instead of $2\times2\times2$) on the value of $\kappa_{zz}$ is shown in \autoref{fig:convergence_FC3} and found to be negligible.


\newpage
\section{Results - bulk, single-crystalline materials}\label{sec:single_crystal_figures}

\begin{figure}[htpb]
\centering
\begin{subfigure}{0.32\textwidth}
\includegraphics[width=\textwidth,trim=0 0 0 0,clip]{plots/single_crystal/id762/ktemp_deb.pdf}
\end{subfigure}
\hfill
\begin{subfigure}{0.32\textwidth}
\caption*{\large AlN}
\includegraphics[width=\textwidth,trim=0 0 0 0,clip]{plots/single_crystal/id762/ktemp_acb.pdf}
\end{subfigure}
\hfill
\begin{subfigure}{0.32\textwidth}
\includegraphics[width=\textwidth,trim=0 0 0 0,clip]{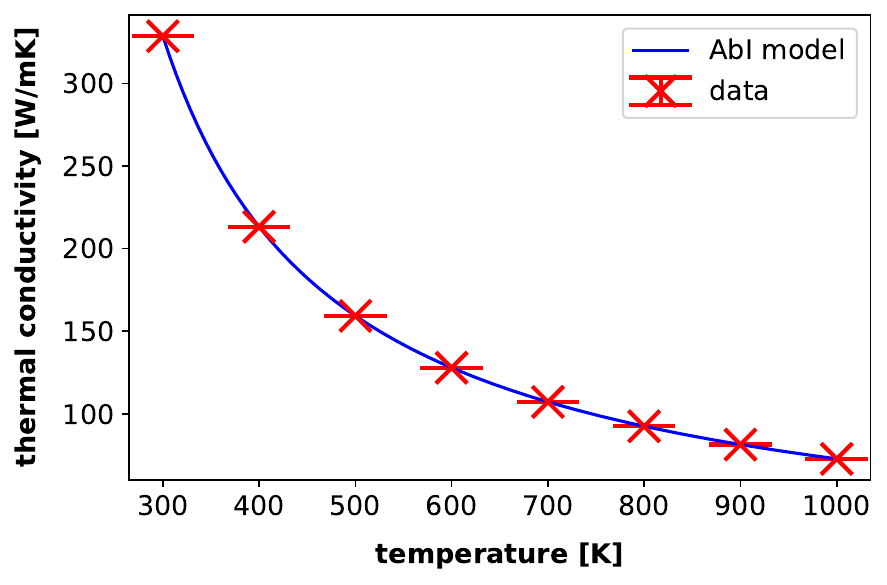}
\end{subfigure}
\begin{subfigure}{0.32\textwidth}
\includegraphics[width=\textwidth,trim=0 0 0 0,clip]{plots/single_crystal/id744/ktemp_deb.pdf}
\end{subfigure}
\hfill
\begin{subfigure}{0.32\textwidth}
\caption*{\large Al$_{0.5}$Ga$_{0.5}$N}
\includegraphics[width=\textwidth,trim=0 0 0 0,clip]{plots/single_crystal/id744/ktemp_acb.pdf}
\end{subfigure}
\hfill
\begin{subfigure}{0.32\textwidth}
\includegraphics[width=\textwidth,trim=0 0 0 0,clip]{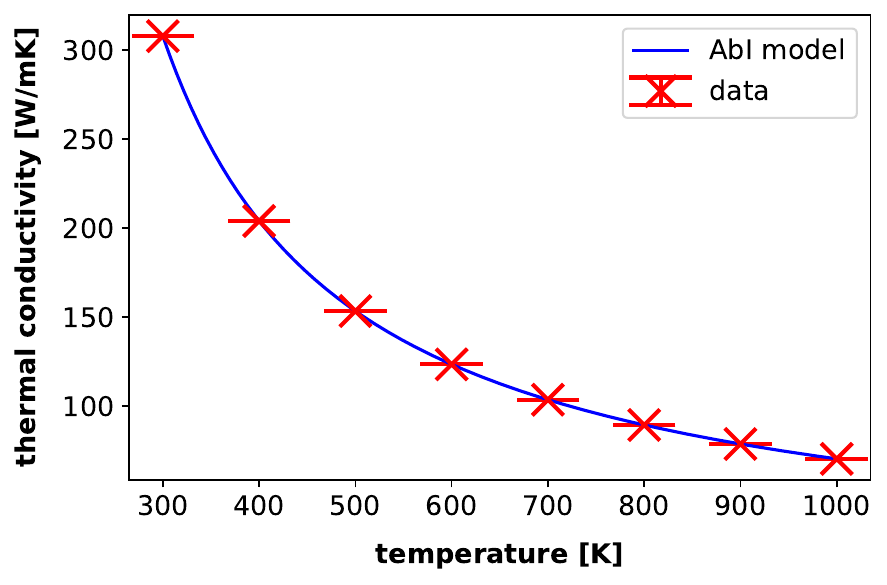}
\end{subfigure}
\begin{subfigure}{0.32\textwidth}
\includegraphics[width=\textwidth,trim=0 0 0 0,clip]{plots/single_crystal/id632/ktemp_deb.pdf}
\subcaption{\large Debye-approximation}
\end{subfigure}
\hfill
\begin{subfigure}{0.32\textwidth}
\caption*{\large Al$_{0.15}$Ga$_{0.85}$N}
\includegraphics[width=\textwidth,trim=0 0 0 0,clip]{plots/single_crystal/id632/ktemp_acb.pdf}
\subcaption{\large acoustic band}
\end{subfigure}
\hfill
\begin{subfigure}{0.32\textwidth}
\includegraphics[width=\textwidth,trim=0 0 0 0,clip]{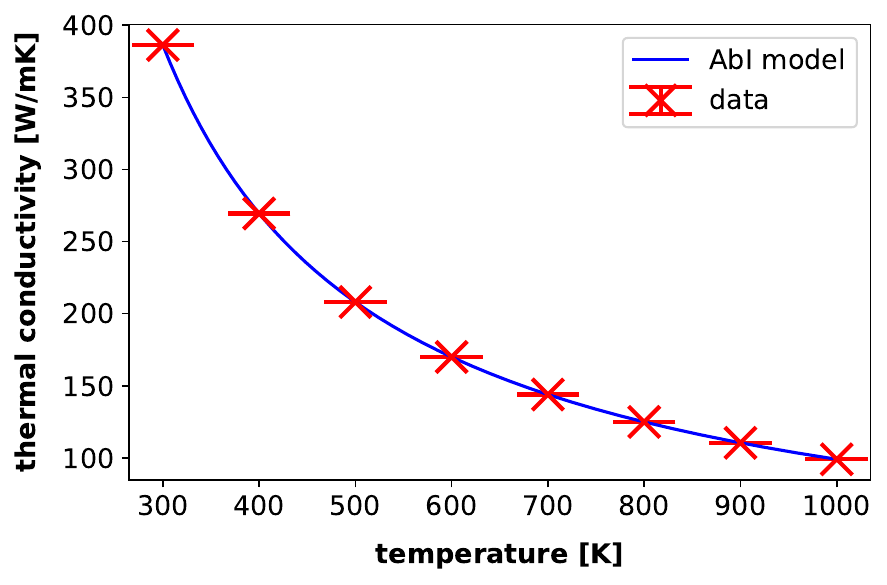}
\subcaption{\large ab-initio-derived}
\end{subfigure}
\caption{Comparison of the temperature-dependent thermal conductivity predicted by our models (blue lines) with first-principles reference data calculated using ShengBTE (red markers) for ideal, bulk, single-crystalline Aluminium Nitride (top row), Al$_{0.5}$Ga$_{0.5}$N (middle row) and Al$_{0.15}$Ga$_{0.15}$N (bottom row). The models from left to right are the Debye-approximation model (Sec. II-D), the acoustic band model (Sec. II-E), and ab-initio-derived model (Sec. II-F). In the case of AlGaN, alloy scattering is neglected, thus these results describe a ficticous, idealized, and effective material.}
\label{fig:res_single_crystal_2}
\end{figure}

\hyperref[fig:res_single_crystal_2]{Figure \ref{fig:res_single_crystal_2}} compares our models with first-principles calculations of the temperature-dependent thermal conductivity of idealized bulk single crystals computed with ShengBTE in the relaxation-time approximation. These calculations consider Umklapp scattering as the sole scattering mechanism (see Sec. 3A of the main manuscript). The Debye-approximation and acoustic-band models were fitted to the first-principles reference data, and the resulting fit parameters (Grüneisen parameter and $b$) are reported in Tab. 2 of the main manuscript. The ab-initio-derived model (right column) is included to demonstrate that it reproduces the first-principles results that rely on the same phonon band structures and lifetimes, but are directly calculated with ShengBTE.


\newpage
\section{Effects of suppressing Umklapp scattering of optical phonons}\label{sec:surpressing_umklapp}

\vspace{10pt}

As discussed in Sec. III-B of the main text, the measured thermal conductivity of the AlN transition layer cannot be adequately described by assuming a single-crystalline thin film. Instead, grain-boundary scattering must be included to account for the polycrystalline nature of the sample. However, combining grain-boundary and Umklapp scattering results in substantially poorer agreement with the experimental data than considering grain-boundary scattering alone. If both scattering contributions are individually described correctly, this suggests that their combined effect may not be adequately captured by simply adding the scattering rates according to Matthiessen's rule and using the resulting total phonon lifetimes in the BTE within the relaxation-time approximation.\\

\begin{figure*}[hpb!]
\centering
\begin{subfigure}{0.40\textwidth}
\centering
\textbf{\large full Umklapp scattering}
\includegraphics[width=\textwidth,trim=0 0 0 0,clip]{plots/id762/with_umklapp/ktemp_abi.pdf}
\includegraphics[width=\textwidth,trim=0 0 0 0,clip]{plots/id762/with_umklapp/w_abi.pdf}
\end{subfigure}
\hfill
\begin{subfigure}{0.40\textwidth}
\centering
\textbf{\large suppressed Umklapp scattering}
\includegraphics[width=\textwidth,trim=0 0 0 0,clip]{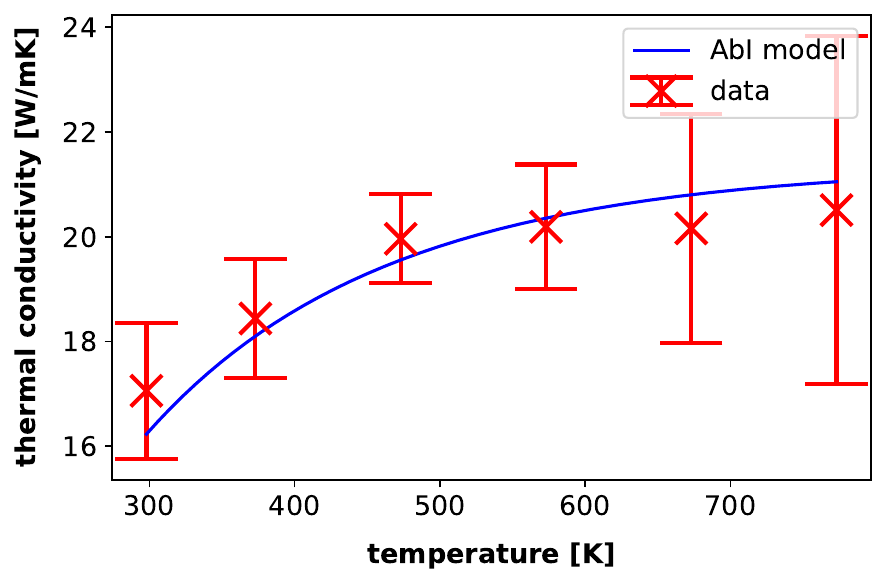}
\includegraphics[width=\textwidth,trim=0 0 0 0,clip]{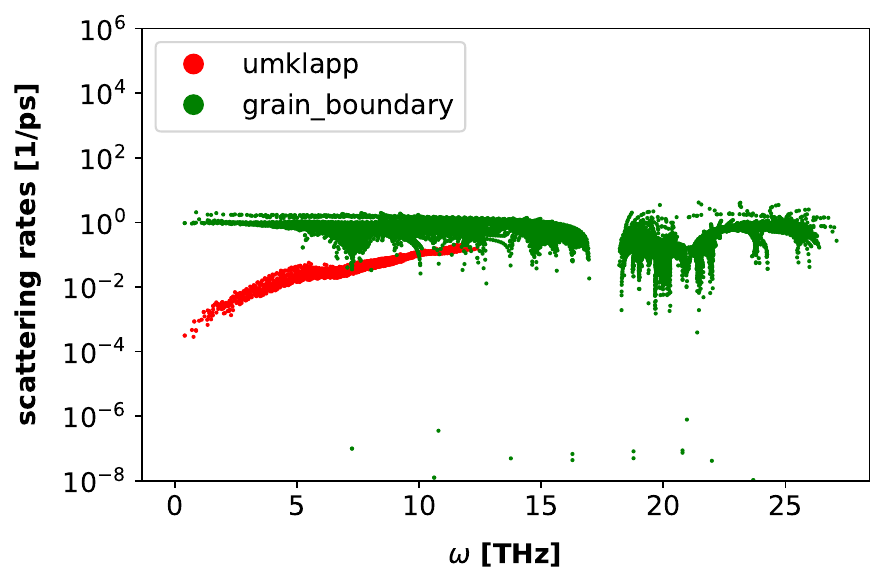}
\end{subfigure}
\caption{The top row shows the measured temperature-dependent thermal conductivity of the AlN transition layer (red error bars), together with the fitted ab-initio-derived model (blue lines). The bottom row shows the corresponding scattering rates as a function of phonon frequency and at a temperature of 300 K. The left and right columns show the results obtained when the full Umklapp scattering is considered and when Umklapp scattering for optical modes suppressed.}
\label{fig:umklapp_suppression}
\end{figure*}

To investigate this further, we examined the effect of neglecting Umklapp scattering for optical modes on the resulting thermal conductivity. These modes were specifically selected because they exhibit the greatest overlap between the Umklapp and grain boudnary scattering contributions and are, therefore, expected to be most affected by a potential overestimation of the total scattering rates. As shown in \autoref{fig:umklapp_suppression}, suppressing the Umklapp scattering contributions from these modes in the ab-initio-derived model substantially improves the agreement with the measured thermal conductivity of the AlN transition layer. Remarkably, the resulting fit is even better than that obtained when considering grain-boundary scattering alone. The improved agreement therefore suggests that the combined effect of these scattering mechanisms may be overestimated for optical modes when their contributions are treated independently within the RTA and Matthiessen's rule. Nevertheless, suppressing Umklapp scattering for optical modes is intended solely as a diagnostic modification and does not imply that these contributions can be physically neglected. The underlying origin of the discrepancy therefore remains unresolved.

\begin{figure*}[h!]
\centering
\begin{subfigure}{0.48\textwidth}
\centering
\textbf{\Large 300 K}
\includegraphics[width=\textwidth,trim=0 0 0 0,clip]{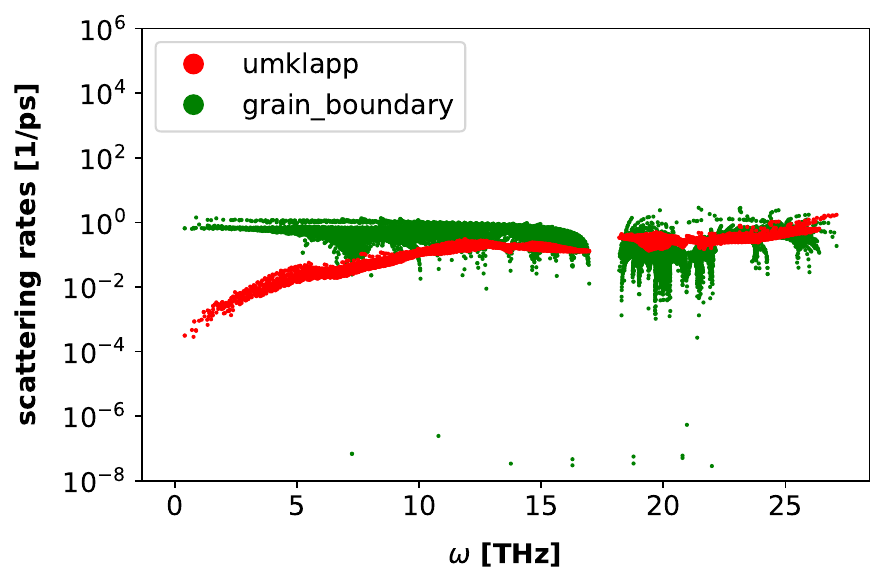}
\end{subfigure}
\hfill
\begin{subfigure}{0.48\textwidth}
\centering
\textbf{\Large 1000 K}
\includegraphics[width=\textwidth,trim=0 0 0 0,clip]{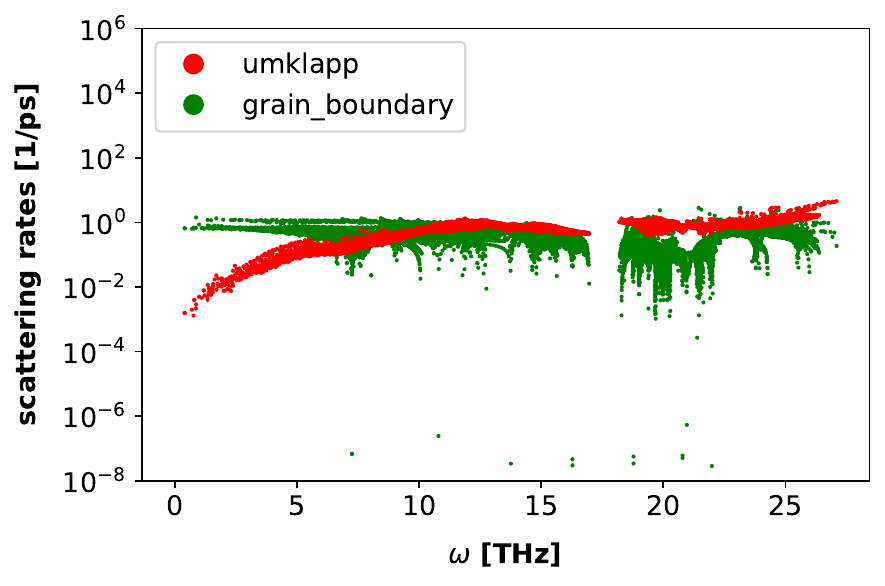}
\end{subfigure}
\caption{Scattering rates as a function of phonon frequency for the ab-initio-derived model fitted to the measured thermal conductivity of the AlN transition layer. The left and right panels show the results at 300 and 1000 K, respectively.}
\label{fig:comaprison_temperature}
\end{figure*}

As shown in \autoref{fig:comaprison_temperature}, Umklapp scattering becomes stronger at higher temperatures, whereas the grain-boundary scattering rates remain constant. This further exacerbates the issue, as the two scattering contributions become comparable in magnitude over a wider frequency range. \vspace{40pt}

\newpage
\section{Grain boundary scattering with frequency-dependent specularity}\label{sec:GB_scattering_nongray}

\vspace{10pt}

\begin{figure*}[h!]
\centering
\begin{subfigure}{0.48\textwidth}
\centering
\textbf{constant specularity}
\includegraphics[width=\textwidth,trim=0 0 0 0,clip]{plots/id762/with_umklapp/ktemp_abi.pdf}
\includegraphics[width=\textwidth,trim=0 0 0 0,clip]{plots/id762/with_umklapp/w_abi.pdf}
\end{subfigure}
\hfill
\begin{subfigure}{0.48\textwidth}
\centering
\textbf{frequency-dependent specularity}
\includegraphics[width=\textwidth,trim=0 0 0 0,clip]{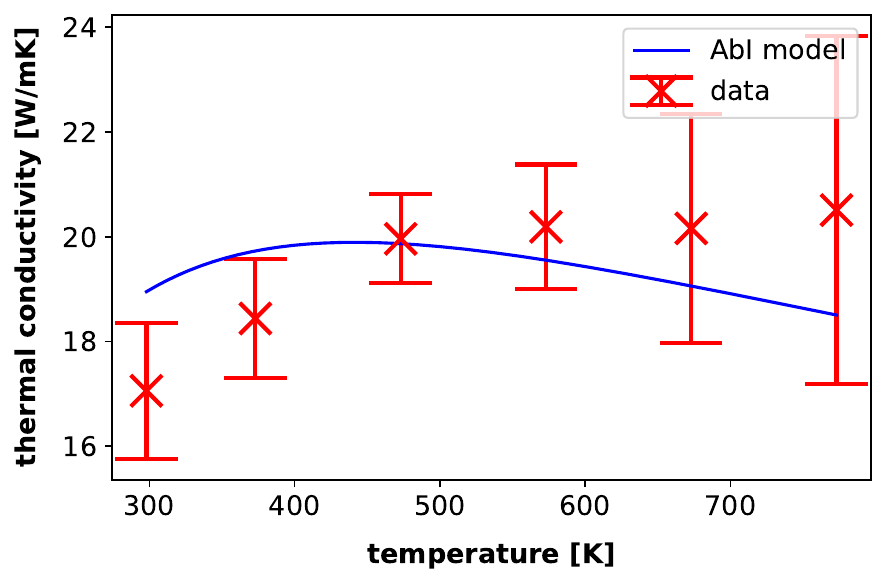}
\includegraphics[width=\textwidth,trim=0 0 0 0,clip]{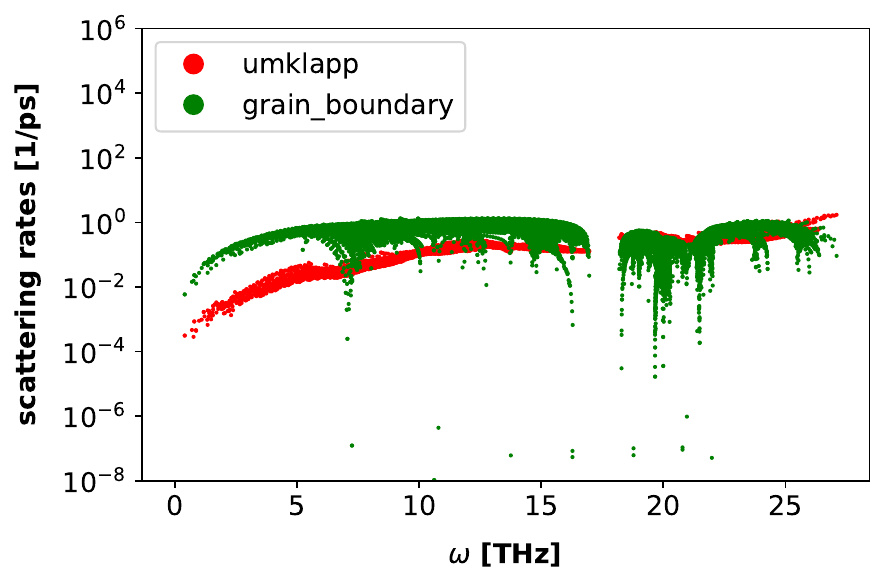}
\end{subfigure}
\caption{Comparison of the thermal conductivity predicted by the ab-initio-derived model (blue lines) for the AlN transition layer with the experimental data (red error bars). The left column shows the results obtained using a constant grain boundary specularity of $p_{\text{\tiny GB}} = 0.5$, as employed in this work, while the right column shows those obtained using a frequency-dependent specularity. The bottom row shows the corresponding phonon scattering rates as a function of frequency at 300 K.}
\label{fig:nongray}
\end{figure*}

To assess whether the difficulties in reproducing the measured thermal conductivity of the AlN transition layer are related to an inadequate description of grain boundary scattering, we also test a frequency-dependent specularity parameter proposed by Ziman \cite{Ziman2001}, which is given by

\begin{equation*}
    p\left(\vec{q}\right) = \exp\left(-4\,\eta^2\vert\vec{q}\vert^2\right),
\end{equation*}

where $\eta$ is the root-mean-square roughness of the grain boundary and $\vec{q}$ is the phonon wave vector. Using this frequency-dependent specularity parameter, we fit the ab-initio-derived model to the experimental data for the AlN transition layer (see Sec. III-B of the main text), including both Umklapp and grain boundary scattering. As demonstrated in \autoref{fig:nongray}, this approach only slightly improves the agreement with the experimental data at elevated temperatures compared to using a constant specularity ($p_{\text{\tiny GB}}=0.5$). The root-mean-square errors of the corresponding fit is $1.35\frac{W}{mK}$ (compared against $1.56\,\frac{W}{mK}$ in the case of $p_{\text{\tiny GB}}=0.5$).

\vspace{20pt}

\end{document}